\documentclass[]{article}

\RequirePackage{amsthm,amsmath,amsfonts,amssymb}
\RequirePackage{graphicx}%

\usepackage[margin=1in]{geometry}

\usepackage{bm}
\usepackage{mathtools}
\usepackage{tabularx}
\usepackage{longtable}
\usepackage[hidelinks]{hyperref}
\usepackage{float}
\usepackage{longtable}
\usepackage{natbib}
\usepackage{authblk}
\usepackage{pdfpages}
\usepackage[defaultcolor=black]{changes}

\mathtoolsset{showonlyrefs}

\defcitealias{un_desa2019}{United Nations 2019}
\defcitealias{un_desa2021}{United Nations 2021}

\newcommand{\MedE}{\mathrm{MedE}}
\newcommand{\MedAE}{\mathrm{MedAE}}

\DeclareMathOperator*{\argmax}{\arg\,\max}

\providecommand{\keywords}[1]{\textbf{\textit{Keywords: }} #1}

\title{Flexible modeling of demographic transition processes with a Bayesian hierarchical B-splines model\footnote{This is the author accepted manuscript of the published manuscript ``Susmann, H., and Alkema, L. (2025). Flexible modeling of demographic transition processes with a Bayesian hierarchical B-splines model. \textit{Journal of the Royal Statistical Society Series C:
Applied Statistics}. \url{https://doi.org/10.1093/jrsssc/qlaf026}".}}

\author[1,$\ast$]{Herbert Susmann}
\author[1]{Leontine Alkema}

\affil[1]{Department of Biostatistics \& Epidemiology, University of Massachusetts Amherst, Massachusetts, USA}

\affil[$\ast$]{\small Corresponding author. \href{email:herbps10@gmail.com}{susmah01@nyu.edu}. Current address: NYU Grossman School of Medicine, 180 Madison Avenue, New York, NY 10016
}

\date{April 7, 2025}

\begin{document}

\maketitle

\abstract{
Several demographic and health indicators, including the total fertility rate (TFR) and modern contraceptive use rate (mCPR), evolve similarly over time, characterized by a transition between stable states. Existing approaches for estimation or projection of transitions in multiple populations have successfully used parametric functions to capture the relation between the rate of change of an indicator and its level. However, incorrect parametric forms may result in bias or incorrect coverage in long-term projections. 
We propose a new class of models to capture demographic transitions in multiple populations. 
Our proposal, the \textit{B-spline Transition Model} (BTM), models the relationship between the rate of change of an indicator and its level using B-splines, allowing for data-adaptive estimation of transition functions. Bayesian hierarchical models are used to share information on the transition function between populations. 
We apply the BTM to estimate and project country-level TFR and mCPR and compare the results against those from extant parametric models. For TFR, BTM projections have generally lower error than the comparison model. For mCPR, while results are comparable between BTM and a parametric approach, the B-spline model generally improves out-of-sample predictions. The case studies suggest that the BTM may be considered for demographic applications.
}


\section*{Introduction}
Projections of demographic and health indicators are of key interest for planning, to track progress towards international targets, or as inputs for projecting other outcomes of interest. For example, long-term projections of the Total Fertility Rate (TFR) to the year 2100 are produced by the United Nations for all countries in the world to provide insights into future outcomes \citep{alkema2011probabilistic,raftery2014bayesian,liu2020,wpp2022methods}. In turn, these projections are used for constructing population projections \citep{wpp2022methods}, and population projections are subsequently used as input for other models, such as projecting warming due to climate change by 2100 \citep{raftery2017less,chen2022long,rennert2022comprehensive}. A second example is the projection of  family planning indicators such as the modern contraceptive prevalence rate (mCPR) \citep{alkema2013national, cahill2018modern, kantorova2020}. Medium term projections of family planning indicators to the year 2030 are used to evaluate progress towards reaching the United Nations Sustainable Development Goals \citep{strong2020monitoring} and to set targets \citep{cahill2020increase}. 

Transition models underlie the projections of a large set of demographic indicators. Transition models refer to models that capture transitions between stable states. Such models have been used to capture changes in indicators including the TFR and mCPR. For the TFR, this is the transition from high to low fertility referred to as the demographic transition \citep{kirk1996demographic}. The mCPR typically follows a transition from low availability and adoption of modern family planning methods to high availability and adoption. Other indicators for which transition models have been used to create projections include life expectancy \citep{raftery2013bayesian}. 
These models have been used by the international community for monitoring trends and generating projections for the indicator of interest.

Transition models aim to capture the relationship between the rate change of the indicator and its level for different populations. For example, for mCPR the rate of change starts slow at low levels of adoption, speeds up as modern methods become increasingly available, and then slows and eventually halts as demand is saturated \citep{alkema2013national, cahill2018modern}. The Family Planning Estimation Model (FPEM) of \cite{cahill2018modern} aims to capture this relation by positing that the rate of change in mCPR is related to its level by a logistic growth equation. The transition model used for projecting the TFR of \cite{alkema2011probabilistic} assumes a double logistic functional form for the relationship between rate of change and level of TFR. In both models, population-specific parameters are used to capture differences in the pace, timing, or stable states of transitions.   

A limitation of these existing transition models is that an assumption is made that the rate versus level relationship for each indicator follows a particular functional form: a double logistic form, for TFR, and a logistic form, for mCPR. While additional model components are included to allow for deviations away from systematic trends, incorrect parametric forms may result in bias or incorrect coverage, in particular for longer term projections. 

In this paper, we propose a new class of transition models in which the relation between rate of change and level is estimated flexibly, allowing for more data-adaptive estimation of transition functions as compared to existing approaches. Our proposed method, which we we refer to as a \textit{B-spline Transition Model} (BTM), models the relationship between the rate of change of an indicator and its level using B-splines \citep{deboor1978practical,eilers1996flexible}. Our approach allows for incorporating general prior knowledge on the shape of this relationship, such as monotonicity constraints, without having to specify a restrictive functional form. Rather, the specific shape of the relationship is learned from the available data, instead of being posited as a strong modeling assumption. The B-spline Transition model generalizes upon the current approaches used for projecting TFR and mCPR. 

\added{Our approach has links to several extant areas of research on time series modeling and projection. The core of the approach is based on a state-space model, that seeks to infer a latent underlying trend based on noisy observed data \citep{durbin2012statespace, triantafyllopoulos2021}. The Kalman filter is a famous example of a state-space model, with an intuitive probabilistic Bayesian interpretation \citep{kalman1960filter, meinhold1983kalman}. More generally, Bayesian forecasting based on state-space models were described in early work by \cite{harrison1971} and \cite{harrison1976}.} We also note that for modeling the latent trend, the proposed transition model can be interpreted as a stochastic difference equation, or as the discretization of a continuous-time stochastic differential equation \citep{oksendal2003stochastic, rodkina2011}. Finally, we draw heavily on applied Bayesian modeling techniques, notably by using hierarchical models to share information between countries \citep{gelman2013bda3}.

\added{We introduce the modeling framework in the context of the Temporal Models for Multiple Populations" (TMMPs) model class \cite{susmann2021temporal}. This model class was introduced to help communicate assumptions underlying models used for producing estimates of demographic and global health indicators in populations with limited data.
The TMMPs model class helps to facilitate both documentation of model assumptions in a standardised way and comparison across models. The class makes a distinction between the process model, which describes latent trends in the indicator interest, and the data model, which describes the data generating process of the observed data. The transition model we describe here falls within the TMMPs class. }  
 
This paper is organized as follows. We first introduce the TFR and mCPR case studies. The methods sections introduce transition models in general, and the proposed B-spline Transition Model. We then discuss customized B-spline Transition Models for both case studies to construct estimates and projections of TFR and mCPR. 
We validate the models through a set of out-of-sample model checks and compare the BTM results to parametric alternatives.  
We end with a discussion of findings.

\section*{Existing Approaches to estimate and project TFR and mCPR}
\label{section:family-planning-estimation}

In this section, we describe existing models for TFR and family planning indicators including mCPR. Each model makes strong functional form assumptions about the shape of  transitions, which motivates our proposed flexible modeling approach based on B-splines.

\subsection*{Total Fertility Rate: BayesTFR}
The Total Fertility Rate (TFR) indicator is defined as the average number of children a woman in a specified population would have if she lives through all of her child-bearing years and experiences the corresponding age-specific fertility rates as she ages \citep{preston2001demography}.  
A Bayesian hierarchical model, BayesTFR, exists for estimating and projecting TFR in countries \citep{alkema2011probabilistic,raftery2014bayesian,liu2020,wpp2022methods}.

\begin{figure}
    \centering
    \includegraphics[width=0.6\columnwidth]{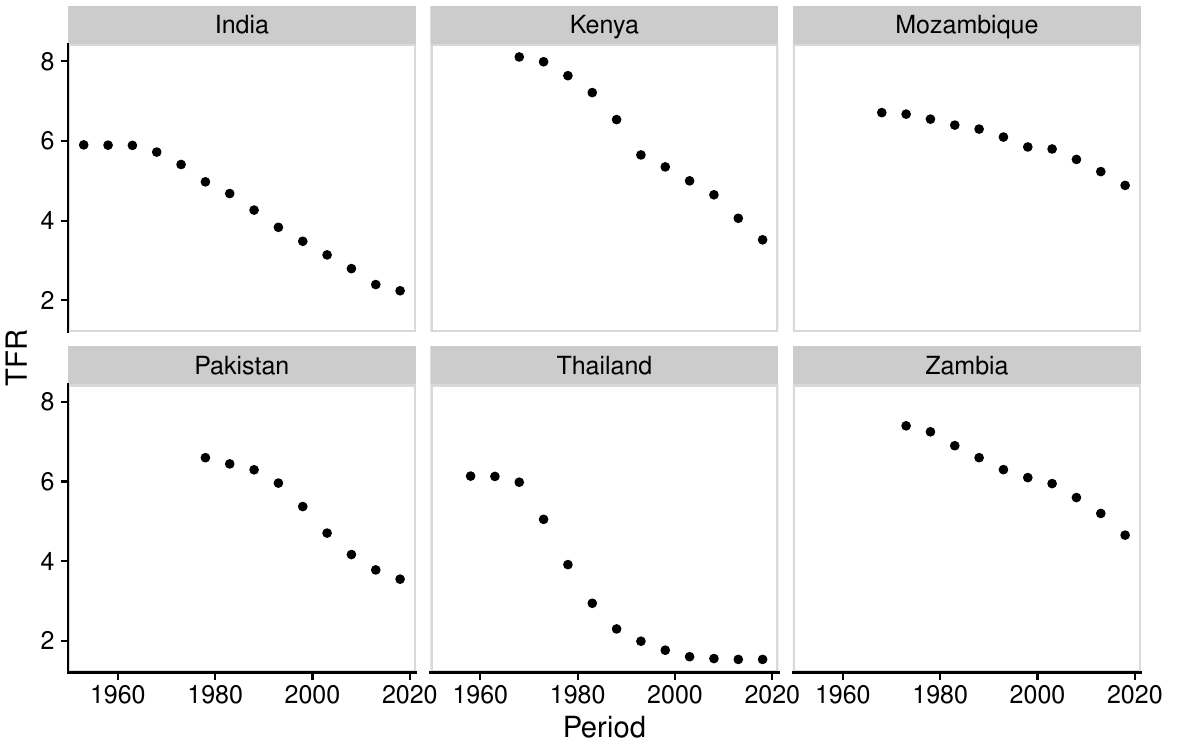}
            \caption{Total Fertility Rate (TFR) over time (5-year periods) in six selected countries.}
    \label{fig:tfr-data}
\end{figure}

The first step of BayesTFR is to apply a deterministic procedure to split the observed TFR time series data into three phases, corresponding to before (Phase I), during (Phase II), and after (Phase III) the demographic transition.  BayesTFR applies different models to each phase as TFR is assumed to evolve differently depending on where a country is in the demographic transition. We will focus on the BayesTFR model for Phase II, as it is during this phase that TFR is expected to follow a transition from high to low. Figure \ref{fig:tfr-data} shows Phase II TFR estimates from six countries, using country estimates of TFR as obtained from the BayesTFR R package \citep{bayestfr2011}, which (per May 2024) contains data from the 2019 edition of the United Nations World Population Prospects \citep{wpp2019}.

\begin{figure}
    \centering
    \includegraphics[width=0.75\columnwidth]{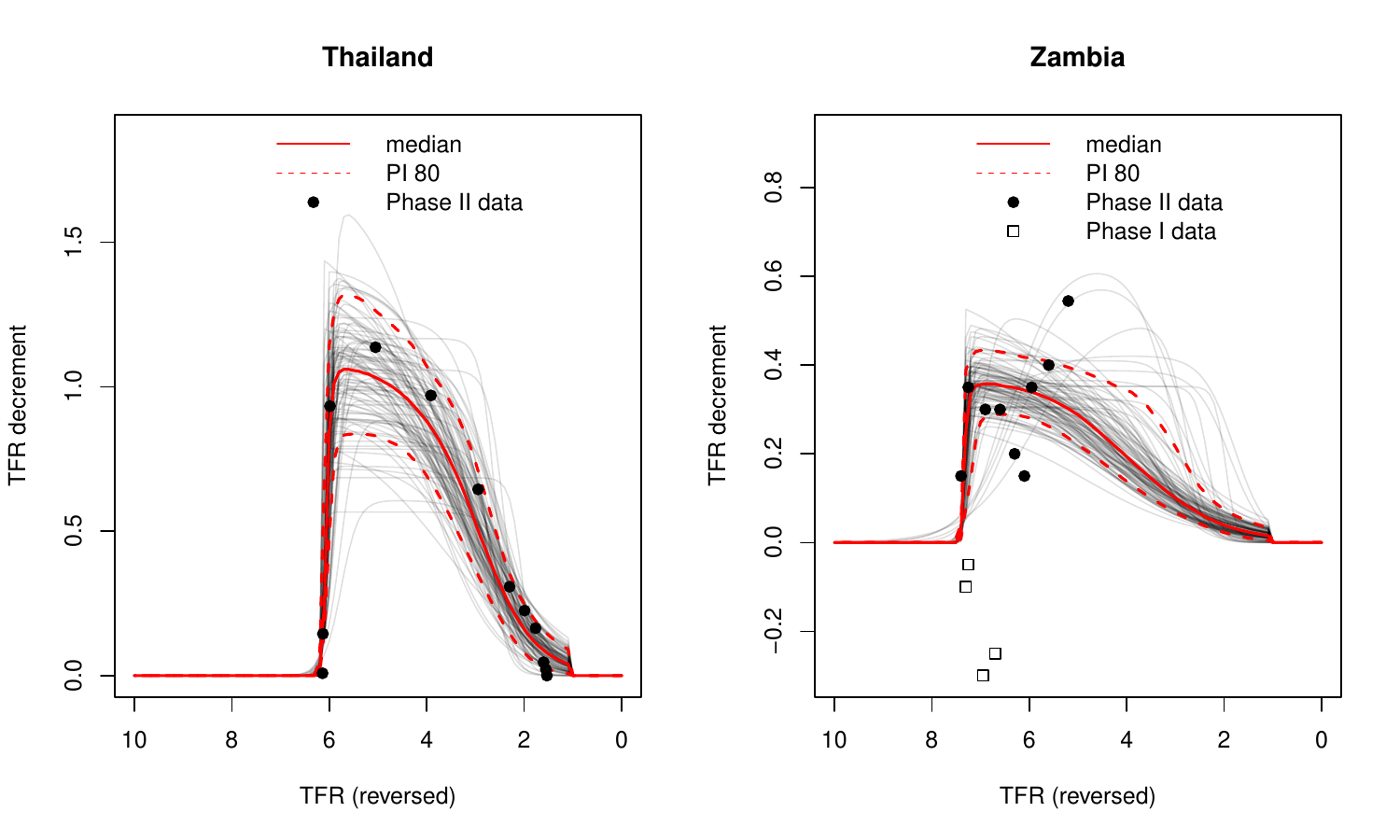}
    \caption{Fitted double logistic transition functions from the BayesTFR model plotted as the period-to-period decrement in TFR versus TFR. The black points show the empirical decrement vs. TFR. Note that the $x$-axis is reversed. }
    \label{fig:bayes-tfr-fit-example}
\end{figure}

BayesTFR models the transition (that is, the relationship between the rate of change and level of TFR) as following a double logistic function. Figure \ref{fig:bayes-tfr-fit-example} shows the posterior distribution of the double logistic curves for Thailand and Zambia from BayesTFR, together with the decrements observed in the TFR estimates from Figure~\ref{fig:tfr-data}. The data from Thailand appear to be reasonably well fit by the model. In Zambia, however, the suitability of the logistic curves is less clear; observed rates of change suggest that decrements have increased in recent years. This pattern is not captured in the median summary of the double logistic functions (displayed in red).
This observation motivates our main contribution, which is to weaken the functional form assumption by modeling the transition using flexible modeling techniques.

\subsection*{Modern Contraceptive Prevalence Rate: the Family Planning Estimation Model}
The Modern Contraceptive Prevalence Rate (mCPR) indicator for married or in-union women is defined as the proportion of married or in-union women of reproductive age within a population who are using (or whose partner is using) a modern contraceptive method. Data on mCPR is typically derived from surveys in which participants report whether they or their partner are currently using a modern contraceptive method. Such surveys may be conducted by international organizations or by local organizations or governments. The UN Population Division has compiled a database of mCPR surveys in countries, subnational regions, and territories \citepalias{un_desa2021}. We will refer to the geographic units included in the database generically as ``countries", although the data includes several subnational regions or territories. Data sources include the Demographic and Health Survey Program (DHS), Performance Monitoring for Action (PMA), UNICEF Multiple Indicator Cluster Surveys (MICS), and other national and international surveys. The UNPD preprocesses the microdata from each survey to yield country-level estimates of mCPR for the time period covered by the survey, taking into account the complex sampling design of each survey. The preprocessing may also yield a measure of the uncertainty in the mCPR estimate owed to the survey design, which we refer to as \textit{sampling error}.

\begin{figure}
    \centering
    \includegraphics[width=0.75\columnwidth]{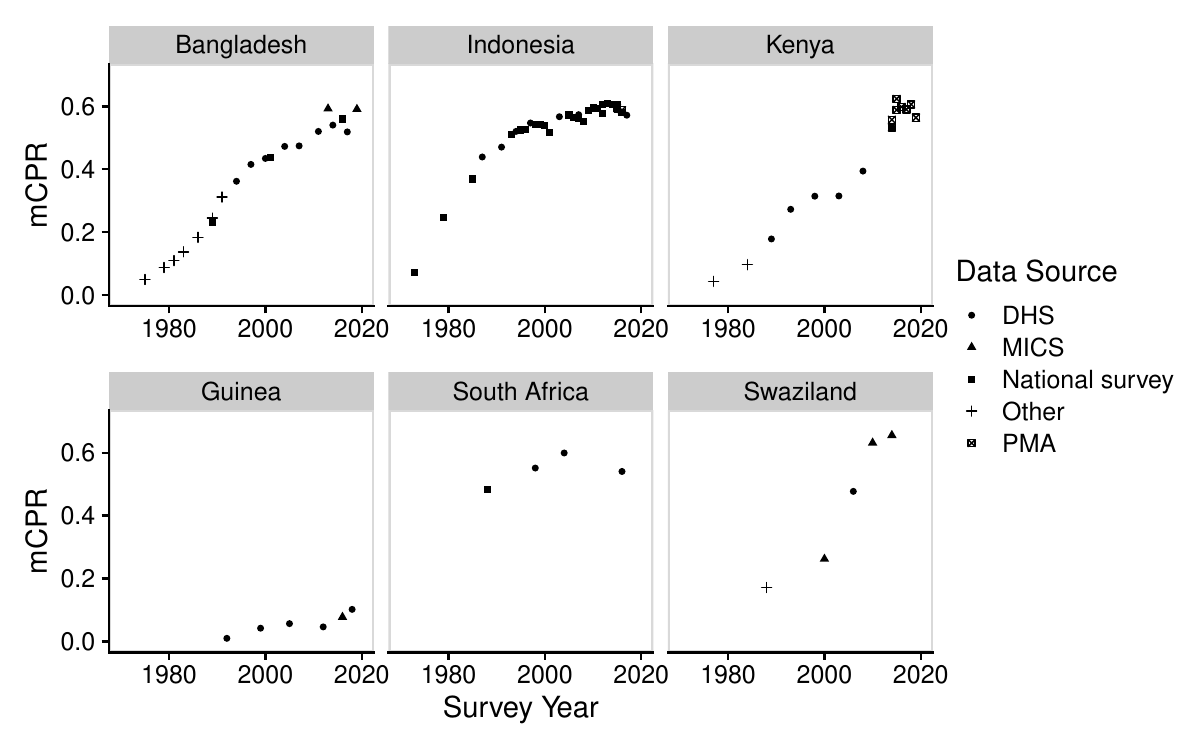}
    \caption{Observations of mCPR in countries with relatively high data availability (top row) vs. relatively low availability (bottom row). }
    \label{fig:data-availability}
\end{figure}

One of the primary challenges in modeling mCPR is posed by the wide range of data availability across countries. In the analysis dataset, availability ranged from a high of $n=34$ observations for Indonesia and a low of $n=1$ observations for $23$ areas. Figure \ref{fig:data-availability} shows data for six areas, including 3 with relatively high availability (Bangladesh, Indonesia, Kenya) and 3 with relatively lower availability (Guinea, South Africa, Swaziland). 

When data availability is high, we are able to see trends in the raw survey data that illustrate the temporal trends that are characteristic of a transition process in which mCPR transitions from between two stable states. In Bangladesh and Indonesia in particular there are sufficient data to see most of the transition from low to high mCPR (Figure \ref{fig:data-availability}). In Indonesia, a plateau in adoption can be seen, and in Bangladesh the data appear to follow a roughly S-shaped curve. In terms of the rate of change of mCPR vs. its level, in Bangladesh we can see that the rate of change is low at low levels, is larger at medium levels, and then becomes smaller again at higher levels as mCPR reaches an asymptote. Kenya provides an example of a country that does not appear to follow an S-shaped transition as neatly, as a temporary stall can be seen around the year 2000.

The Family Planning Estimation Model (FPEM) is a Bayesian hierarchical model that has been used to generate projections of family planning indicators, including mCPR \citep{alkema2013national, cahill2018modern}. FPEM draws on the observed transitions in family planning indicators over time in its modeling approach. Specifically, FPEM produces transitions in mCPR based on the assumptions of (i) logistic growth in use of any method of contraceptives (modern as well as traditional methods) and (ii) logistic growth in the ratio of mCPR over total use. The model is fitted to data on mCPR and traditional contraceptive use, accounting for correlation between the two indicators. Given the focus of this work on transition models, we discuss a simplified version of FPEM, using its assumption of logistic growth for capturing the transition in mCPR directly. This simplification is accurate for many countries with low levels of use of traditional methods. Assuming a logistic growth equation to capture long term changes in mCPR corresponds to assuming that the rate of change of mCPR as a function of its level is given by a parametric function derived from a logistic growth equation. This logistic growth assumption at the core of FPEM may not always be justified. In particular, the symmetry of logistic growth implies that the pace of the contraceptive transition at the beginning of the transition is the same as at the end of the transition. Practically, this means that, modulo other model components that capture other dynamics, such as stochastic smoothing components, countries starting their transition are locked into a pace that carries forward into projections. While this pace may be informed by those of other countries farther in their transition via hierarchical models (a strength of the Bayesian hierarchical modeling approach), the model nevertheless imposes a symmetry assumption on the level of the transition function.

\section*{Transition Model}
\label{section:transition-model-definition}
We begin by formalizing the inferential problem. Let $\eta_{c,t} \in \mathcal{E}$ be the true, latent value of an indicator in country $c$ at time $t$. The set $\mathcal{E}$ is left arbitrary for full generality; it is typically $(0, 1)$ or $[0, 1]$ for indicators that are proportions. We seek to estimate $\eta_{c,t}$ for countries $c = 1, \dots, C$ and time points $t = 1, \dots, T$ conditional on observations $y_i$, $i = 1, \dots, n$, where $y_i$ is an observation of the indicator in country $c[i]$ and time $t[i]$.

Next, we introduce a model for indicators that are assumed to follow a transition process: that is, indicators assumed to follow a transition between two stable states, and where the rate of change is assumed to be a function of level. Following the framework of Temporal Models for Multiple Populations (TMMPs), we separate the model into a \textit{process model} and a \textit{data model}  \citep{susmann2021temporal}.
The process model defines the evolution of $\eta_{c,t}$ over time and the data model encodes the relationship between $\eta_{c,t}$ and observations $y_i$. In this section, we propose a process model based on estimating the relationship between rate and level with B-splines. Where applicable, we adopt the same notation as the TMMP model class. 

Formally, the process model is given by
\begin{align} \label{eq:transition-systematic-component}
    g_1(\eta_{c,t}) = \begin{cases}
        \Omega_c, & t = t^*, \\
        g_1(\eta_{c,t-1}) + f(\eta_{c,t-1}, \bm{\lambda}_c, \bm{\beta}_c) + \epsilon_{c,t}, & t > t^*, \\
        g_1(\eta_{c,t+1}) - f(\eta_{c,t+1},\bm{\lambda}_c, \bm{\beta}_c) - \epsilon_{c,t + 1}, & t < t^*,
    \end{cases}
\end{align}
where $g_1 : \mathcal{E} \to \mathbb{R}$ is a deterministic link function,  $\Omega_c \in \mathbb{R}$ is the level of the indicator at a fixed reference year $t_c^*$, and $\epsilon_{c,t}$ is a residual. The function $f$ is a \textit{transition function} depending on parameters $\bm{\lambda}_c \in \mathcal{E}^2$ and $\bm{\beta}_c \in \mathbb{R}^p$. The transition function $f$ predicts the rate of change of an indicator (on a possibly transformed scale, depending on $g_1$) as a function of its level (on the original scale). The parameter vector $\bm{\lambda}_c = \{ \lambda_c^l, \lambda_c^u \}$, $\lambda_c^u > \lambda_c^l$, controls the asymptotes of the systematic component: that is, $\lambda_c^l$ is the minimum $\eta_{c,t}$ and $\lambda_c^u$ the maximum $\eta_{c,t}$ attainable through the systematic component. The parameter $\bm{\beta}_c$ controls the rate of the transition. 

We introduce an additional  constraint that $f$ is either non-negative or non-positive; that is, $f(\eta, \bm{\lambda}_c, \bm{\beta}_c) \geq 0$ or $f(\eta, \bm{\lambda}_c, \bm{\beta}_c) \leq 0$ for all $\eta \in \mathcal{E}$. This constraint ensures that 
$f$ encodes a transition between two stable states. The manner in which $\eta_{c,t}$ moves between these two states will be determined by the particular form of $f$.

The deviations $\epsilon_{c,t}$ are intended to incorporate data-driven trends in the observed data that are not well captured by the systematic component. To prevent overfitting, a model is placed on the deviations to enforce some degree of smoothness in the model fit and referred to as a stochastic smoothing model. Aligned with the definition of TMMPs, we consider smoothing models in which the deviation terms (or some linear combination of its differenced version) is assumed to be normally distributed with mean zero \citep{susmann2021temporal}.
The choice of smoothing model will likely depend on the indicator of interest, so the general class of transition models given by 
\eqref{eq:transition-systematic-component} does not make additional a-priori restrictions on $\epsilon_{c,t}$.

\subsection*{Logistic Transition Model}
\begin{figure}
    \centering
    \includegraphics[width=0.65\columnwidth]{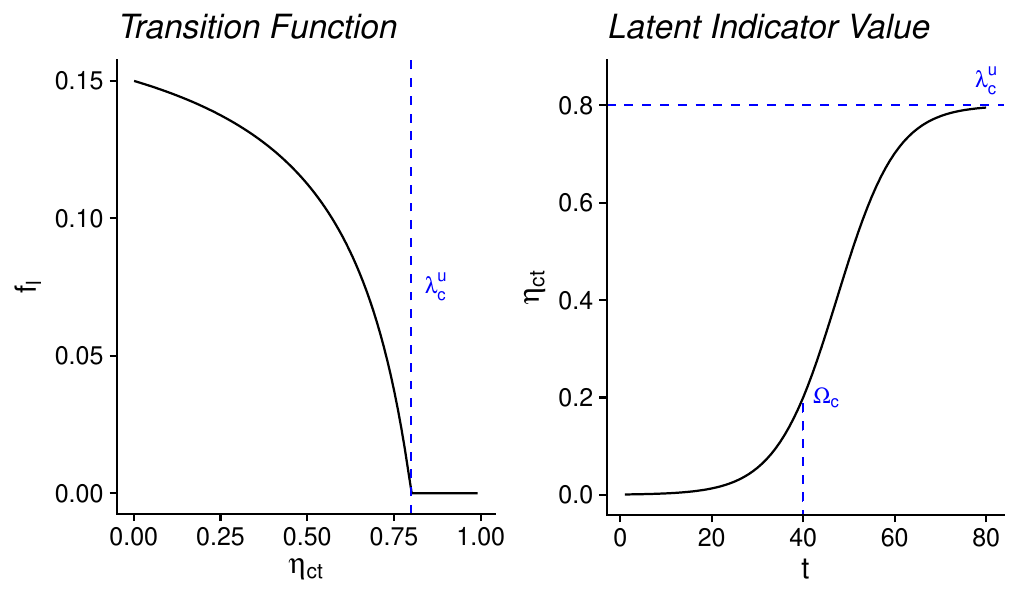}
    \caption{Left: an example of a logistic transition function $f_{l}$ for capturing a transition from zero to high $\eta_{c,t}$. Right: the evolution of $\eta_{c,t}$ over time that follows from the pictured transition function. The parameter $\Omega_c$ controls the level of the indicator at a reference time point and $P_c^u$ controls the asymptote. }
    \label{fig:fpem-transition-function}
\end{figure}

One way of defining the transition function $f$ is to choose a particular parametric form. For example, if the transition is assumed to follow a logistic growth curve, starting at zero and reaching an asymptote $P_c^u$, then a logistic transition function $f_l$ can be chosen:
\begin{align}
    \label{eq:logistic-transition-function}
    f_{l}(\eta_{c,t}, \bm{\lambda}_c,
    \bm{\beta}_c) = \begin{cases}
     \mathrm{logit}\left( \lambda^u_c \cdot \mathrm{logit}^{-1}\left( \mathrm{logit}\left(\frac{\eta_{c, t}}{\lambda^u_c}  \right) + \omega_c \right) \right) - \mathrm{logit}(\eta_{c,t}), & \eta_{c,t} < \lambda^u_c \\
       0, & \text{otherwise,}
    \end{cases}
\end{align}
where $\bm{\beta}_c = \{ \omega_c \}$ for $\omega_c \geq 0$, the pace of the logistic curve. The form of $f_l$ can be derived algebraically as the rate of change on the logit scale of logistic growth as a function of level, and it can be seen that $f_l$ satisfies the non-negativity constraint. This form implies an asymptote in the transition when $\eta_{c,t} = \lambda_c^u$ and logistic growth when $\eta_{c,t} < \lambda^u_c$. An example logistic transition function is shown in Figure \ref{fig:fpem-transition-function}. The Family Planning Estimation Model (FPEM) is an example of a transition model that uses this logistic transition function \citep{cahill2018modern}.

\label{section:double-logistic-transition-model: BayesTFR}
\subsection*{Double Logistic Transition Model}
\begin{figure}
    \centering
    \includegraphics[width=0.75\columnwidth]{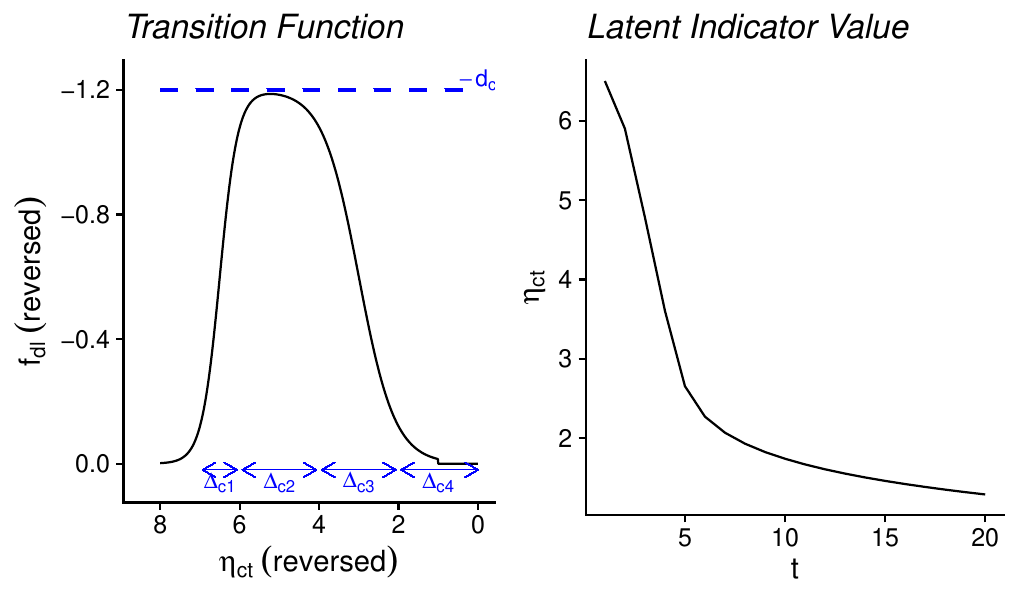}
    \caption{Left: an example of the double logistic transition function $f_{dl}$ for capturing a transition from high to low TFR $\eta_{c,t}$. Note that the $x$-axis and $y$-axis have been reversed, following convention. Right: the evolution of $\eta_{c,t}$ over time that follows from the pictured transition function. }
    \label{fig:bayestfr-transition-function}
\end{figure}

The BayesTFR model provides a second example of a transition function defined by a specific functional form. Define a transition function $f_{dl}$ as
\begin{align}
    f_{dl}(\eta_{c,t}, \bm{\lambda}_c, \bm{\beta}_c) =& \begin{cases}
        \frac{-d_c}{1+\exp\left( -\frac{2\ln(9)}{\Delta_{c,1}} (\eta_{c,t} - \sum_i \Delta_{c,i} + 0.5 \Delta_{c,1}) \right)} \\ \qquad + \frac{d_c}{1+\exp\left( -\frac{2\ln(9)}{\Delta_{c,3}} (\eta_{c,t} - \Delta_{c,4} - 0.5 \Delta_{c,3}) \right)}, & \eta_{c,t} > 1, \\
        0, & \text{otherwise},
    \end{cases}
\end{align}
where $\bm{\beta}_c = \left\{ d_c, \Delta_{c,1}, \Delta_{c,2}, \Delta_{c,3}, \Delta_{c,4} \right\}$. The parameter $d_c$ controls the maximum decrement in TFR during the transition, and the parameters $\Delta_{c,1}, \dots, \Delta_{c,4}$ determine the shape of the double logistic curve. An example double logistic transition function is shown in Figure \ref{fig:bayestfr-transition-function}.

\label{section:b-spline-transition-model}
\subsection*{B-spline Transition Model}
Modeling $f$ using flexible modeling techniques allows capturing a wide variety of transitions.
In this section we propose a flexible form for $f$ based on B-splines  \citep{deboor1978practical,eilers1996flexible, kharratzadeh2017splines}. A B-spline function is defined by choosing a sequence $t_1 < t_2 < \cdots < t_K$ of knots and a spline degree $d$. Then the B-spline function is defined as a linear combination of basis functions $B_{d, j}(x)$, where $d$ is the basis degree and $j = 1, \dots, K + d - 1$. The basis functions are defined recursively, with $d=0$ yielding piecewise constant bases:
\begin{align}
    B_{0, j}(x) = \begin{cases}
      1, & t_j \leq x \leq t_{j+1}, \\
      0, & \text{otherwise}.
    \end{cases}
\end{align}
Higher degree basis functions are defined recursively on an extended sequence of knots, in which $t_1$ is repeated $d$ times at the beginning of the sequence, and $t_K$ is repeated $d$ times at the end of the sequence. Then 
\begin{align}
    B_{d, j}(x) = w_{d, j} B_{d - 1, j}(x) + (1 - w_{i+1, k}) B_{d - 1, j + 1}(x),
\end{align}
where the weights are given by
\begin{align}
    w_{d, j} = \begin{cases}
        \frac{x - t_j}{t_{j + d - 1} - t_j}, & t_j \neq t_{j + d - 1}, \\
        0, & \text{otherwise.}
    \end{cases}
\end{align}
The smoothness of the B-spline function is controlled by the degree of basis functions (higher degrees yields a smoother functions) and the number of knots (fewer knots yields a smoother function). A degree $d=2$ is the smallest degree for which the spline function has first-order derivatives that exist everywhere, because $d=0$ leads to a piecewise constant function and $d=1$ leads to a piecewise linear function.

We define the B-spline transition function as
\begin{align}
    \label{eq:f-spline}
    f_b(\eta_{c,t}, \bm{\lambda}_c, \bm{\beta}_c) = \sum_{j=1}^{J}  h_j(\beta_{c, j}) B_{d,j}\left(\frac{\eta_{c,t} - \lambda_c^l}{\lambda_c^u - \lambda_c^l} \right),
\end{align}
where $B_{d,j}(\cdot)$ are spline basis functions as defined above, $h_j(\beta_{c,j})$ refers to the $j$th spline coefficient for area $c$, where $h_j : \mathbb{R} \to \mathbb{R}^+$ is a transformation of the parameter $\beta_{c,j}$, and parameters $\lambda_c^l, \lambda_c^u \in \mathcal{E}$ are asymptote parameters that can be used to control the minimum and maximum of the range of $\eta_{c,t}$ values for which the rate of change is non-zero. The flexibility of $f_b$ can be tuned through the number of knots and the degree of the spline basis functions. The spline coefficients $\bm{\beta}_c$ control the shape of the transition function.

\begin{figure}
    \centering
    \includegraphics[width=0.40\columnwidth]{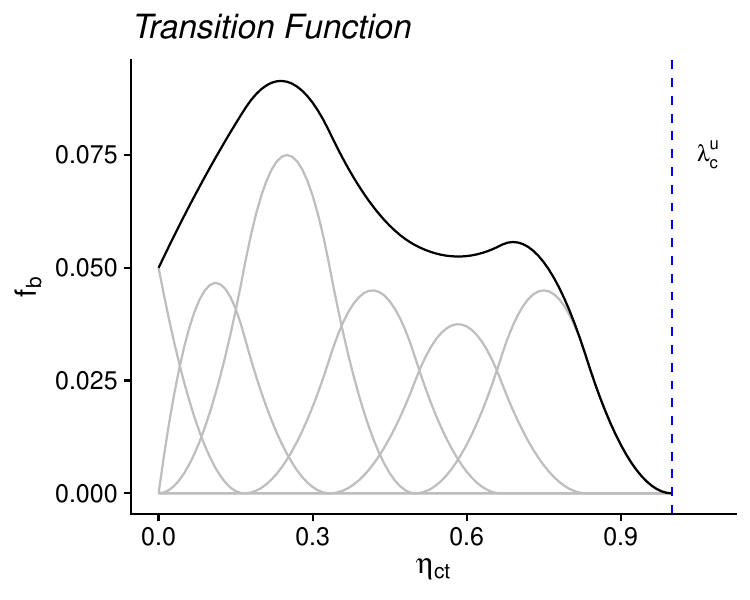}
    \includegraphics[width=0.40\columnwidth]{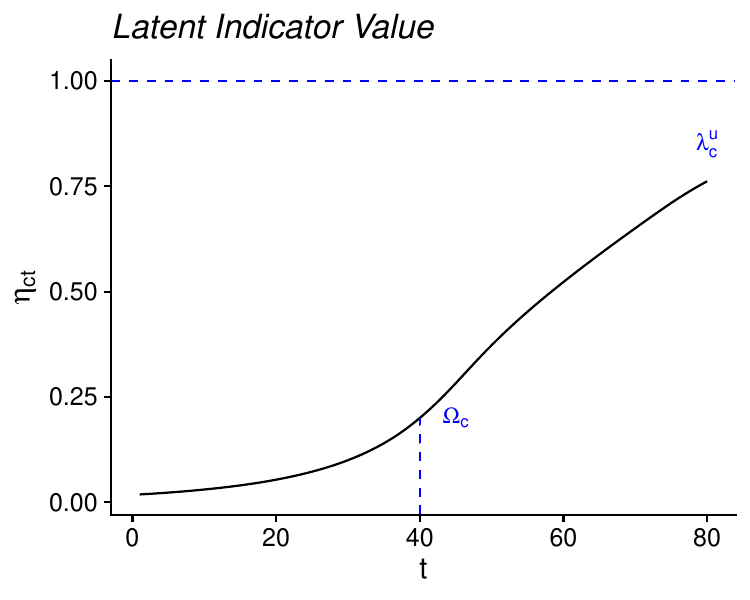}
    \caption{Left: an example of the B-spline transition function $f_b$ for capturing a transition from zero to high mCPR $\eta_{c,t}$. The gray lines illustrate a scaled version of the non-zero spline basis functions. Right: the evolution of $\eta_{c,t}$ over time that follows from the pictured transition function. The parameter $\Omega_c$ controls the level of the indicator at a reference time point, and the parameter $P_c$ controls the asymptote.}
    \label{fig:mcpr-transition-function}
\end{figure}

An example B-spline transition function is shown in Figure \ref{fig:mcpr-transition-function}. For $f_b$ to be a well-defined transition function it is necessary that $f_b(\eta, \bm{\lambda}_c, \bm{\beta}_c) \geq 0$ (or $\leq 0$) for all $\eta \in \mathcal{E}$. This constraint can be satisfied through careful choice of the functions $h_j$. For example, it is sufficient to set $h_j(x) = \exp(x)$ for all $j$ to obtain $f_b(\eta, \bm{\lambda}_c, \bm{\beta}_c) \geq 0$.

If we set constraints on the spline coefficients such that $f_b$ is positive when $\eta_{c,t} / \lambda^u_c < 1$ and zero when $\eta_{c,t} / \lambda^u_c \geq 1$ then the transition function dictates that the indicator will increase and have an asymptote at $\eta_{c,t} = \lambda_c^u$. 

The logistic transition function  $f_l$ \eqref{eq:logistic-transition-function} serves as a useful point of comparison to the B-spline transition function $f_b$. Because $f_l$ is monotonically decreasing, it cannot capture accelerations or stalls in the transition\added{; such accelerations or stalls would need to be captured using a stochastic smoothing component, as is done in the FPEM model}. In contrast, $f_b$ need not have such a monotonicity constraint, although one can be imposed if desired. In addition, a consequence of the logistic growth assumption is that the rate of change given by $f_l$ is symmetric around the midpoint of the transition. The B-spline function $f_b$, on the other hand, does not necessarily exhibit any such symmetry. 

The B-spline transition function can be tailored to approximate any parametric transition function. To do so, the B-spline coefficients can be set in such a way that the resulting function $f_b$ approximates the parametric function. As such, any parametric transition function can be thought of as a special case of the B-spline transition function (in an approximate sense). In Appendix~\ref{section:appendix-logistic} we present the approximation of the logistic transition function with a B-spline representation.  

\subsubsection*{Computation}
 We have developed subroutines in Stan \citep{stan2019, cmdstanr2022} that can be used as building blocks for implementing custom B-spline Transition Models. At its core is a custom B-spline implementation based on De Boor's algorithm \citep{deboor1971subroutine}. Sampling from the posterior distribution of model parameters is performed using Hamiltonian Markov Chain Monte-Carlo (MCMC) and the No-U-Turn sampler \citep{hoffman2014no}.
 We synthesized our software into an open-source package for fitting B-spline Transition Models called \texttt{BayesTransitionModels} (\url{https://github.com/AlkemaLab/BayesTransitionModels}) implemented in the R programming language \citep{r2022}. An additional feature of the software is that hierarchical structures on model parameters $\bm{\lambda}_c$ and $\bm{\beta}_c$ can be easily incorporated. Additional R scripts for reproducing the results of this paper are available at \url{https://github.com/herbps10/spline_rate_model_paper}.

\section*{B-spline Transition Model for TFR}
 Model specifications related to the use of data and the phases associated with the TFR transitions followed those made by \citep{alkema2011probabilistic}. Let $\eta_{c,t} \in \mathbb{R}^+$ be the true (latent) TFR in country $c$ at time period $t$, with countries and time periods indexed by $c = 1, \dots, C$ and $t = 1, \dots, T$. As per \cite{alkema2011probabilistic} specifications, United Nations Population Division (UNPD) TFR estimates are assumed to equal the true TFR. The process model is given by 
\begin{align}
    \eta_{c,t} = \begin{cases}
        \Omega_c, & t = t_c^*, \\
        \eta_{c, t - 1} + f_b(\eta_{c,t-1}, \bm{\lambda}_c, \bm{\beta}_c) + \epsilon_{c,t}, & t > t_c^*, \\
        \eta_{c, t + 1} - f_b(\eta_{c,t+1}, \bm{\lambda}_c, \bm{\beta}_c) - \epsilon_{c,t+1}, & t < t_c^*, \\
    \end{cases}
\end{align}
The reference year $t_c^*$ is set to the first time point of Phase II in country $c$, and the parameter $\Omega_c$ is fixed to the value of the first observed data point in Phase II. The upper asymptote is set to $\lambda^u_c = \Omega_c$ and the lower to $\lambda^l_c = 1$. That is, the transition is assumed to start at the level of the first data point in Phase II and is lower bounded by a TFR of 1.

We set up the splines to encode the assumption that the demographic transition is from high to low TFR, and that the rate of change of TFR should approach zero smoothly when TFR approaches one. The implementation is as follows: 
\begin{itemize}
\item We chose $K + 1$ knots of order $d$ such that the first knot is at $-\infty$ (for computation, it is set to -100) and the remaining $K$ knots are spaced evenly between 0 and 1.
\item $f_b(0, \bm{\lambda}_c, \bm{\beta}_c) = 0$ by constraining the first $d + 1$ coefficients to be zero: $h_0(x), \dots, h_{d+1}(x) = 0$ and $h_{J - d - 1}(x), \dots, h_{J}(x) = 0$.
    \item $f_b$ is constrained to be negative by constraining the middle $J - d - 1$ coefficients to be between $-0.01$ and $-2.49$:
    \begin{align}
        h_j(x) = -0.01 - 2.49 \cdot \mathrm{logit}(x), j = d + 1, \dots, J.
    \end{align}
\end{itemize}

\subsection*{Hierarchical Models}
Aligned with BayesTFR, a hierarchical model is applied to the transition parameters $\bm{\beta}_c$ such that information on the rate and shape of the transition is shared between countries:
\begin{align}
    \label{eq:tfr-beta-hierarchical-model}
    \beta_{c,j} \mid \beta^{(r)}_{s[c], j}, \sigma_{\beta, j}^{(c)} &\sim N\left(\beta^{(w)}_{j}, \left(\sigma_{\beta, j}^{(c)}\right)^2\right).
\end{align}
That is, each $\beta_{c,j}$, $j = 1, \dots, J$ is normally distributed around a world mean $\beta_j^{(w)}$ with standard deviation $\sigma_{\beta,j}^{(c)}$.
The following priors are used for the hyperparameters:
\begin{align*}
    \sigma^{(c)}_{\beta,j} &\sim N_+\left(0, 1\right), \\
    \beta^{(w)}_j &\sim N\left(0, 1\right),
\end{align*}
where $j = d, \dots, J - d - 1$.

\subsection*{Smoothing Component}
Again aligned with BayesTFR, a white noise smoothing model is assumed: $\epsilon_{c,t} \sim N(0, \tau)$, where $\tau$ is given the prior $\tau \sim N_+(0, 1)$.

\subsection*{Data}
\label{section:tfr-data}
We use data from the 2019 United Nations World Population Prospects \citep{wpp2019}. These data can be found in the BayesTFR R package \citep{bayestfr2011}. The analysis dataset included $n = 2231$ TFR observations from $C = 211$ UN divisions (referred to generically as countries). Formally, the observed data denoted $y_i$, $i = 1, \dots, n$ for the estimated TFR in country $c[i]$ and period $t[i]$. 

\section*{B-spline Transition Model for mCPR}
\label{section:b-spline-transition-model-for-tfr-mcpr}

In this section, we show how to apply our proposed modeling approach to build a transition model for mCPR. 
The description in this section focuses on the process model, with information on the data and data model given in section \ref{section:mcpr-data}. The full model specification is given in Appendix A.

Let $\eta_{c,t} \in (0, 1)$ be the true (latent) mCPR in country $c$ at time $t$, with countries and years indexed by $c = 1, \dots, C$ and $t = 1, \dots, T$. Let $s[c]$ index the subregion of area $c$ and $r[s]$ index the region of subregion $s$ (with $S$ total subregions and $R$ total regions). The process model is given by the Transition Model \eqref{eq:transition-systematic-component}, with the choice of transformation $g_1 = \mathrm{logit}$ and the choice of the B-spline Transition Function $f_b$ yielding
\begin{align}
    \label{eq:mcpr-transition-systematic-component}
  \mathrm{logit}(\eta_{c,t}) = \begin{cases}
        \Omega_c, & t = t_c^*, \\
        \mathrm{logit}(\eta_{c, t - 1}) + f_b(\eta_{c,t-1}, \bm{\lambda}_c, \bm{\beta}_c) + \epsilon_{c,t}, & t > t_c^*, \\
        \mathrm{logit}(\eta_{c, t + 1}) - f_b(\eta_{c,t+1}, \bm{\lambda}_c, \bm{\beta}_c) - \epsilon_{c,t+1}, & t < t_c^*. \\
    \end{cases}
\end{align}
and $f_b$ is defined as before \eqref{eq:f-spline}. The reference year $t^*_c$ is set to 1990. Because we know that mCPR starts at zero in all countries, the parameter $\lambda_c^l$ is fixed to zero. The following splines set up is introduced to ensure that $f_b$ is non-negative and induces an asymptote at $\lambda^u_c$:
\begin{itemize}
\item We chose $K + 1$ knots of order $d$ such that the first $K$ knots are spaced evenly between $0$ and $1$ and the final knot is at $+\infty$ (for computation, it is set to $1000$). 

    \item The first $J - d - 1$ coefficients are constrained to be positive:
    \begin{align}
        h_{j}(x) &= 0.01 + 0.29 \cdot \mathrm{logit}^{-1}(x), j = 1, \dots, J - d - 1.
    \end{align}
    The upper bound is introduced only to improve the computational performance of the model, and is chosen such that the particular value of the bound is not informative of the model results. The lower bound on the first $J-d-1$ coefficients is introduced so that $\lambda^u_c$ is identifiable. Without a lower bound, spline coefficients at or near zero will induce an asymptote below $\lambda^u_c$ and $\lambda^u_c$ will not be well identified. 
    \item The final $d + 1$ coefficients are constrained to be zero: $h_J(x)  = h_{J - 1}(x) = \cdots = h_{J - d}(x) = 0$. This implies that $f_b(\eta,  \bm{\lambda}_c, \bm{\beta}_c) = 0$ for all $\eta \geq \lambda^u_c$ and that $f^\prime_b(x, \bm{\lambda}_c, \bm{\beta}_c) = 0$.
\end{itemize}
Given these constraints, the parameter $\lambda^u_c$ can be interpreted as the upper asymptote of mCPR in country $c$. In other words, $\lambda^u_c$ is the highest value that $\eta_{c,t}$ can achieve solely through the systematic component. An example of the B-spline transition function with the stated constraints is shown in Figure \ref{fig:mcpr-transition-function}.

\subsection*{Hierarchical Models}
Estimating the spline coefficients may be difficult in countries where \added{few} data are available. We make use of hierarchical models to share information on the shape of the transition function between countries, such that countries with few observations can borrow information from countries where more data are available. We follow the hierarchical structures used in \cite{cahill2018modern}. 

The following hierarchical model is set on the spline coefficients $\beta_{c,j}$ for $j = 1, \dots, J$:
\begin{align}
    \label{eq:beta-hierarchical-model}
    \beta_{c,j} \mid \beta^{(r)}_{s[c], j}, \sigma_{\beta, j}^{(c)} &\sim N\left(\beta^{(s)}_{s[c], j}, \left(\sigma_{\beta, j}^{(c)}\right)^2\right), \\
    \beta^{(s)}_{s,j} \mid \beta^{(r)}_{r[s], j}, \sigma_{\beta, j}^{(s)} &\sim N\left(\beta^{(r)}_{r[s], j}, \left(\sigma_{\beta, j}^{(s)}\right)^2\right), \\
    \beta^{(r)}_{r,j} \mid \beta^{(w)}_{j}, \sigma_{\beta, j}^{(r)} &\sim N\left(\beta^{(w)}_{j}, \left(\sigma_{\beta, j}^{(r)}\right)^2\right).
\end{align}
That is, the country-specific spline parameters $\beta_{c,j}$ are distributed around a subregion mean $\beta^{(s)}_{s[c], j}$, the subregion mean is distributed around a region mean $\beta^{(r)}_{r[s], j}$, and the regional mean is distributed around a world mean $\beta^{(w)}_{j}$. The standard deviations at the country, subregion, and region level are given by $\sigma^{(c)}_{\beta, j}$, $\sigma^{(s)}_{\beta, j}$, and $\sigma^{(r)}_{\beta, j}$. In this way, the hierarchical variance is allowed to differ for each coefficient. 

In addition, a hierarchical model on $\lambda^u_c$ is used to share information about the mCPR asymptote between countries, with a transformation such that $\lambda_c$ is constrained to be between 0.5 and 0.95 \citep{cahill2018modern}. Let
\begin{align}
    \lambda^u_c = 0.5 + 0.45 \cdot \mathrm{logit}^{-1}\left( \tilde{\lambda}^u_c \right),
\end{align}
where $\tilde{\lambda}^u_c \in \mathbb{R}$. Then a hierarchical model is applied to $\tilde{\lambda}^u_c$
\begin{align}
   \tilde{\lambda}^u_c \mid \tilde{\lambda}^{u,(w)}, \sigma_{\tilde{\lambda}}^{(c)} &\sim N\left(\tilde{\lambda}^{u,(w)},  \left(\sigma^{(c)}_{\tilde{\lambda}^u}\right)^2\right),
\end{align}
such that $\tilde{\lambda}^u_c$ is distributed around a world mean $\tilde{\lambda}^{u,(w)}$ with standard deviation $\sigma_{\tilde{\lambda}}^{u,(c)}$. 

Finally, the parameter $\Omega_{c}$ gives the level of the process model at time $t = t_c^*$. To share information about the level of the indicator across countries, we assign $\Omega_c$ a hierarchical model such that $\Omega_c$ is distributed around a subregion, region, and world mean:
\begin{align}
    \label{eq:etabar-hierarchical-model}
    \Omega_c \mid \Omega^{(r)}_{s[c]}, \sigma^{(c)}_{\Omega} &\sim N\left(\Omega^{(r)}_{r[c]}, (\sigma^{(c)}_{\Omega})^2\right), \\
    \Omega^{(s)}_s \mid \Omega^{(r)}_{r[s]}, \sigma^{(s)}_{\Omega} &\sim N\left(\Omega^{(r)}_{r[s]}, (\sigma^{(r)}_{\Omega})^2\right), \\
    \Omega^{(r)}_r \mid \Omega^{(w)}, \sigma^{(r)}_{\Omega} &\sim N\left(\Omega^{(w)}, (\sigma^{(r)}_{\Omega})^2\right).
\end{align}
The hyperparameters for all the hierarchical models are assigned vague priors (see Appendix \ref{section:appendix-mcpr-model}).

\subsection*{Smoothing Component}
An AR(1) model is used to smooth the deviation terms $\epsilon_{c,t}$, with global autocorrelation parameter $\rho$ and variance parameter $\tau$:
\begin{alignat}{3}
    \epsilon_{c,t} &\mid \rho, \tau \sim N\left(0, \rho^2 / (1 - \tau^2)\right),\,  & t = t_c^*, \\
    \epsilon_{c,t} &\mid \epsilon_{c, t - 1}, \rho, \tau \sim N\left(\rho \cdot \epsilon_{c,t-1}, \tau^2\right),\,  & t > t_c^*, \\
    \epsilon_{c,t} &\mid \epsilon_{c,t+1}, \rho, \tau \sim N\left(\rho \cdot \epsilon_{c,t+1}, \tau^2\right),\, & t < t_c^*.
\end{alignat}
The hyperparameters are assigned the following priors:
\begin{align}
    \rho &\sim \mathrm{Uniform}(0, 1), \\
    \tau &\sim N_+(0, 2^2),
\end{align}
\added{where the prior for $\rho$ is non-informative (flat) and the prior for $\tau$ is intended to be weakly informative by favoring small deviation term variance.}

\subsection*{Data and Data Model}
\label{section:mcpr-data}
We use the UNPD database to produce estimates of mCPR  \citep{un_desa2021}, available in the \texttt{fpemlocal} R package \citep{fpemlocal2019}. We apply several additional preprocessing steps to the UNPD database to construct an analysis dataset. We exclude two observations from the area ``Other non-specified areas". We also exclude $96$ observations marked in the database as exhibiting geographical region bias (e.g. the survey only sampled a subregion of the country) or bias related to measurement of modern method use. The analysis dataset after exclusions consisted of $n=1020$ observations. Because the Northern America region only includes The United States of America and Canada, it was combined with the Europe region. We impute sampling errors that are missing or reported as zero following the method described in \citep{cahill2018modern}. Finally, although most surveys covered a period of one year or less, 31 observations covered a period of more than one year, including one that covered a period of approximately five years. We took the simplifying step of defining the survey year for each observation to be the midpoint of the start and end year, rounded down to the nearest year.

We define a simple data model. Let $y_i$ be the observed mCPR in country $c[i]$ and time $t[i]$ for $i = 1, \dots, n$.  Let $d[i]$ index the data source type of observation $i$ (with $D$ total data source types). Let $s_i^2$ be the fixed sampling variance for observation $i$.  We adopt a truncated normal data model:
\begin{align}
    \label{eq:mcpr-data-model}
    y_i \mid \eta_{c[i], t[i]}, \sigma_{d[i]} \sim N_{[0, 1]}(\eta_{c[i], t[i]}, s_i^2 + \sigma_{d[i]}^2),
\end{align}
where  $\sigma_{d[i]}^2$ is a non-sampling error estimated for each data source type. The following prior is placed on the non-sampling error:
\begin{align}
    \sigma_{d} \sim N_+(0, 0.1^2).
\end{align}

\subsection*{Comparison: Approximate Logistic Transition Model}
We use a B-spline approximation of the logistic transition model as a baseline comparison model. The approximate logistic transition function is designed to follow the logistic transition function closely while still observing the constraint that the derivative of the spline function be zero at the asymptote $P_c$. We chose to approximate the logistic transition function using splines both for computational reasons, as the non-differentiability of the logistic transition function at $P_c$ can cause problems for gradient-based sampling algorithms such as Hamiltonian Markov Chain Monte-Carlo, and so that all parameters of the comparison model can be interpreted in the same way as the proposed model. As such, the approximate logistic model can be seen as a simplified version of the Family Planning Estimation Model, which models multiple family planning indicators \citep{cahill_modern_2018}. Full details on the specification of the comparison model can be found in Appendix~A (Section \ref{section:appendix-logistic}).

\section*{Model Validation}
\label{section:validations}

In this section, we describe a set of out-of-sample model checks that can be used to evaluate the performance of a transition model, following those used to evaluate FPEM \citep{alkema2013national}. The model checks follow a similar structure which we describe here in the general case. Split the observations into a training set $\bm{y}^{\mathrm{train}}$ and validation set $\bm{y}^{\mathrm{val}}$.  Fit the full model on $\bm{y}^{\mathrm{train}}$, yielding a set of posterior draws $\eta_{c,t}^{(\ell)}$ for $c = 1, \dots, C$ and $t = 1, \dots, T$, with $\ell$ indexing the draw. Compute the posterior predictive distribution for each of the held out data points, following the data model. 
For example, for the data model used for mCPR, we compute 
\begin{align}
    \hat{y}^{(\ell)}_i \mid \eta_{c[i], t[i]}^{(\ell)}, \sigma_{s[i]}^{(\ell)} \sim N_{[0, 1]}(\eta_{c[i], t[i]}^{(\ell)}, s_i^2 + (\sigma^{(\ell)}_{d[i]})^2)
\end{align}
for all $i$ with $y_i \in \bm{y}^{\mathrm{val}}$. Let $\hat{y}_i^{q}$ be the empirical $q\%$ quantile of $\hat{y}_i^{(\ell)}$. Let $\bm{y}^{\mathrm{error}} \subset \bm{y}^{\mathrm{val}}$ be a subset of the held-out observations used for calculating coverage and error measures. We define the proportion included within the $(1 - \alpha)\%$ prediction interval as
\begin{align}
    PPC = \frac{1}{|\bm{y}^\mathrm{error}|} \sum_{y_i \in \bm{y}^\mathrm{error}} \mathbb{I}[\hat{y}^{\alpha / 2}_i \leq y_i \leq \hat{y}^{1 - \alpha / 2}_i].
\end{align}
We also compute the proportion of observations $y_i \in \bm{y}^{\mathrm{error}}$ that are below and above the 95\% uncertainty interval given by $[\hat{y}^{0.025}, \hat{y}^{0.975}]$, and the 95\% credible interval width, given by $\hat{y}^{0.975} - \hat{y}^{0.025}$. Define the posterior predictive median error (\added{MedE}) and median absolute error (\added{MedAE}) as
\begin{align}
    \MedE &= \mathrm{median}_{y_i \in \bm{y}^{\mathrm{error}}}\left\{ y_i - \hat{y}^{0.5}_i \right\} \\ 
    \MedAE &= \mathrm{median}_{y_i \in \bm{y}^{\mathrm{error}}}\left\{ |y_i - \hat{y}^{0.5}_i| \right\}.
\end{align}

The validation is designed to evaluate the performance of the model in medium-term projections. First, let $L$ define a cutoff point, in years. Let $\bm{y}^\mathrm{train} = \{ y_i : t[i] < L \}$, and $\bm{y}^\mathrm{val} = \{ y_i : t[i] \geq L \}$. Countries that do not have any observations before the cutoff $L$ are not included in the summary measures. Because errors are likely to be correlated within countries, only the most recent held-out observation for each country is used to compute the validation measures.

\section*{TFR Case Study}
\label{section:results-tfr}
In this section, we present results from the B-spline Transition Model applied to estimating and projecting TFR. 

\subsection*{Computation}
All models were fit with spline degree $d=2$. The MCMC algorithm was run for $500$ iterations after $500$ warmup iterations. The maximum treedepth was set to $12$ and the \texttt{adapt delta} tuning parameter to $0.999$. Convergence of the MCMC algorithm was assessed with the improved split-chain $\hat{R}$ diagnostic and effective sample size (ESS) estimators described in \cite{vehtari2021rhat}. 
For the final selected model with $K = 4$, no transitions were divergent or exceeded the maximum treedepth. The $\hat{R}$ for all parameters was between 0.998 and 1.018, and the bulk ESS was between $504$ and $5521$.

\subsection*{Validation Results}
\added{As a benchmark we compare our results to a simplified version of BayesTFR that generates projections only using the Phase II model}. This is done to highlight the long-term behavior of the B-spline and double logistic transition models in projections. However, we emphasize that both the BayesTFR and B-spline results presented here are therefore not comparable to published projections from the full BayesTFR model and should be interpreted solely in the context of understanding the possible benefits and drawbacks of the proposed B-spline transition model for Phase II estimation. \added{We refer the interested reader to extant work validating the forecasting performance of the full BayesTFR model including all phases, as can be found in  \cite{alkema2011probabilistic, raftery2014bayesian}, and \cite{liu2020}.}

For the spline model specification, we tested differing numbers of knots $K = \{ 2, \dots, 7 \}$ in order to investigate how the flexibility of the transition function effects model performance. 
Model checks, \added{based only on countries that have not yet entered Phase III as of the 2019 UNPD dataset used}, are presented in Table \ref{tab:tfr-validation-results-splines}. The performance of the B-spline model varies with to the number of knots (Table \ref{tab:validation-results-splines}). The highest median absolute errors are seen with $K=2$, as the transition model is not flexible enough to capture trends in the observed data. For the checks with cutoff set to $L = 2003$, generally as more knots are added the median error increases, suggesting a tradeoff between additional flexibility of the transition model and empirical performance. 

We further investigated the performance of the B-spline model by disaggregating the validation results by world region, focusing on Africa, Asia, Europe, and Latin America and the Caribbean. Results for cutoff year $L = 2003$ are shown in Table~\ref{tab:tfr-validation-by-country}, and results for cutoff years $L = 2008$ and $L = 2013$ are included in Appendix D. For $L = 2003$, the B-spline model with $K=3$ knots outperforms \added{the benchmark} in terms of MedAE in Africa, Asia, and Europe. In particular, in African countries the B-spline model has lower median and median absolute error than the \added{benchmark}. The B-spline model also improves upon \added{the benchmark} with respect to calibration, as measured through the proportion of left-out observations that fall above or below their respective prediction intervals. Notably, the B-splines model reduces the proportion of observations that fall above the 80\%PI for Africa from 19.3 to 12.3\%.

Based on the validation results, we considered using $K = 3$ or $K = 4$ to report substantive results from the spline model specification, as both settings had comparable model validation results, with errors that were typically smaller than for other $k$. Given similar validation results for both $K$, we chose to report the results for $K=4$ to allow for more flexibility and uncertainty in longer term projections.

\begin{table}[ht]
    \centering
    \begin{tabular}{|lrrrrrr|}
        \hline
        & \multicolumn{4}{c}{80\% UI} & \multicolumn{2}{c|}{Error}  \\
        & \% Below & \% Included & \% Above & CI Width $\times 100$ & $\MedE$ $\times 100$ & $\MedAE$ $\times 100$ \\ 
        \hline
        \multicolumn{7}{|l|}{Model Check: $L = 2003$} \\
        Spline (K = 2) & 14.91\% & 83.23\% & 1.86\% & 145.0 & -25.6 & 30.8\\
        Spline (K = 3) & 4.35\% & 85.71\% & 9.94\% & 113.5 & 5.5 & 21.0\\
        Spline (K = 4) & 3.73\% & 85.09\% & 11.18\% & 115.7 & 8.5 & 21.4\\
        Spline (K = 5) & 2.48\% & 86.96\% & 10.56\% & 116.7 & 11.9 & 21.6\\
        Spline (K = 6) & 1.86\% & 85.09\% & 13.04\% & 115.8 & 12.1 & 23.2\\
        Spline (K = 7) & 1.24\% & 85.71\% & 13.04\% & 117.1 & 10.5 & 24.3\\
        \added{Benchmark} & 0.62\% & 84.47\% & 14.91\% & 97.6 & 15.1 & 24.7\\
        \multicolumn{7}{|l|}{Model Check: $L = 2008$} \\
        Spline (K = 2) & 9.94\% & 89.44\% & 0.62\% & 116.3 & -18.8 & 22.6\\
        Spline (K = 3) & 3.11\% & 88.20\% & 8.70\% & 88.6 & -0.1 & 12.7\\
        Spline (K = 4) & 3.73\% & 87.58\% & 8.70\% & 88.6 & -0.1 & 12.1\\
        Spline (K = 5) & 2.48\% & 88.20\% & 9.32\% & 88.7 & 3.6 & 10.0\\
        Spline (K = 6) & 3.11\% & 87.58\% & 9.32\% & 88.5 & 3.1 & 10.8\\
        Spline (K = 7) & 3.11\% & 87.58\% & 9.32\% & 89.3 & 2.1 & 11.9\\
        \added{Benchmark} & 1.86\% & 87.58\% & 10.56\% & 77.1 & 4.9 & 11.1\\
        \multicolumn{7}{|l|}{Model Check: $L = 2013$} \\
        Spline (K = 2) & 4.97\% & 95.03\% & 0.00\% & 82.2 & -10.1 & 11.4\\
        Spline (K = 3) & 1.24\% & 96.27\% & 2.48\% & 61.4 & -1.4 & 6.1\\
        Spline (K = 4) & 2.48\% & 95.03\% & 2.48\% & 60.1 & -1.1 & 5.4\\
        Spline (K = 5) & 1.24\% & 97.52\% & 1.24\% & 60.3 & 0.2 & 5.6\\
        Spline (K = 6) & 2.48\% & 95.03\% & 2.48\% & 59.3 & -0.6 & 5.4\\
        Spline (K = 7) & 1.24\% & 96.27\% & 2.48\% & 59.3 & -0.6 & 5.7\\
        \added{Benchmark} & 2.48\% & 93.79\% & 3.73\% & 52.7 & -0.1 & 3.5\\
        \hline
    \end{tabular}
    \caption{Validation results for TFR summarizing the posterior predictive distribution of the held-out data points. The validation metrics are 80\% interval score, empirical coverage (\% of held-out observations below, included, and above the 80\% credible interval), 80\% credible interval (CI) width, median error (MedE), and median absolute error (MedAE).}
    \label{tab:tfr-validation-results-splines}
\end{table}

\begin{table}[]
    \centering
    \begin{longtable}{|llllllll|}
\hline
& & \multicolumn{4}{c}{80\% UI} & \multicolumn{2}{c|}{Error}  \\
Region & Model & \% Below & \% Included & \% Above & CI Width $\times 100$ & $\MedE$ $\times$ 100 & $\MedAE$ $\times 100$ \\
\hline
\multicolumn{8}{|l|}{Model Check: $L = 2003$} \\
Africa & Spline (K = 2) & 14.04\% & 84.21\% & 1.75\% & 147.7 & -32.9 & 39.5\\
 & Spline (K = 3) & 0.00\% & 87.72\% & 12.28\% & 141.5 & 18.5 & 25.7\\
 & Spline (K = 4) & 1.75\% & 85.96\% & 12.28\% & 140.0 & 20.8 & 29.4\\
 & Spline (K = 5) & 1.75\% & 85.96\% & 12.28\% & 142.8 & 18.8 & 27.0\\
 & Spline (K = 6) & 0.00\% & 89.47\% & 10.53\% & 145.3 & 21.8 & 27.0\\
 & Spline (K = 7) & 0.00\% & 89.47\% & 10.53\% & 147.7 & 22.8 & 29.8\\
 & \added{Benchmark} & 0.00\% & 80.70\% & 19.30\% & 120.0 & 30.8 & 35.9\\
Asia & Spline (K = 2) & 27.91\% & 67.44\% & 4.65\% & 144.9 & -33.3 & 54.8\\
 & Spline (K = 3) & 11.63\% & 69.77\% & 18.60\% & 101.1 & -3.1 & 27.8\\
 & Spline (K = 4) & 6.98\% & 72.09\% & 20.93\% & 108.0 & -4.7 & 32.0\\
 & Spline (K = 5) & 6.98\% & 72.09\% & 20.93\% & 107.9 & 0.0 & 31.3\\
 & Spline (K = 6) & 6.98\% & 67.44\% & 25.58\% & 105.2 & 2.1 & 31.6\\
 & Spline (K = 7) & 4.65\% & 69.77\% & 25.58\% & 105.4 & 2.8 & 35.9\\
 & \added{Benchmark} & 0.00\% & 74.42\% & 25.58\% & 90.5 & 11.1 & 29.1\\
Europe & Spline (K = 2) & 0.00\% & 100.00\% & 0.00\% & 143.0 & -3.1 & 10.5\\
 & Spline (K = 3) & 0.00\% & 100.00\% & 0.00\% & 93.7 & 5.6 & 11.3\\
 & Spline (K = 4) & 0.00\% & 100.00\% & 0.00\% & 92.1 & 4.2 & 7.4\\
 & Spline (K = 5) & 0.00\% & 100.00\% & 0.00\% & 87.2 & 12.8 & 12.8\\
 & Spline (K = 6) & 0.00\% & 100.00\% & 0.00\% & 83.1 & 19.8 & 19.8\\
 & Spline (K = 7) & 0.00\% & 100.00\% & 0.00\% & 82.8 & 16.4 & 16.4\\
 & \added{Benchmark} & 0.00\% & 90.91\% & 9.09\% & 67.3 & 12.0 & 12.0\\
Latin America and the Caribbean & Spline (K = 2) & 11.11\% & 88.89\% & 0.00\% & 142.8 & -25.9 & 25.9\\
 & Spline (K = 3) & 5.56\% & 94.44\% & 0.00\% & 95.5 & -3.2 & 15.4\\
 & Spline (K = 4) & 5.56\% & 91.67\% & 2.78\% & 99.0 & -7.8 & 16.5\\
 & Spline (K = 5) & 0.00\% & 100.00\% & 0.00\% & 98.1 & 3.4 & 16.4\\
 & Spline (K = 6) & 0.00\% & 94.44\% & 5.56\% & 95.6 & 5.8 & 13.4\\
 & Spline (K = 7) & 0.00\% & 94.44\% & 5.56\% & 95.5 & 4.6 & 11.1\\
 & \added{Benchmark} & 2.78\% & 97.22\% & 0.00\% & 82.2 & 5.0 & 13.2\\
\hline
\end{longtable}
    \caption{Validation results for Africa, Asia, Europe, and Latin America and the Caribbean for TFR summarizing the posterior predictive distribution of the held-out data points separated. The validation metrics are 80\% interval score, empirical coverage (\% of held-out observations below, included, and above the 80\% credible interval), 80\% credible interval (CI) width, median error (MedE), and median absolute error (MedAE).}
    \label{tab:tfr-validation-by-country}
\end{table}

\subsection*{TFR Results}

\begin{figure}
    \centering
    \includegraphics[width=0.85\columnwidth]{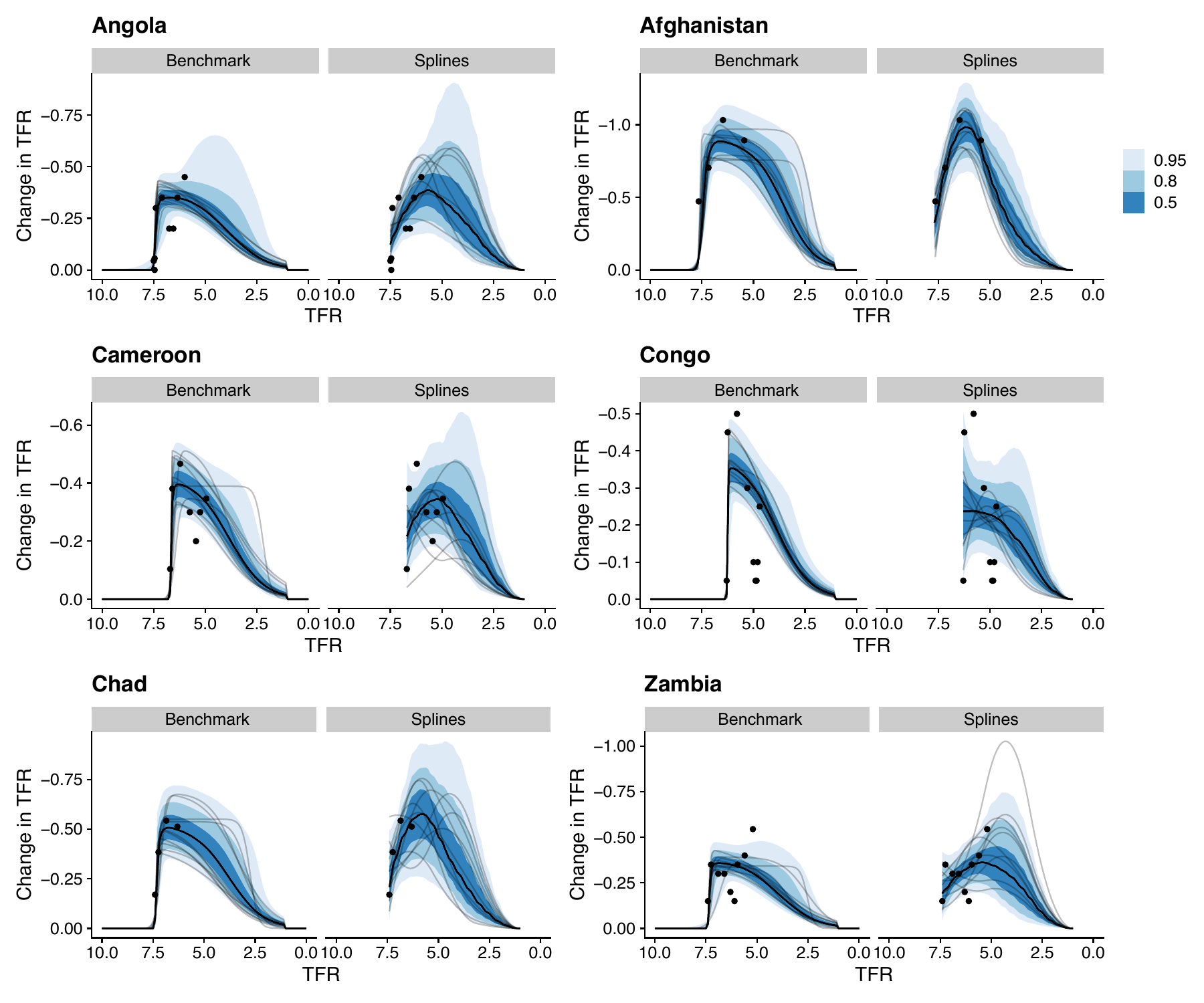}
    \caption{Estimated transition functions from the \added{benchmark} and B-spline Transition Model for TFR (K = 4) giving the estimated rate of change in TFR versus level of TFR. Black lines are posterior medians, and shaded regions show 50\%, 80\%, and 95\% credibility intervals. A sample of 10 draws from the posterior distribution of the transition functions are shown as gray lines. The points show the observed declines in TFR as a function of TFR.}
    \label{fig:tfr_transition_examples}
\end{figure}

The estimated transition functions are shown in Figure \ref{fig:tfr_transition_examples} for six countries. In several countries we can see that transition functions as estimated by the spline model are more variable than those from \added{the benchmark}. In the case of Zambia, we see that the B-splines function fits the observed declines in TFR more closely, as compared to the double logistic function. 
The spline transition functions for all of the pictured countries include the possibility of larger declines in TFR than is found in the double logistic fits. 

\begin{figure}
    \centering
    \includegraphics[width=0.75\columnwidth]{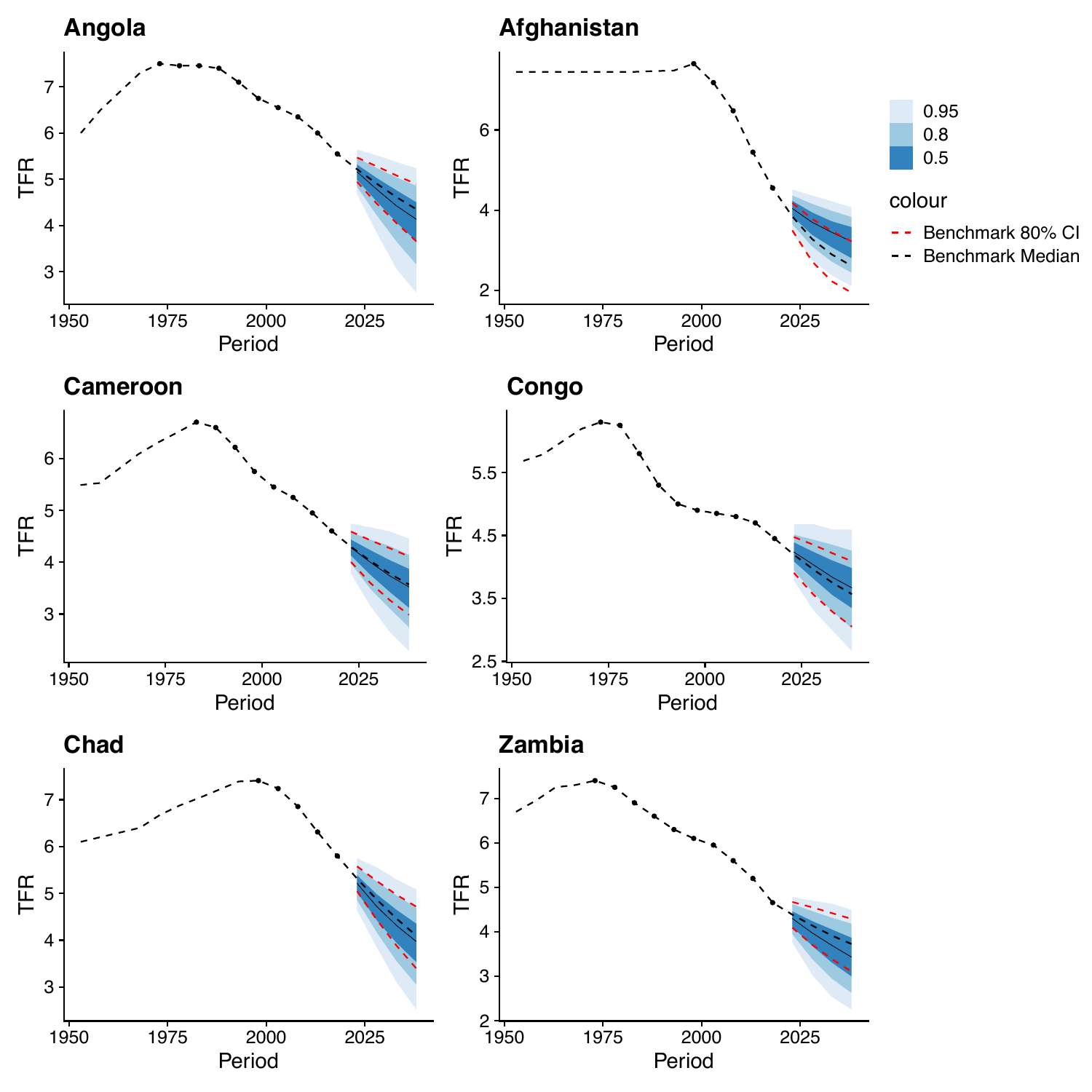}
    \caption{Posterior projections of the TFR from the \added{benchmark model} and B-spline Transition Model for TFR. The black dotted line shows the \added{benchmark} posterior median, and the red dotted line the \added{benchmark} 80\% credibility interval. The shaded regions show the 50\%, 80\%, and 95\% credibility intervals from the B-spline Transition Model. Note that estimation is performed for $5$-year intervals. The continuous lines on the plot are meant only to aid the eye and should not be interpreted as yearly estimates.}
    \label{fig:tfr_fit_examples}
\end{figure}

The posterior distribution of $\eta_{c,t}$ with projections to the 2035-2040 5-year period for the same six countries is shown in Figure \ref{fig:tfr_fit_examples}. The associated transition functions for each country are shown in Figure \ref{fig:tfr_transition_examples}. The B-spline based projections generally exhibit higher uncertainty than those based on the double logistic functions, which is not surprising given the additional flexibility of the splines. The median spline projections for Angola and Zambia are lower than those from the double logistic transition, reflecting how the spline transition functions include the possibility of larger declines. In Afghanistan, splines projections suggest a more modest decline in TFR, as compared to the results based on double logistic transitions. In Cameroon, Chad, and Congo, the median projections are remarkably similar, although the spline projections exhibit higher uncertainty. 

\section*{mCPR Case Study}
\label{section:results-mcpr}

\subsection*{Computation}
\label{section:splines-computation}
Tuning parameters differed for model validation runs and for a set of final models results. For model validations, 500 joint posterior distribution samples were drawn per chain from 4 chains after 250 warmup iterations. The max treedepth was set to $15$ and \texttt{adapt delta} to $0.999$.  Based on the results of the model validations, two model configurations were chosen to generate a set of final results: the B-spline model with $d=2$, $K=5$ and the approximate logistic transition model. For the final results, the max treedepth was set to $14$, \texttt{adapt delta} to 0.999, and $1500$ samples were drawn after $500$ warmup iterations, reflecting an increased emphasis on sampling accuracy.

For the final B-spline model ($d = 2$, $K = 5$), $\hat{R}$ for all parameters was between $0.999$ and $1.0220$, below the recommended threshold of $1.05$. The estimated bulk ESS ranged from $197.6$ to $15019.3$.  Finally, the diagnostics from the No-U-Turn sampler were optimal, with no divergent transitions and no transitions exceeding the maximum treedepth.

\subsection*{Validation Results}
The model validations were conducted for several model specifications and tuning parameter values. For the base model specification, we tested every combination of spline degree $d=\{ 2, 3 \}$ and number of knots $K = \{ 5, 7 \}$ in order to investigate how the smoothness of the transition function effects model performance. 
As a comparison model, we used the approximate logistic transition model with $d=2$, $K=7$. Model checks are presented in Table \ref{tab:validation-results-splines}.

No single set of tuning parameters clearly outperforms the others (Table \ref{tab:validation-results-splines}), although each either matches or beats the performance of the logistic comparison model. We are generally cautious to over-interpret the differences in validation measures between the models due to the relatively small number of held-out data points in the validation exercises, although some broad trends can be observed. A general pathology in all the models can be seen in their negative median errors, suggesting they tend to over-predict mCPR in projections.  However, the spline methods have median errors closer to zero than the comparison logistic model. Similarly, several of the base models have near-optimal empirical coverage of the 95\% credible interval. 

Based on the validation results, particularly in view of the similarity between the specifications, we chose to report substantive results from the base model specification with $d=2$, $K = 5$ as it exhibited good performance in both the $L=2010$ and $L=2015$ model checks, \textit{and achieved the lowest MedAE among the B-spline models in the $L=2015$ model check}.

\begin{table}[ht]
    \centering
    \begin{tabular}{|lrrrrrr|}
        \hline
        & \multicolumn{4}{c}{95\% UI} & \multicolumn{2}{c|}{Error}  \\
        & \% Below & \% Included & \% Above & CI Width $\times 100$ & $\MedE$ $\times 100$ & $\MedAE$ $\times 100$ \\ 
        \hline
        \multicolumn{7}{|l|}{Model Check: $L = 2010$} \\
        B-spline ($d=2$, $K=5$) & 3.76\% & 94.7\% & 1.50\% & 32.1 & -1.47 & 4.62\\
        B-spline ($d=2$, $K=7$) & 3.76\% & 94.0\% & 2.26\% & 32.2 & -0.97 & 4.99\\
        B-spline ($d=3$, $K=5$) & 3.76\% & 94.7\% & 1.50\% & 31.6 & -1.42 & 4.59\\
        B-spline ($d=2$, $K=7$) & 3.76\% & 94.7\% & 1.50\% & 32.5 & -1.84 & 4.78\\
        Logistic & 6.02\% & 93.2\% & 0.752\% & 33.0 & -3.17 & 4.49\\
        \multicolumn{7}{|l|}{Model Check: $L = 2015$} \\
        B-spline ($d=2$, $K=5$) & 5.05\% & 92.9\% & 2.02\% & 23.5 & -0.808 & 3.66\\
        B-spline ($d=2$, $K=7$) & 5.05\% & 92.9\% & 2.02\% & 23.7 & -0.534 & 3.71\\
        B-spline ($d=3$, $K=5$) & 5.05\% & 92.9\% & 2.02\% & 23.5 & -1.080 & 3.91\\
        B-spline ($d=2$, $K=7$) & 6.06\% & 91.9\% & 2.02\% & 23.8 & -1.030 & 3.74\\
        Logistic & 6.06\% & 91.9\% & 2.02\% & 25.0 & -2.120 & 3.28\\
        \hline
    \end{tabular}
    \caption{Validation results for mCPR summarizing the posterior predictive distribution of the held-out data points. The validation metrics are empirical coverage (\% of held-out observations below, included, and above the 95\% credible interval), 95\% credible interval (CI) width, median error (MedE), and median absolute error (MedAE).}
    \label{tab:validation-results-splines}
\end{table}

\subsection*{mCPR Results}

Posterior estimates of $\eta_{c,t}$ for the six selected countries of Figure \ref{fig:data-availability}, chosen to represent countries with high and low data availability, are shown in Figure \ref{fig:mcpr-results}. Results for all countries are included in the Supplementary Material \citep{supplemental2022}. The estimates of $\eta_{c,t}$ tend to follow the observed data closely, although with added uncertainty derived from the non-sampling error term in the data model. 

\begin{figure}
    \centering
    \includegraphics[width=0.75\columnwidth]{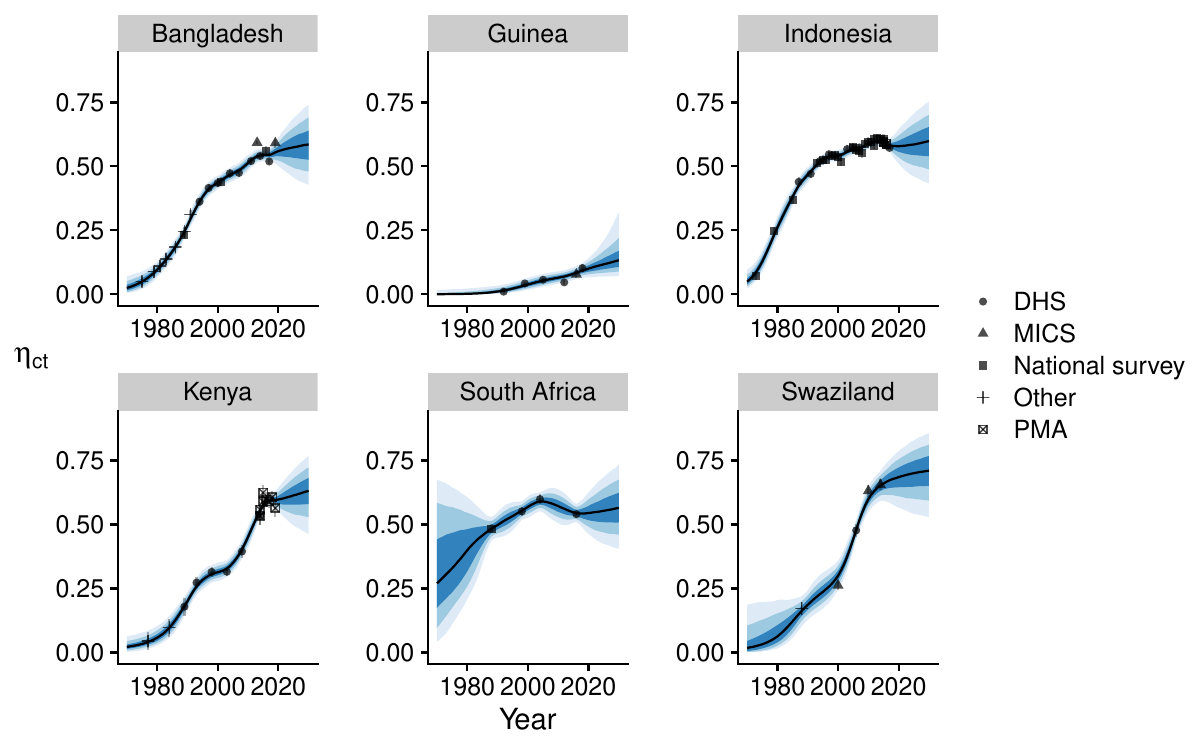}
    \caption{Posterior median (black line) and 50\%, 80\%, and 95\% credible intervals (shaded regions) of $\eta_{c,t}$ (latent mCPR) in six countries from a B-spline Transition Model ($d = 2$, $K = 5$). The points show the observations, with vertical lines indicating the 95\% confidence interval based on the sampling error.}
    \label{fig:mcpr-results}
\end{figure}

\begin{figure}
    \centering
    \includegraphics[width=0.70\columnwidth]{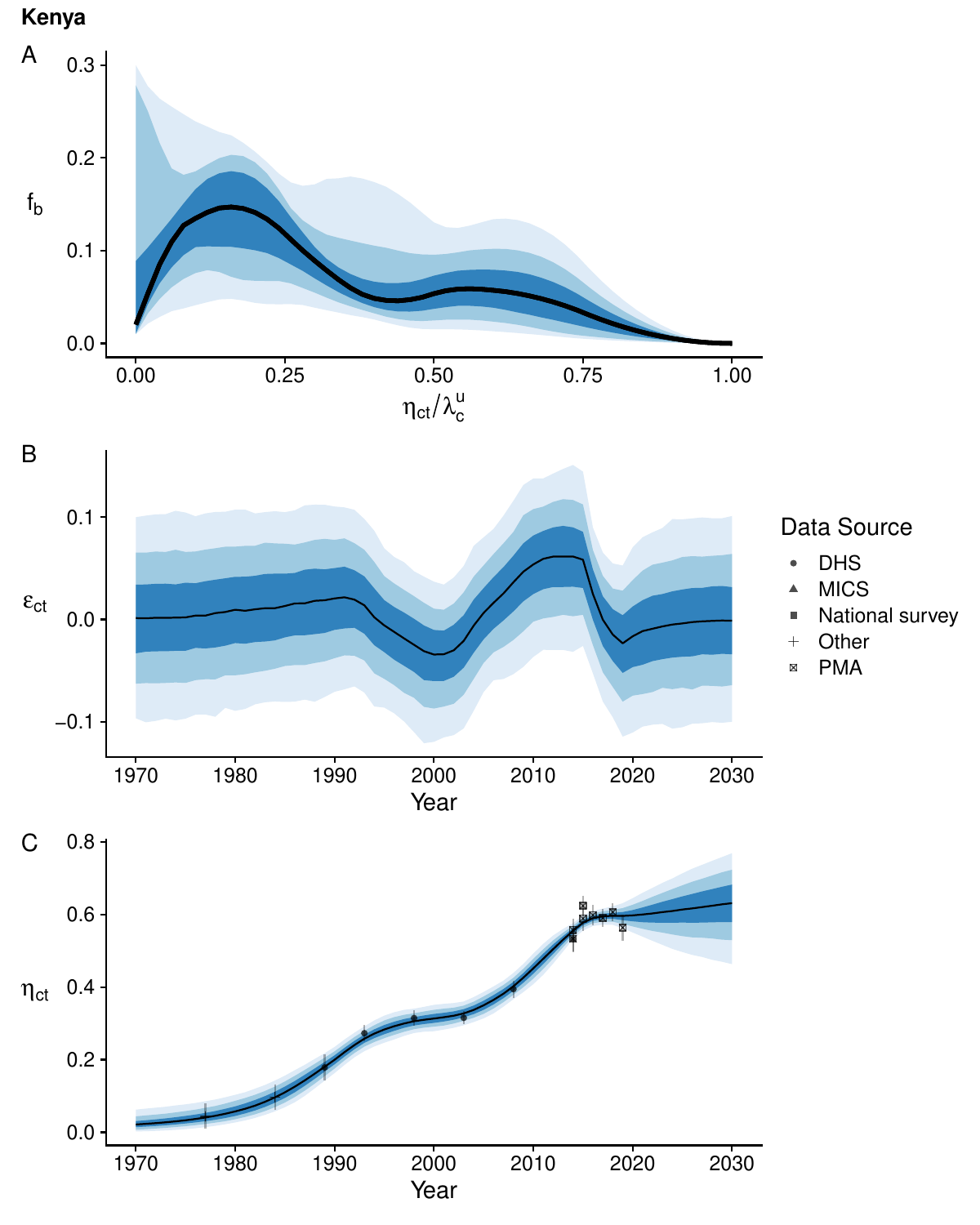}
    \caption{Posterior median (black lines) and 50\%, 80\%, and 95\% credible intervals (shaded bands) of B-spline ($d = 2, K = 5$) model results from Kenya showing (A) transition function $f_b$, (B) smoothing component $\epsilon_{c,t}$, and (C) latent mCPR $\eta_{c,t}$. The points show the observations, with vertical lines indicating the 95\% confidence interval based on the sampling error.}
    \label{fig:spline-example}
\end{figure}

The systematic and smoothing components of the process model interact to form the estimates of $\eta_{c,t}$ for each country. As an illustrative example, Figure \ref{fig:spline-example} shows posterior estimates of the transition function, smoothing component, and $\eta_{c,t}$ for Kenya. The observed data from Kenya suggest that mCPR growth slowed around the year 2000. The transition function captures this by having a ``double-peak" shape in which the rate of change is high, decreases in order to capture the stall, and then increases again to capture the observed increase in growth after the stall. The smoothing component also contributes to capturing the stall and subsequent increase, as the posterior median $\epsilon_{c,t}$ tends negative during the stall before turning positive.

\begin{figure}
    \centering
    \includegraphics[width=1\columnwidth]{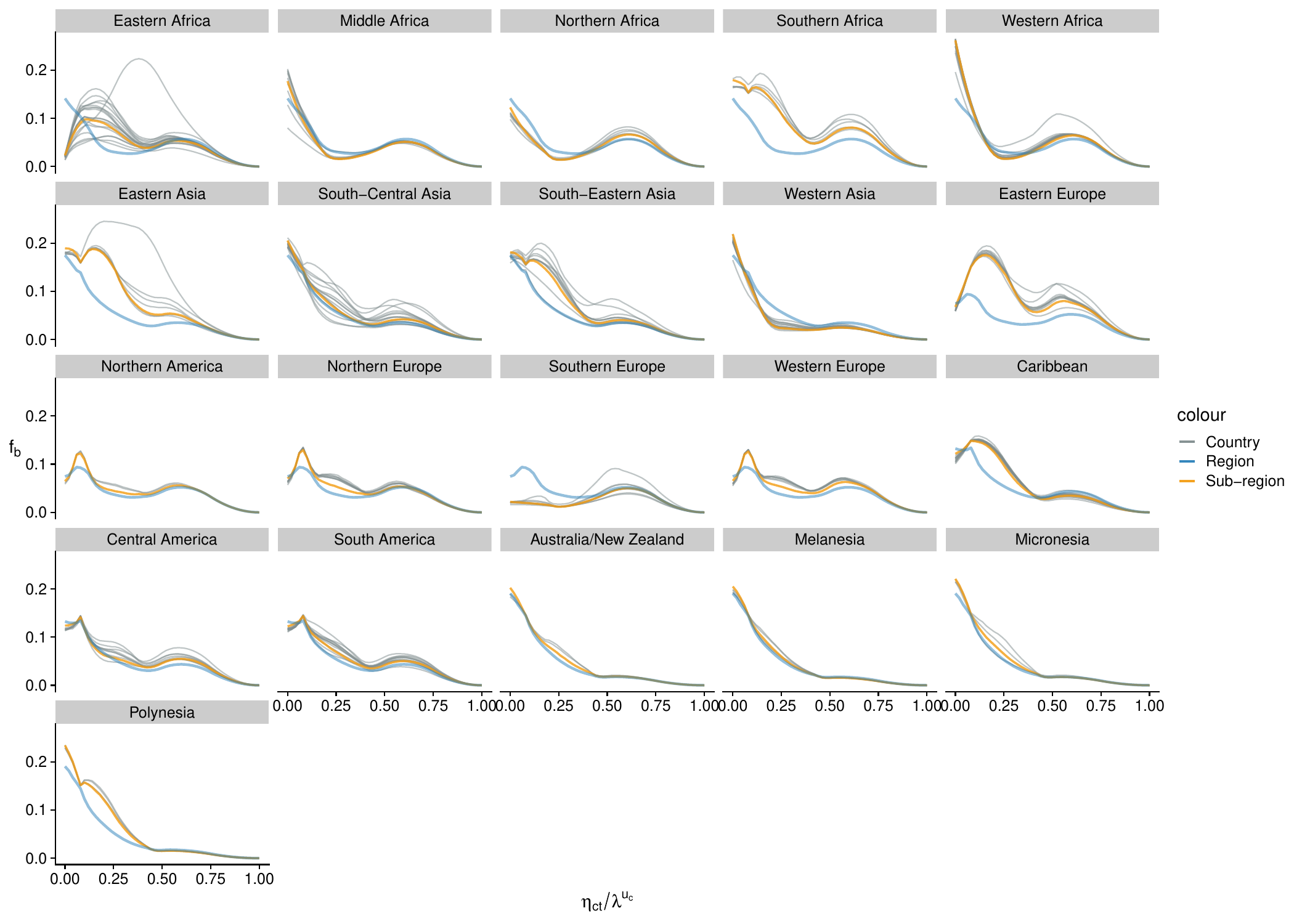}
    \caption{Posterior median transition functions $f_b$ for countries, sub-regions, and regions from the B-spline transition model ($d = 2$, $K = 5$).}
    \label{fig:transition-function-hierarchy}
\end{figure}

The shape of the transition function is informed by country-level data, when available, and by the transitions of other countries via the hierarchical model placed on the spline coefficients. Figure \ref{fig:transition-function-hierarchy} illustrates the hierarchical distribution by showing the posterior median of the region, sub-region, and country-specific transition functions. The sub-regional transition function for Eastern Africa, for example, has a  distinct ``double-peak" shape, perhaps reflecting a slow-down in adoption in the 1990s and 2000s. 
The sub-regional transition function for Eastern Asia is quite different than for Asia as a whole, due in part to the steep increase in mCPR in China and the Republic of Korea from 1970 to 2000.

\subsection*{Comparison with Logistic model}
The added flexibility of the B-spline model relative to the logistic model leads to differences in estimates in some countries. Illustrative fits from the logistic model for the six selected countries are shown in Appendix Figure \ref{fig:mcpr-logistic-results}. The posterior estimates of mCPR in the year 2030 for the 15 countries with the largest absolute difference in posterior median between the B-spline and logistic models are shown in Appendix Figure \ref{fig:absolute-differences}. For all of these countries, the B-spline model is more conservative than the logistic model, projecting lower mCPR in 2030 (in terms of the posterior median).

\begin{figure}
    \centering
    \includegraphics[width=0.85\columnwidth]{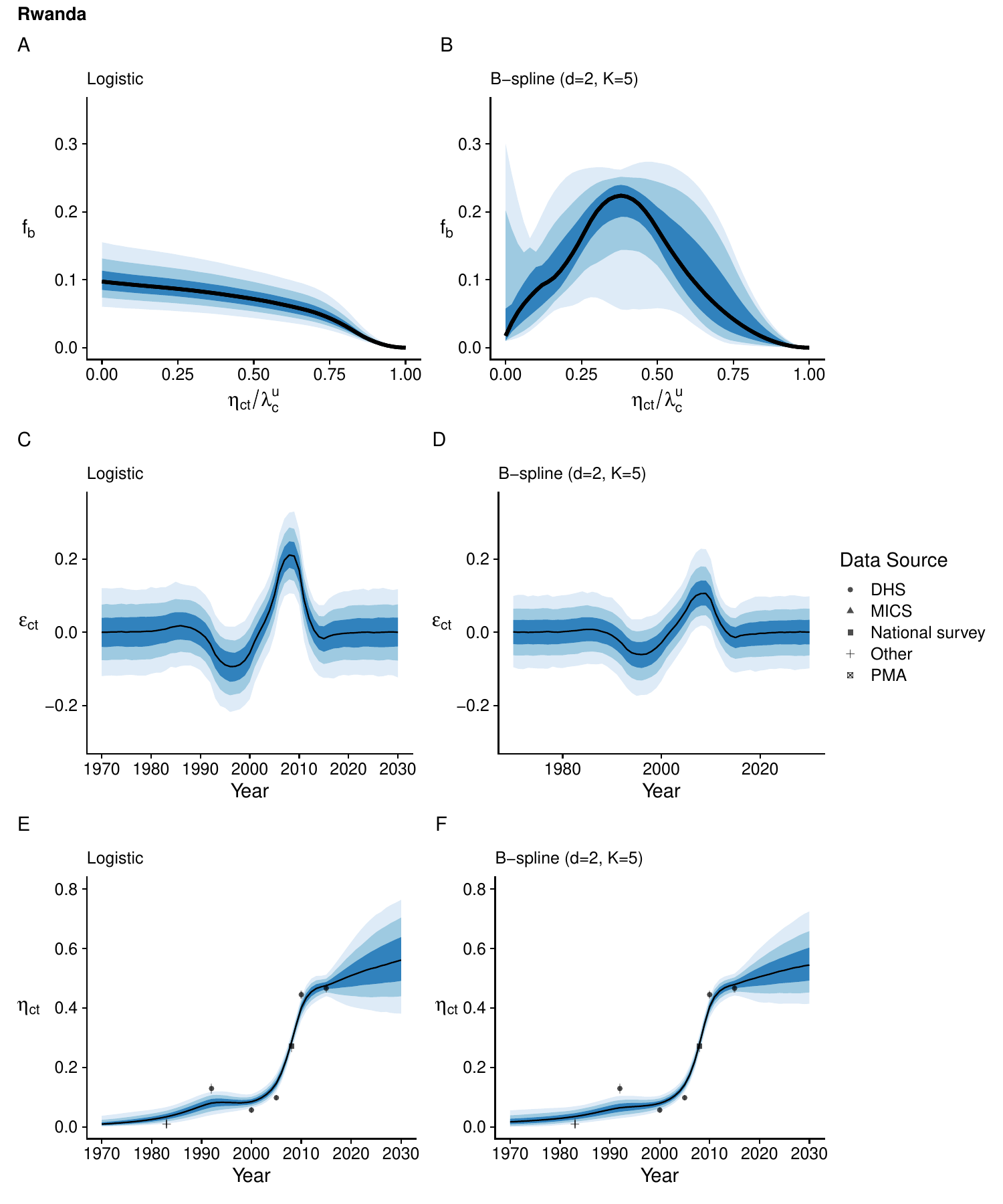}
    \caption{Posterior median (black lines) and 50\%, 80\%, and 95\% credible intervals (shaded bands) of (A, B) transition functions $f_b$ for the logistic and B-spline ($d=2,K=5$) models, (C, D) smoothing component $\epsilon_{c,t}$, and (E, F) latent mCPR $\eta_{c,t}$. The points show the observations, with vertical lines indicating the 95\% confidence interval based on the sampling error. }
    \label{fig:rwanda-example}
\end{figure}

\begin{figure}[H]
    \centering
    \includegraphics[width=0.65\columnwidth]{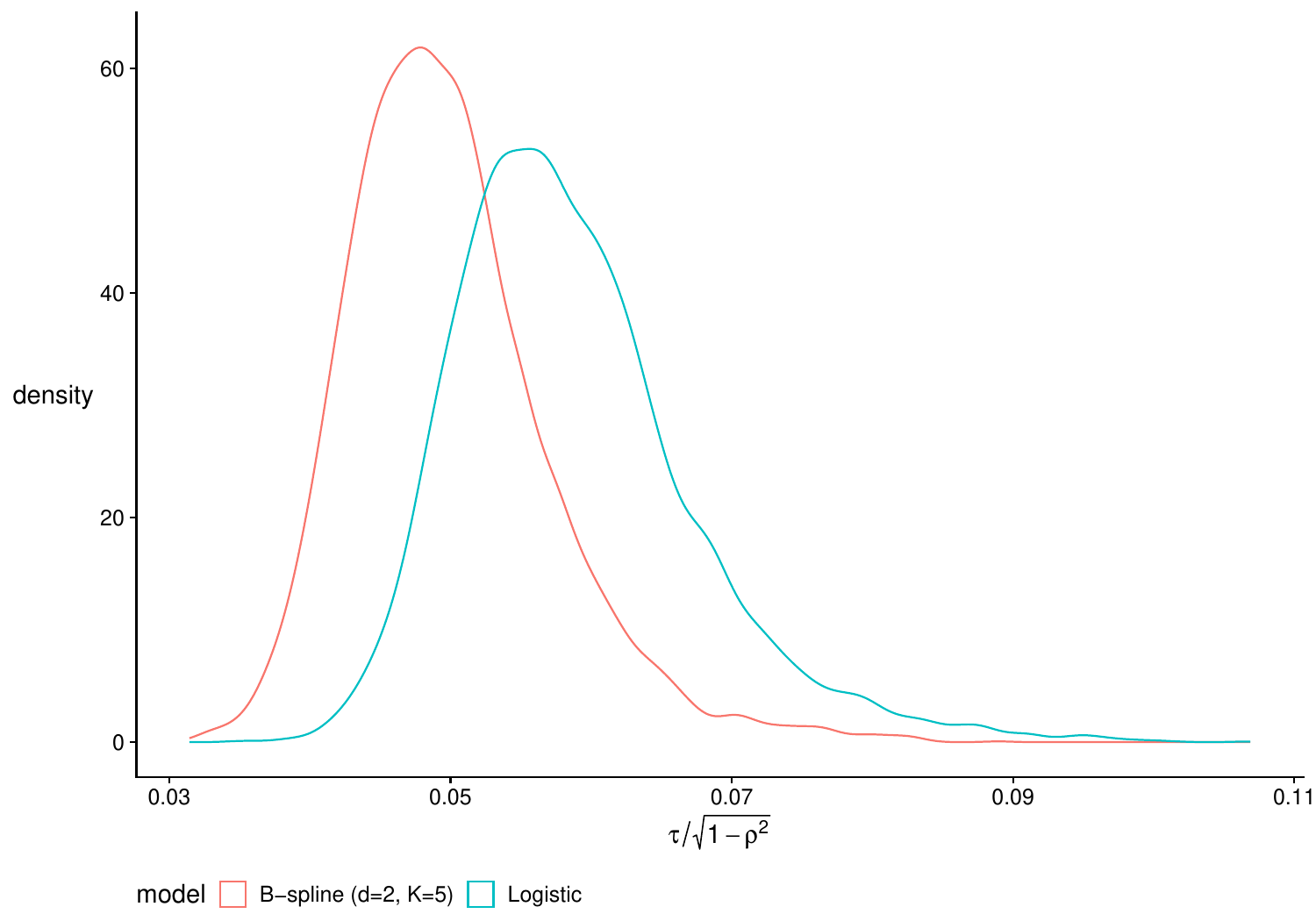}
    \caption{Posterior distribution of $\tau \slash \sqrt{1 - \rho^2}$, the unconditional standard deviation of the smoothing component, for the B-spline ($d=2$, $K=5$) and Logistic models.}
    \label{fig:ar-logistic-spline-comparison}
\end{figure}

The smoothing components of the B-spline and logistic models also exhibit differences. To summarize the hyperparameters of the smoothing component, we calculated the unconditional standard deviation of $\epsilon_{c,t}$, given by $\tau / \sqrt{1 - \rho^2}$. The posterior distribution for the unconditional standard deviation for the B-spline model is generally lower than that of the Logistic model (Appendix Figure \ref{fig:ar-logistic-spline-comparison}). Rwanda provides an illustrative example of the difference in the smoothing component between the two models: the $\epsilon_{c,t}$ are smaller in the B-spline model because the sharp increase in adoption starting around 2005 can be in part captured by the more flexible B-spline transition function (Figure \ref{fig:rwanda-example}). However, the model estimates for $\eta_{c,t}$ for Rwanda are remarkably similar between the two models, suggesting that the added uncertainty from allowing the transition function to be more flexible balances the lower variance in the smoothing component.

\section*{Discussion}
\label{section:discussion}
The B-spline Transition Model proposed in this paper demonstrates that it is possible to estimate and project demographic and health indicators using flexible estimation techniques that eschew strong functional form assumptions. Using TFR and mCPR as our applications, we found it is possible to weaken assumptions on the shape of the relationship between the rate of change and level of an indicator without sacrificing out-of-sample predictive performance. These results demonstrate that the B-spline Transition Model can be considered more generally for projecting indicators that follow transition processes.

In addition to providing a flexible estimation approach, our model also yields a novel way to summarize and communicate trends in indicators through the estimated shape of the transition functions. For example, in our mCPR case study, the regional and sub-regional transition functions, derived from the hierarchical distribution placed on the spline coefficients reveals systematic differences in the transitions between groups of countries. 

Substantively, in our applications, we found that projections from the B-spline model were often similar to those from more parameterized models.  Despite the added flexibility inherent in the B-spline Transition Model, in many countries the mCPR estimates of $\eta_{c,t}$ are quite similar to the comparison Logistic model. This could be seen as evidence that the logistic assumption made by earlier models is a sound one. The similarities may also reflect the data-sparse setting in the case-study, in which models of a similar structure may converge on similar results because they reach the limit of what is possible to infer given the limited data. For TFR projections, we did obtain outcomes that differed more substantially between the spline and double-logistic models. 

The results from the TFR case-study indicate there may be a bias-variance trade-off between strongly parametric models and more flexible approaches. For example, the TFR validations for $2003$ and $2008$ show that \added{the benchmark model}, the more parametric approach, has higher median error (bias) but lower credible interval widths (related to variance) than the B-spline model. We emphasize that having more uncertainty in projections is not inherently negative if it accurately reflects true uncertainty. For example, countries that have not yet completed the demographic transition will not have many data points for low TFR; extrapolating into this region based on parametric assumptions is naturally only valid if the parameterization is correct, while using more flexible approaches allows for higher uncertainty to be reflected in the projections. Indeed, we find that the B-splines approach improve calibration as compared to \added{a benchmark based on a parametric transition function}. \added{However, this is a delicate trade-off, as we see the B-splines model can yield overly conservative prediction intervals in some regions. Additional research is needed to investigate the effect of the data model on forecast uncertainty, and if improved model selection and tuning can lead to a better tuned bias-variance tradeoff.}

The B-spline transition model can be extended to capture more complex behaviors. For example, a transition function that is allowed to vary over time would allow the systematic component to fit temporal trends in the transition. For example, the mCPR indicator in some subregions appears to have gone through a slowdown in the 1990s and 2000. In our current model, this is captured in the shape of the (time-invariant) transition functions and by the smoothing component. However, if there is a shared, temporally localized slowdown then it may be more effective to explicitly capture it via a time-varying transition function. In addition, analyzing the posterior summaries of the time-varying transition functions would provide another method to summarize and communicate trends, such as regional or subregional slowdowns in adoption. Finally, while the focus of this article is to propose a flexible estimation strategy for the transition function, an alternative specification for the smoothing term may also yield benefits. 

Our approach has several limitations. First, our approach does not account for uncertainty in the model specification, in that the results are conditional on a specific spline configuration. One possibility for future work is to use Bayesian model averaging or model stacking to combine predictions from multiple candidate specifications \cite{yao2018stacking}. In addition, adaptive methods for choosing the knot placements may make the model specification more robust and adaptive to data availability in each country. {In this work, we chose a model specification based on validation results; however, doing so may overstate the performance of our method.}  Second, for TFR we focused on only estimating transitions from high to low TFR (referred to as Phase II of the demographic transition; \citealt{alkema2011probabilistic}), where the segmentation of TFR data into phases is done with deterministic rules. Our work could be possibly extended to model all phases of the demographic transition, obviating then need for segmentation into phases.  Finally, in this article we chose to focus on the process model specification, leaving the data model fixed. It is possible that alternative data models would yield better performance in the case studies; we leave further investigation to future work. 

The B-spline Transition Model is a flexible model for indicators that follow transition processes and is extendable to capture trends exhibited by other indicators. As interest in generating estimates and projections of health indicators grows, the B-Spline Transition Model provides a useful framework for developing models that relax strong functional form assumptions in favor of learning relationships from the data.

\section*{Competing interests}
No competing interest is declared.

\section*{Author contributions statement}
H.S. and L.A. conceived the model and H.S. implemented the model in software and conducted the experiments. H.S. and L.A. wrote and reviewed the manuscript.

\section*{Data Availability}
The data that support the findings of this study  are openly available from the United Nations, Department of Economic and Social Affairs, Population Division at \url{https://www.un.org/development/desa/pd/data/world-contraceptive-use} for \citep{un_desa2021} 
and \url{https://population.un.org/wpp/} for \citep{un_desa2019}.

\section*{Acknowledgements}
This work was supported, in whole or in part, by the Gates Foundation (INV-00844). The conclusions and opinions expressed in this work are those of the author(s) alone and shall not be attributed to the Foundation. Under the grant conditions of the Foundation, a Creative Commons Attribution 4.0 License has already been assigned to the Author Accepted Manuscript. The  published version of this manuscript can be found at \url{https://doi.org/10.1093/jrsssc/qlaf026}.

\bibliography{bibliography}       


\newpage
\section*{Appendix A: Full mCPR B-Spline Transition Model Specification}\label{section:appendix-mcpr-model}
This appendix collects the equations that describe the mCPR B-spline transition model. The process model is given by:
\begin{align}
    \label{eq:mcpr-transition-systematic-component}
  \mathrm{logit}(\eta_{c,t}) = \begin{cases}
        \Omega_c, & t = t_c^*, \\
        \mathrm{logit}(\eta_{c, t - 1}) + f_b(\eta_{c,t-1}, \bm{\lambda}_c, \bm{\beta}_c) + \epsilon_{c,t}, & t > t_c^*, \\
        \mathrm{logit}(\eta_{c, t + 1}) - f_b(\eta_{c,t+1}, \bm{\lambda}_c, \bm{\beta}_c) - \epsilon_{c,t+1}, & t < t_c^*. \\
    \end{cases}
\end{align}
Hierarchical model and priors for spline coefficients, for $j = 1, \dots, J - d - 1$:
\begin{align}
    \label{eq:beta-hierarchical-model}
    \beta_{c,j} \mid \beta^{(r)}_{s[c], j}, \sigma_{\beta, j}^{(c)} &\sim N\left(\beta^{(s)}_{s[c], j}, \left(\sigma_{\beta, j}^{(c)}\right)^2\right), \\
    \beta^{(s)}_{s,j} \mid \beta^{(r)}_{r[s], j}, \sigma_{\beta, j}^{(s)} &\sim N\left(\beta^{(r)}_{r[s], j}, \left(\sigma_{\beta, j}^{(s)}\right)^2\right), \\
    \beta^{(r)}_{r,j} \mid \beta^{(w)}_{j}, \sigma_{\beta, j}^{(r)} &\sim N\left(\beta^{(w)}_{j}, \left(\sigma_{\beta, j}^{(r)}\right)^2\right), \\
    \sigma^{(c)}_{\beta,j} &\sim N_+\left(0, 5^2\right), \\
    \sigma^{(s)}_{\beta,j} &\sim N_+\left(0, 5^2\right), \\
    \sigma^{(r)}_{\beta,j} &\sim N_+\left(0, 5^2\right), \\
    \beta^{(w)}_j &\sim N\left(0, 5\right).
\end{align}
Hierarchical models and priors for asymptote and starting level parameters:
\begin{align}
    \lambda^u_c &= 0.5 + 0.45 \cdot \mathrm{logit}^{-1}\left( \tilde{\lambda}^u_c \right), \\
   \tilde{\lambda}^u_c \mid \tilde{\lambda}^{u,(w)}, \sigma_{\tilde{\lambda}}^{(c)} &\sim N\left(\tilde{\lambda}^{u,(w)},  \left(\sigma^{(c)}_{\tilde{\lambda}^u}\right)^2\right), \\
    \tilde{\lambda}^{u,(w)} &\sim N\left(0, 3\right), \\ \sigma^{(c)}_{\tilde{\lambda}^u} &\sim N_+\left(0, 3\right), \\
    \Omega^{(r)}_r \mid \Omega^{(w)}, \sigma^{(r)}_{\Omega} &\sim N\left(\Omega^{(w)}, (\sigma^{(r)}_{\Omega})^2\right), \\
    \Omega_c \mid \Omega^{(r)}_{s[c]}, \sigma^{(c)}_{\Omega} &\sim N\left(\Omega^{(r)}_{r[c]}, (\sigma^{(c)}_{\Omega})^2\right), \\
    \Omega^{(s)}_s \mid \Omega^{(r)}_{r[s]}, \sigma^{(s)}_{\Omega} &\sim N\left(\Omega^{(r)}_{r[s]}, (\sigma^{(r)}_{\Omega})^2\right), \\
    \Omega^{(w)} &\sim N\left(0, 3\right) \\
    \sigma^{(c)}_{\Omega} &\sim N_+\left(0, 3\right), \\
    \sigma^{(r)}_{\Omega} &\sim N_+\left(0, 3\right).
\end{align}
AR(1) smoothing component and priors:
\begin{align*}
    \epsilon_{c,t} \mid \rho, \tau &\sim N\left(0, \rho^2 / (1 - \tau^2)\right), & t = t_c^*, \\
    \epsilon_{c,t} \mid \epsilon_{c, t - 1}, \rho, \tau &\sim N\left(\rho \cdot \epsilon_{c,t-1}, \tau^2\right), & t > t_c^*, \\
    \epsilon_{c,t} \mid \epsilon_{c,t+1}, \rho, \tau &\sim N\left(\rho \cdot \epsilon_{c,t+1}, \tau^2\right), & t < t_c^*, \\
    \rho &\sim \mathrm{Uniform}(0, 1), \\
    \tau &\sim N_+(0, 2^2).
\end{align*}
Data model:
\begin{align}
    y_i \mid \eta_{c[i], t[i]}, \sigma_{d[i]} &\sim N_{[0, 1]}(\eta_{c[i], t[i]}, s_i^2 + \sigma_{d[i]}^2), \\
    \sigma_{d} &\sim N_+(0, 0.1^2).
\end{align}

\section*{Appendix B: Approximate Logistic Transition Function}\label{section:appendix-logistic}
This appendix describes an approximation of the logistic transition function using the B-spline transition function. To enforce the logistic growth assumption within our model framework, we introduce a new set of constraints on the first $J - 3$ spline coefficients.  Let $x^*_j = \argmax_{x \in [0, 1]} B_j(x)$ be the point at which the $j$th spline basis function is maximized. Set $\beta_{c,j} = l(x^*_j, \lambda^u_c, \omega_c)$ for $j = 1, \dots, J - 3$, where the parameter $\omega_c$ is a pace parameter controlling the rate of the logistic growth curve and $l(x, \lambda^u, \omega) = ((x - \lambda^u)\omega)\slash(\lambda^u(x - 1))$ gives the rate of change for logistic growth for a level $x$, upper asymptote $\lambda^u$, and pace parameter $\omega$. 

\begin{figure}[H]
    \centering
    \includegraphics[width=0.8\columnwidth]{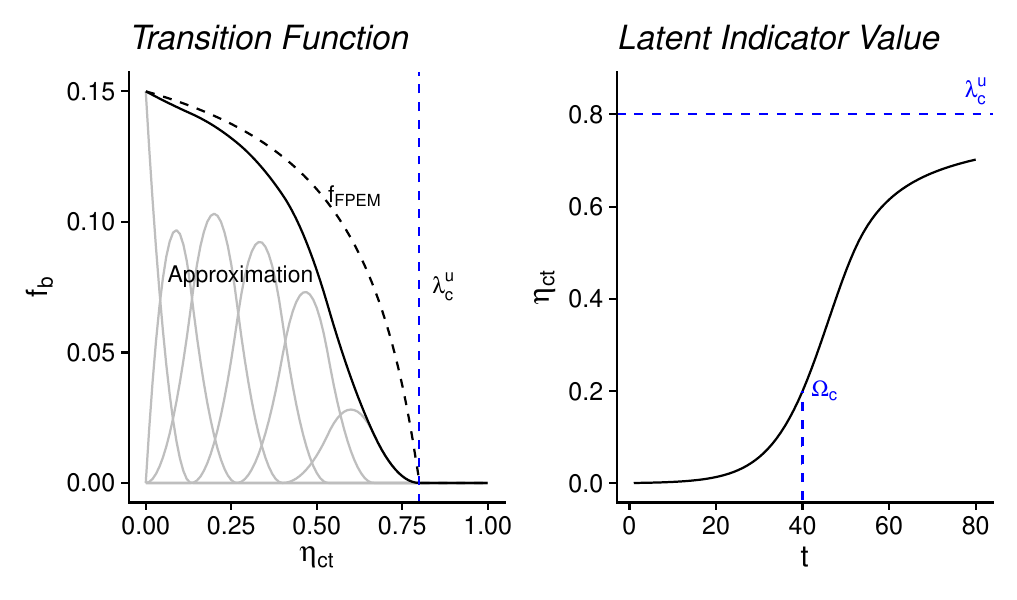}
    \caption{Left: an example of a constrained B-spline transition function $f_b$ for the approximate logistic transition model. The gray lines illustrate a scaled version of the non-zero spline basis functions. Right: the curve implied by the pictured transition function.}
    \label{fig:rate_vs_level_logistic_example}
\end{figure}

Figure \ref{fig:rate_vs_level_logistic_example} shows an example B-spline transition function with this constraint applied. The difference between the approximation and the true $f_{\textrm{FPEM}}$ is caused by the constraint inherited from the B-spline setup that the transition function must have a derivative of zero at the asymptote. 

We place a hierarchical model on $\omega_c$ to share information between countries, nested within subregion, region, and world. Let
\begin{align}
    \omega_c = 0.5 \cdot \mathrm{logit}^{-1}(\tilde{\omega}_c),
\end{align}
for $\tilde{\omega}_c \in \mathbb{R}$ such that $\omega_c$ is constrained to be between $0$ and $0.5$. We use the following hierarchical model:
\begin{align}
    \tilde{\omega}_c \mid \tilde{\omega}^{(s)}_{s[c]}, \sigma^{(c)}_{\tilde{\omega}}  &\sim N\left(\tilde{\omega}^{(s)}_{s[c]}, \left(\sigma^{(c)}_{\tilde{\omega}}\right)^2\right), \\
    \tilde{\omega}^{(s)}_{s} \mid \tilde{\omega}^{(s)}_{r[s]}, \sigma^{(s)}_{\tilde{\omega}} &\sim N\left(\tilde{\omega}^{(r)}_{r[s]}, \left(\sigma^{(s)}_{\tilde{\omega}}\right)^2\right), \\
    \tilde{\omega}^{(r)}_{r} \mid \tilde{\omega}^{(w)}, \sigma^{(r)}_{\tilde{\omega}} &\sim N\left(\tilde{\omega}^{(w)}, \left(\sigma^{(r)}_{\tilde{\omega}}\right)^2\right).
\end{align}
The hyperparameters are assigned the following priors:
\begin{align}
    \sigma^{(c)}_{\tilde{\omega}} &\sim N_+\left(0, 5\right), \\
    \sigma^{(s)}_{\tilde{\omega}} &\sim N_+\left(0, 1\right), \\
    \sigma^{(r)}_{\tilde{\omega}} &\sim N_+\left(0, 5\right), \\
    \tilde{\omega}^{(w)} &\sim N\left(0, 5\right).
\end{align}

\section*{Appendix C: Additional TFR results}
\begin{longtable}{|llllllll|}
\hline
& & \multicolumn{4}{c}{80\% UI} & \multicolumn{2}{c|}{Error}  \\
Region & Model & \% Below & \% Included & \% Above & CI Width $\times 100$ & $\MedE$ $\times$ 100 & $\MedAE$ $\times 100$ \\
\hline
\multicolumn{8}{|l|}{Model Check: $L = 2008$} \\
Africa & Spline (K = 2) & 12.28\% & 87.72\% & 0.00\% & 117.2 & -26.6 & 30.2\\
 & Spline (K = 3) & 3.51\% & 87.72\% & 8.77\% & 101.5 & 6.6 & 14.4\\
 & Spline (K = 4) & 3.51\% & 89.47\% & 7.02\% & 100.3 & 7.7 & 13.1\\
 & Spline (K = 5) & 3.51\% & 89.47\% & 7.02\% & 101.3 & 4.3 & 11.5\\
 & Spline (K = 6) & 3.51\% & 91.23\% & 5.26\% & 103.5 & 6.9 & 11.3\\
 & Spline (K = 7) & 1.75\% & 92.98\% & 5.26\% & 107.0 & 6.5 & 11.7\\
 & \added{Benchmark} & 1.75\% & 87.72\% & 10.53\% & 87.6 & 9.3 & 14.0\\
Asia & Spline (K = 2) & 16.28\% & 81.40\% & 2.33\% & 116.5 & -19.8 & 34.5\\
 & Spline (K = 3) & 4.65\% & 76.74\% & 18.60\% & 83.8 & -1.2 & 20.2\\
 & Spline (K = 4) & 6.98\% & 72.09\% & 20.93\% & 85.4 & -1.8 & 20.5\\
 & Spline (K = 5) & 2.33\% & 74.42\% & 23.26\% & 84.4 & 1.3 & 17.4\\
 & Spline (K = 6) & 4.65\% & 72.09\% & 23.26\% & 83.6 & -1.3 & 20.1\\
 & Spline (K = 7) & 6.98\% & 69.77\% & 23.26\% & 81.9 & -4.0 & 19.8\\
 & \added{Benchmark} & 2.33\% & 74.42\% & 23.26\% & 75.8 & 5.5 & 20.4\\
Europe & Spline (K = 2) & 0.00\% & 100.00\% & 0.00\% & 113.9 & -2.1 & 5.7\\
 & Spline (K = 3) & 0.00\% & 100.00\% & 0.00\% & 79.7 & 1.1 & 5.5\\
 & Spline (K = 4) & 0.00\% & 100.00\% & 0.00\% & 76.8 & 4.0 & 5.1\\
 & Spline (K = 5) & 0.00\% & 100.00\% & 0.00\% & 74.3 & 7.5 & 7.7\\
 & Spline (K = 6) & 0.00\% & 100.00\% & 0.00\% & 72.8 & 9.5 & 9.5\\
 & Spline (K = 7) & 0.00\% & 100.00\% & 0.00\% & 71.7 & 11.9 & 11.9\\
 & \added{Benchmark} & 0.00\% & 100.00\% & 0.00\% & 59.2 & 3.6 & 5.1\\
Latin America and the Caribbean & Spline (K = 2) & 5.56\% & 94.44\% & 0.00\% & 115.8 & -17.5 & 17.5\\
 & Spline (K = 3) & 2.78\% & 97.22\% & 0.00\% & 79.5 & -3.9 & 10.2\\
 & Spline (K = 4) & 2.78\% & 97.22\% & 0.00\% & 80.4 & -4.7 & 9.7\\
 & Spline (K = 5) & 2.78\% & 97.22\% & 0.00\% & 80.1 & 0.7 & 7.1\\
 & Spline (K = 6) & 2.78\% & 94.44\% & 2.78\% & 77.5 & 1.3 & 6.8\\
 & Spline (K = 7) & 2.78\% & 94.44\% & 2.78\% & 77.7 & 0.1 & 6.0\\
 & \added{Benchmark} & 2.78\% & 97.22\% & 0.00\% & 68.6 & 0.6 & 7.1\\
\multicolumn{8}{|l|}{Model Check: $L = 2013$} \\
Africa & Spline (K = 2) & 5.26\% & 94.74\% & 0.00\% & 82.9 & -17.5 & 17.5\\
 & Spline (K = 3) & 0.00\% & 98.25\% & 1.75\% & 64.8 & -1.0 & 6.6\\
 & Spline (K = 4) & 1.75\% & 96.49\% & 1.75\% & 63.5 & 0.3 & 4.9\\
 & Spline (K = 5) & 0.00\% & 98.25\% & 1.75\% & 63.9 & -3.7 & 6.3\\
 & Spline (K = 6) & 0.00\% & 98.25\% & 1.75\% & 63.6 & -3.3 & 5.7\\
 & Spline (K = 7) & 0.00\% & 98.25\% & 1.75\% & 65.5 & -2.1 & 5.7\\
 & \added{Benchmark} & 0.00\% & 98.25\% & 1.75\% & 57.6 & -0.5 & 1.7\\
Asia & Spline (K = 2) & 11.63\% & 88.37\% & 0.00\% & 81.9 & -9.2 & 12.6\\
 & Spline (K = 3) & 4.65\% & 88.37\% & 6.98\% & 60.4 & 0.4 & 9.0\\
 & Spline (K = 4) & 6.98\% & 86.05\% & 6.98\% & 59.3 & -0.9 & 10.7\\
 & Spline (K = 5) & 4.65\% & 93.02\% & 2.33\% & 60.1 & 2.0 & 11.0\\
 & Spline (K = 6) & 9.30\% & 83.72\% & 6.98\% & 58.4 & 1.3 & 10.8\\
 & Spline (K = 7) & 4.65\% & 88.37\% & 6.98\% & 57.3 & 0.5 & 11.1\\
 & \added{Benchmark} & 4.65\% & 86.05\% & 9.30\% & 52.9 & 1.0 & 8.3\\
Europe & Spline (K = 2) & 0.00\% & 100.00\% & 0.00\% & 81.7 & -0.8 & 3.4\\
 & Spline (K = 3) & 0.00\% & 100.00\% & 0.00\% & 57.9 & 1.5 & 3.4\\
 & Spline (K = 4) & 0.00\% & 100.00\% & 0.00\% & 56.4 & 1.8 & 3.1\\
 & Spline (K = 5) & 0.00\% & 100.00\% & 0.00\% & 55.6 & 3.8 & 5.0\\
 & Spline (K = 6) & 0.00\% & 100.00\% & 0.00\% & 54.2 & 3.6 & 4.2\\
 & Spline (K = 7) & 0.00\% & 100.00\% & 0.00\% & 53.5 & 3.6 & 5.5\\
 & \added{Benchmark} & 0.00\% & 100.00\% & 0.00\% & 40.8 & 1.1 & 4.9\\
Latin America and the Caribbean & Spline (K = 2) & 0.00\% & 100.00\% & 0.00\% & 82.0 & -8.3 & 8.4\\
 & Spline (K = 3) & 0.00\% & 100.00\% & 0.00\% & 58.8 & -2.9 & 4.2\\
 & Spline (K = 4) & 0.00\% & 100.00\% & 0.00\% & 57.8 & -2.4 & 3.8\\
 & Spline (K = 5) & 0.00\% & 100.00\% & 0.00\% & 57.3 & 0.3 & 3.7\\
 & Spline (K = 6) & 0.00\% & 100.00\% & 0.00\% & 56.0 & -0.8 & 1.9\\
 & Spline (K = 7) & 0.00\% & 100.00\% & 0.00\% & 54.8 & -1.0 & 3.6\\
 & \added{Benchmark} & 5.56\% & 94.44\% & 0.00\% & 47.9 & -0.2 & 1.7\\
\hline
    \caption{Validation results for Africa, Asia, Europe, and Latin America and the Caribbean for TFR summarizing the posterior predictive distribution of the held-out data points separated. The validation metrics are 80\% interval score, empirical coverage (\% of held-out observations below, included, and above the 80\% credible interval), 80\% credible interval (CI) width, median error (MedE), and median absolute error (MedAE).}
    \label{tab:tfr-validation-by-country-appendix}
\end{longtable}

\section*{Appendix E: Additional mCPR Results}
\label{section:additional-results}

\begin{figure}[H]
    \centering
    \includegraphics[width=1\columnwidth]{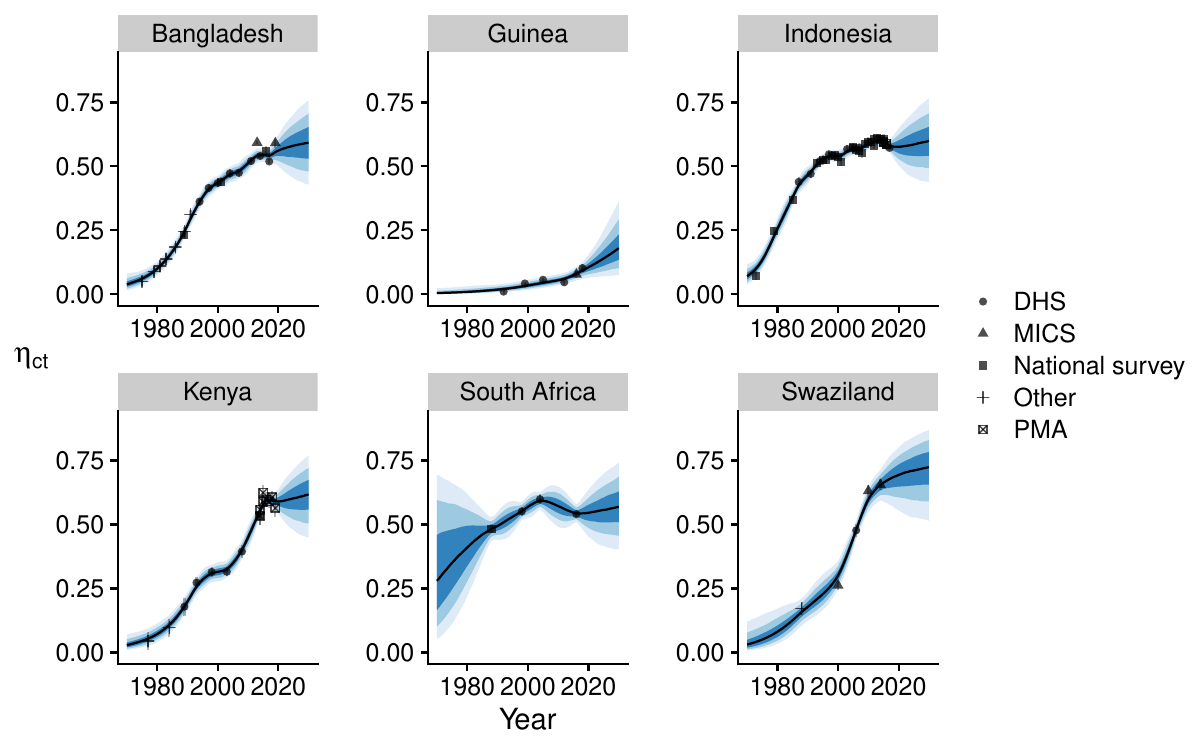}
    \caption{Posterior median (black line) and 50\%, 80\%, and 95\% credible intervals (shaded regions) of $\eta_{c,t}$ (latent mCPR) in six countries from the logistic approximation to the B-spline Transition Model. The points show the observations, with vertical lines indicating the 95\% confidence interval based on the sampling error.}
    \label{fig:mcpr-logistic-results}
\end{figure}

\begin{figure}[H]
    \centering
    \includegraphics[width=0.7\columnwidth]{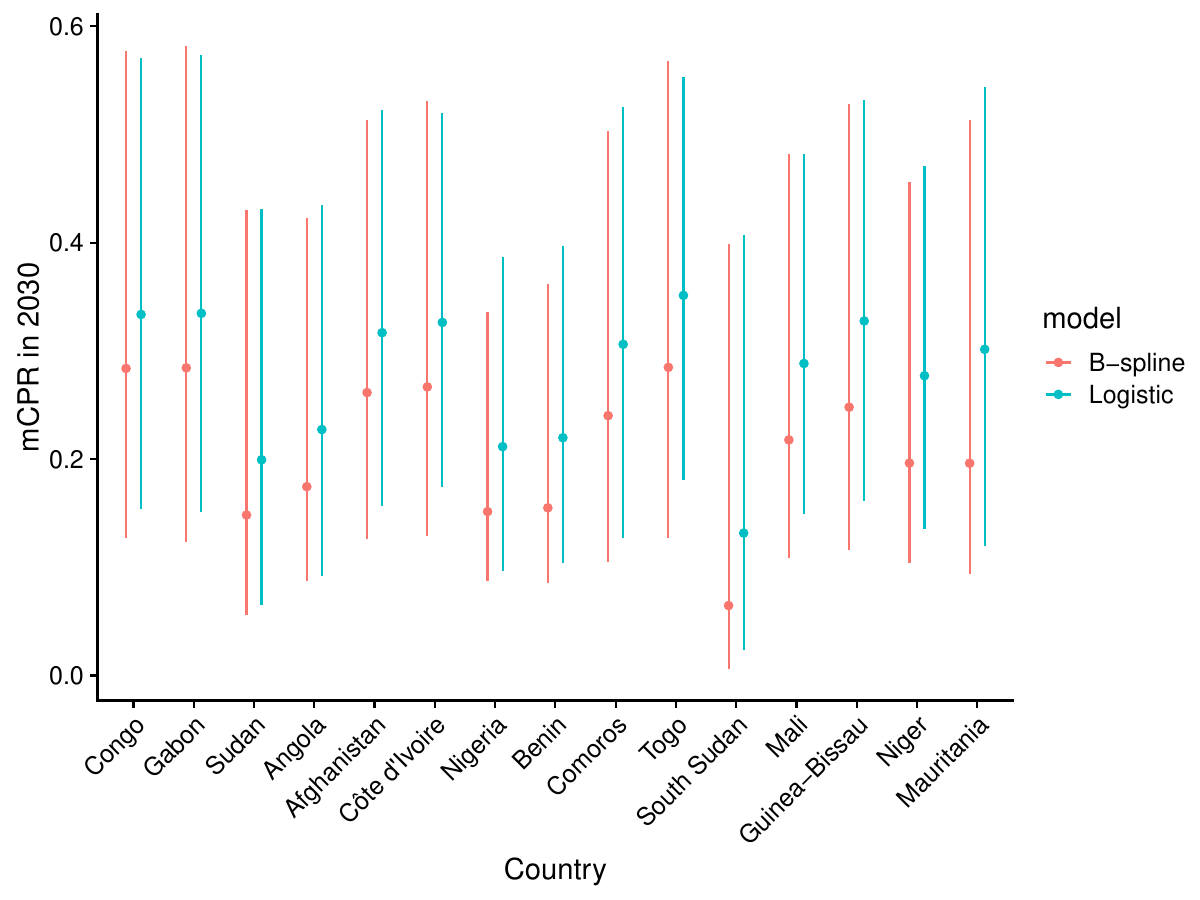}
    \caption{Posterior median and 95\% credibility interval of $\eta_{c,2030}$ from the B-spline ($d = 2$, $K = 5$) and logistic models for the 15 countries where the absolute differences between the posterior median from each model are largest.}
    \label{fig:absolute-differences}
\end{figure}

\section*{Supplementary Material: Estimates and projections of mCPR for all countries}
Model fits for all countries from the B-spline transition model for mCPR ($d = 2$, $K = 5$) are presented below.



\section*{Supplementary material}

**Afghanistan**

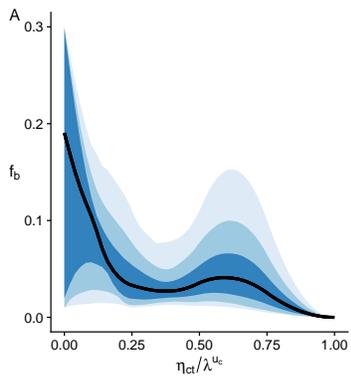 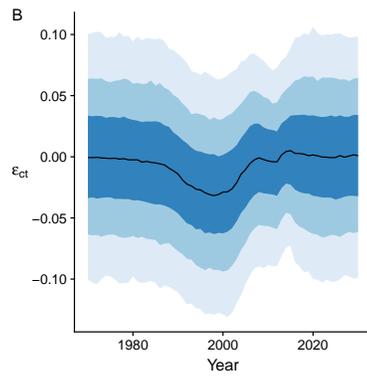 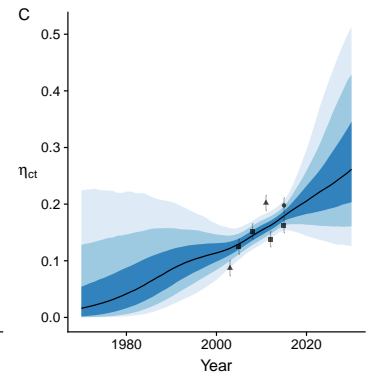

**Albania**

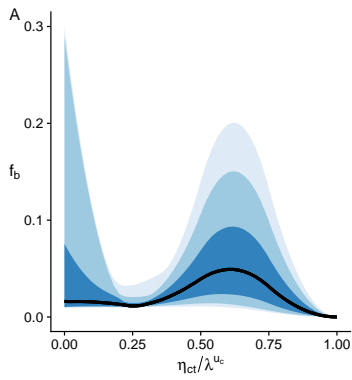 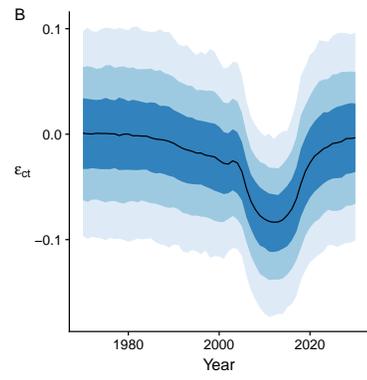 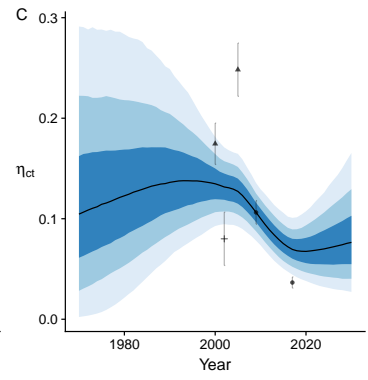

**Algeria**

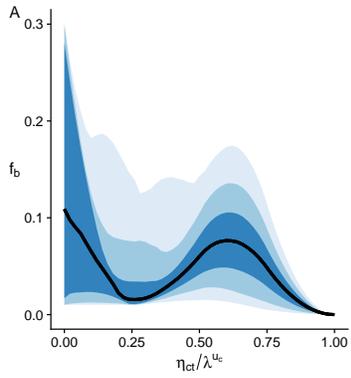 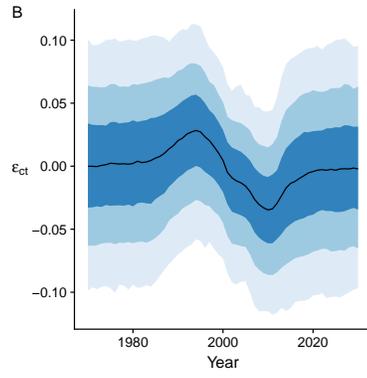 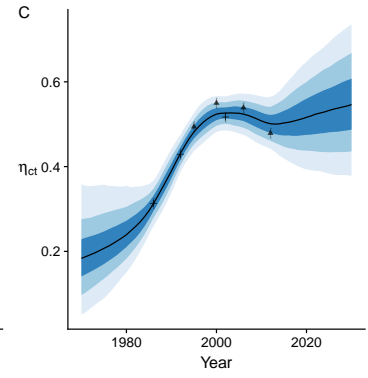

**Angola**

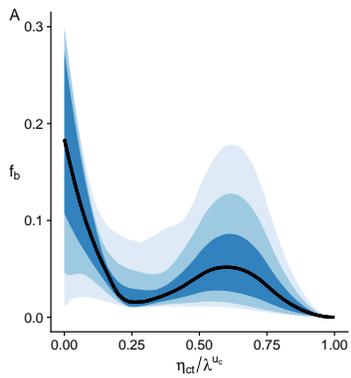 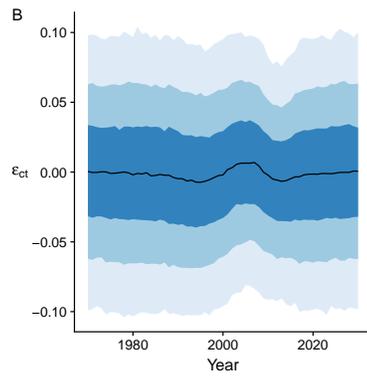 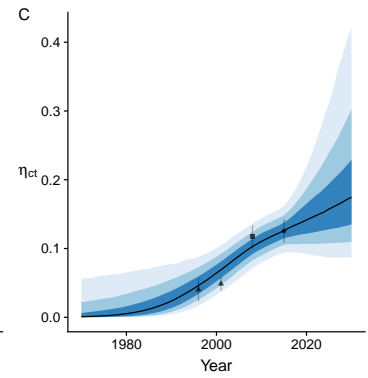

**Antigua and Barbuda**

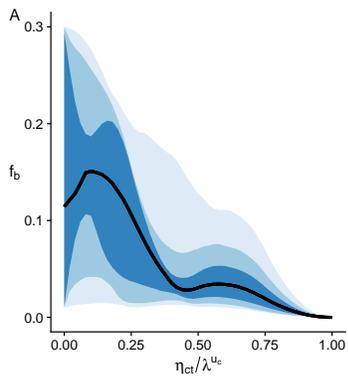 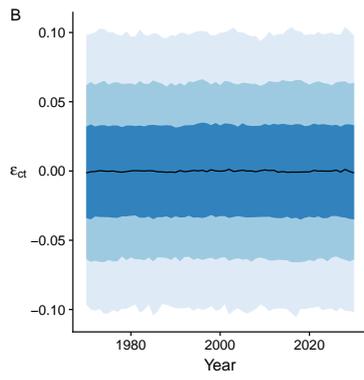 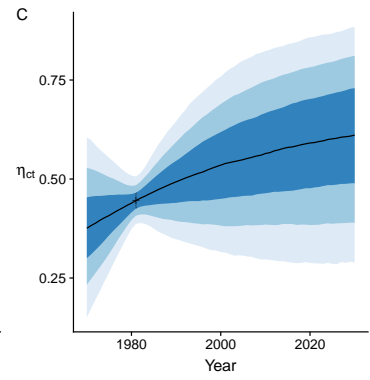

**Armenia**

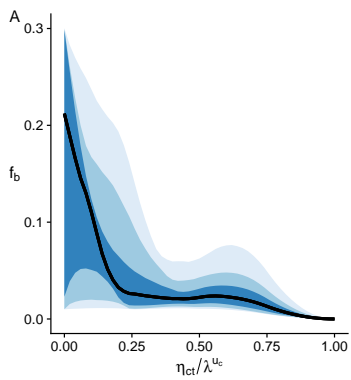 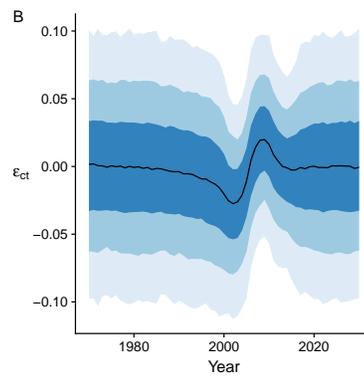 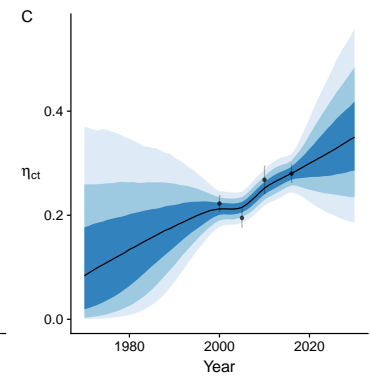

**Australia**

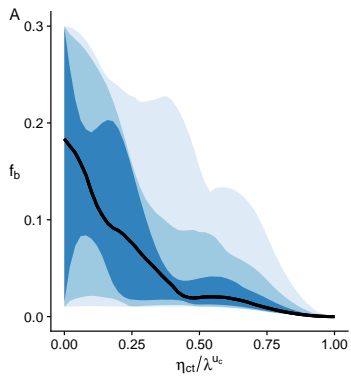 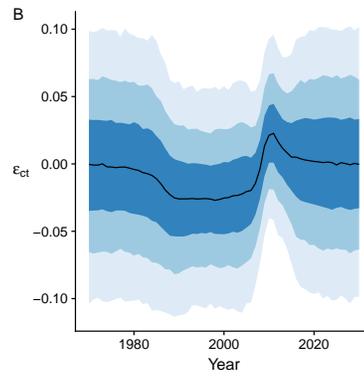 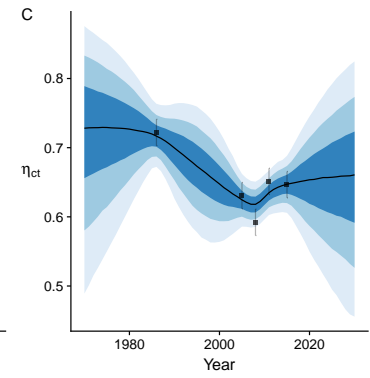

**Austria**

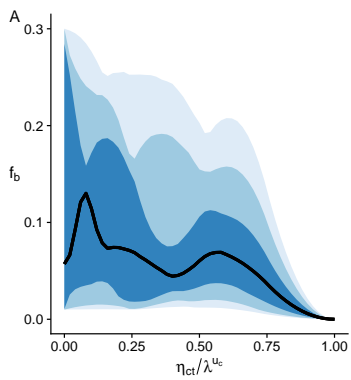 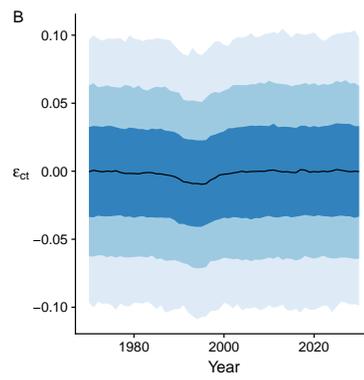 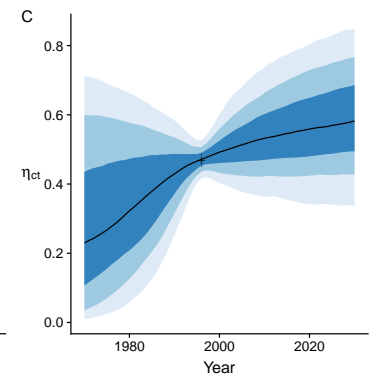

### Azerbaijan

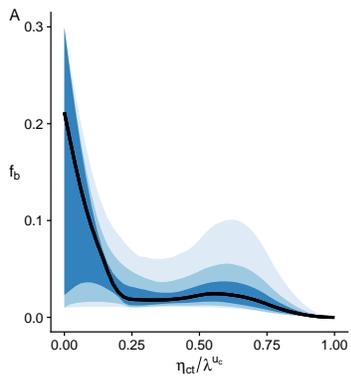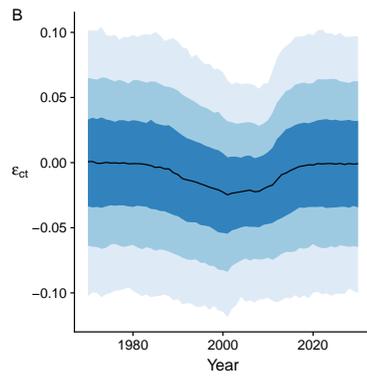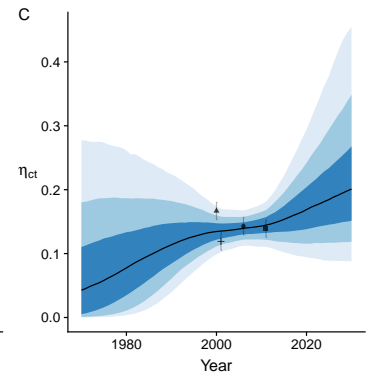

### Bahamas

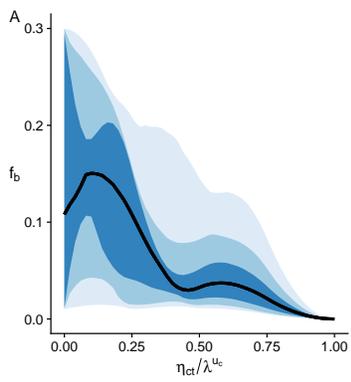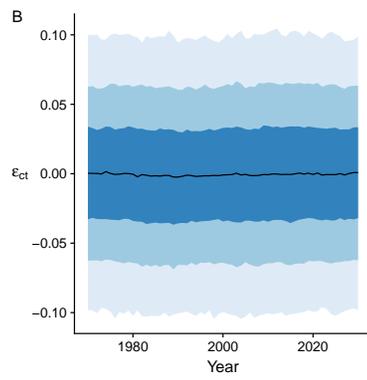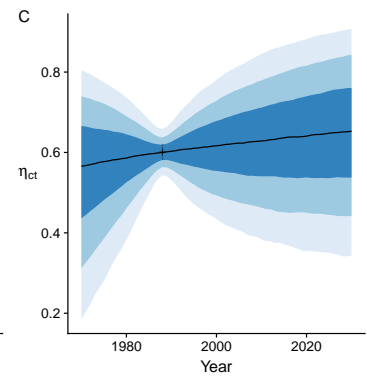

### Bangladesh

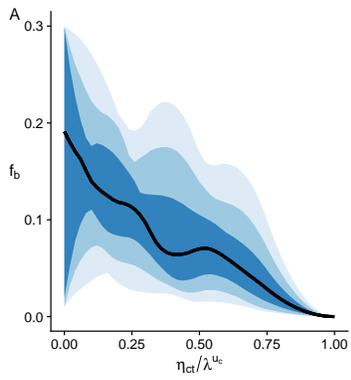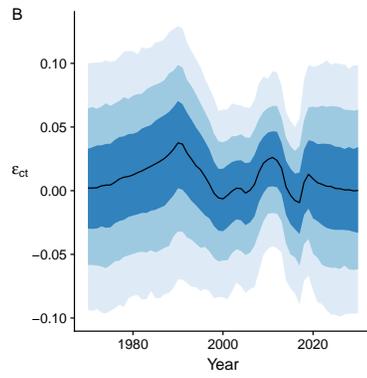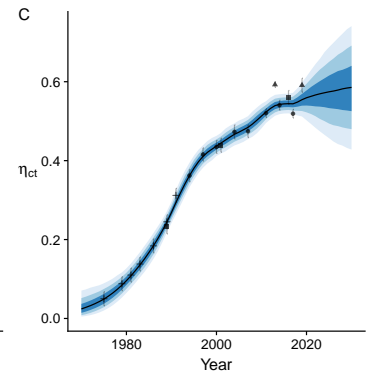

### Barbados

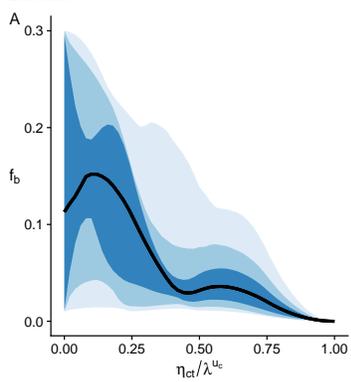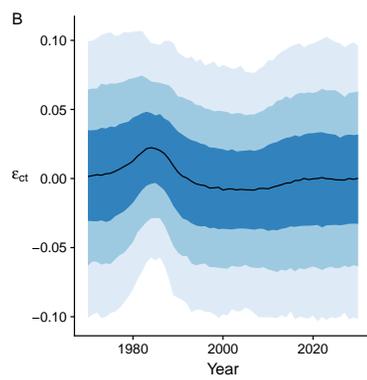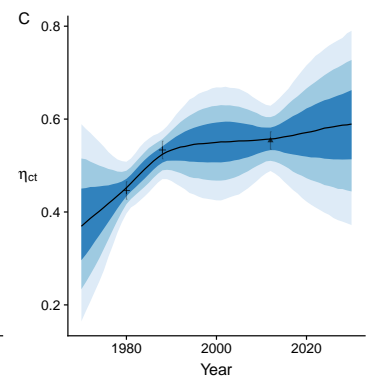

**Belarus**

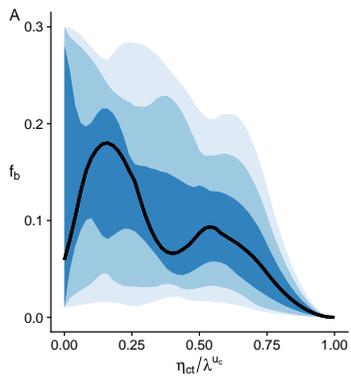 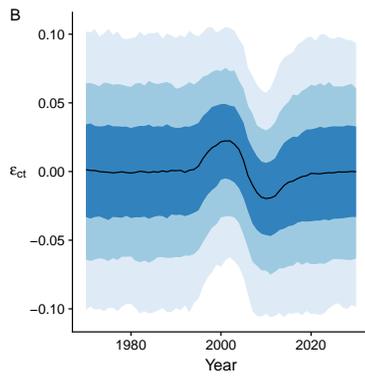 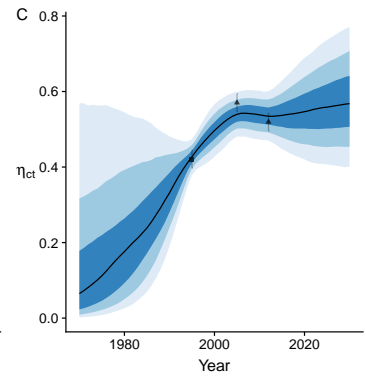

**Belgium**

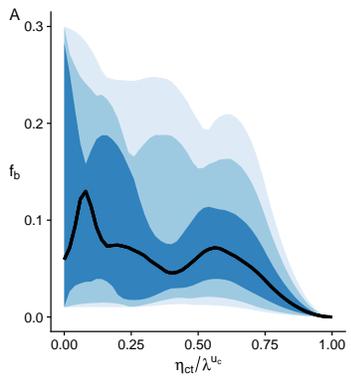 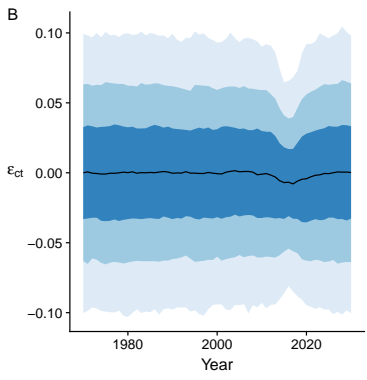 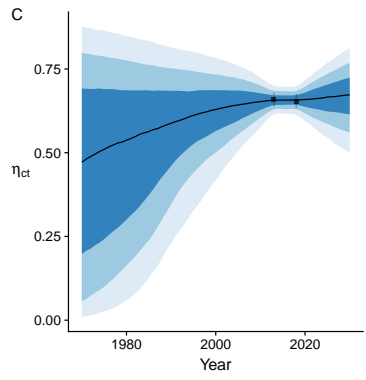

**Belize**

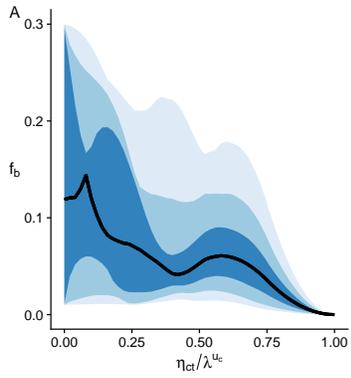 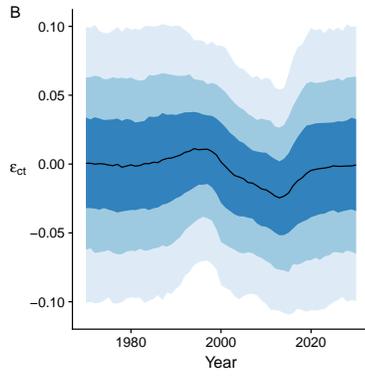 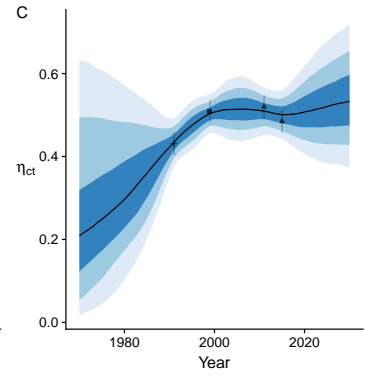

**Benin**

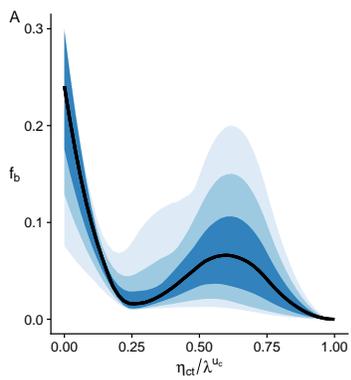 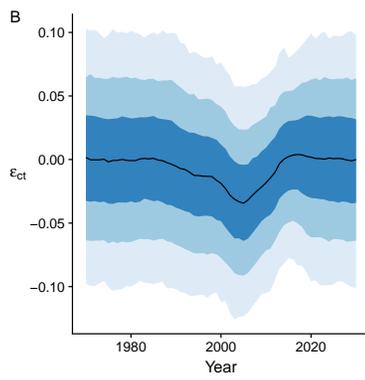 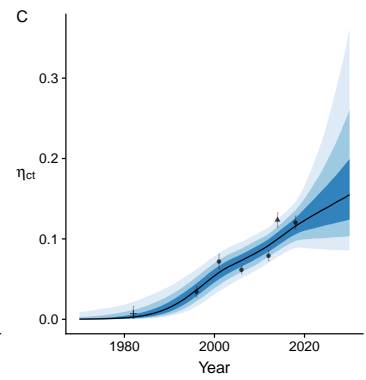

**Bhutan**

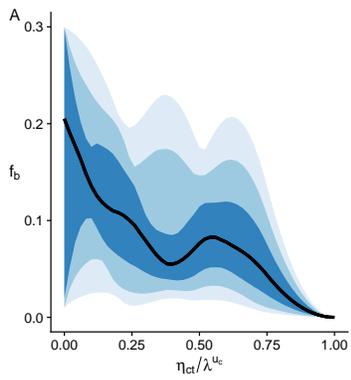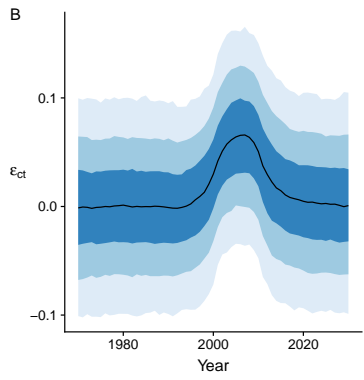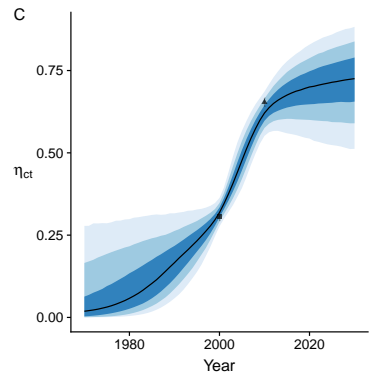

**Bolivia (Plurinational State of)**

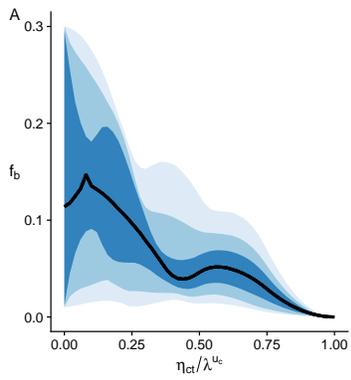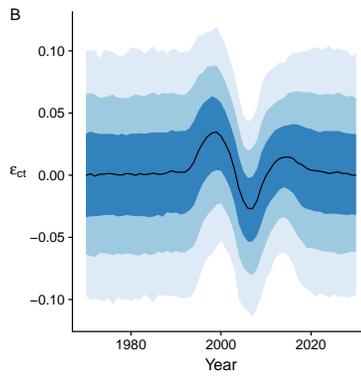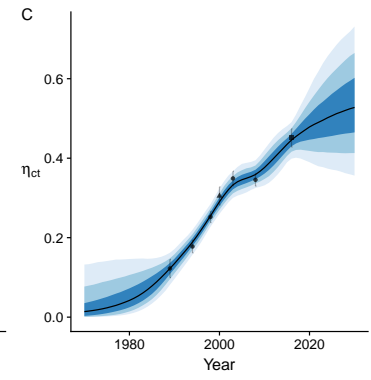

**Bosnia and Herzegovina**

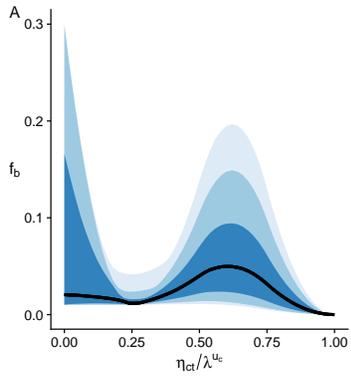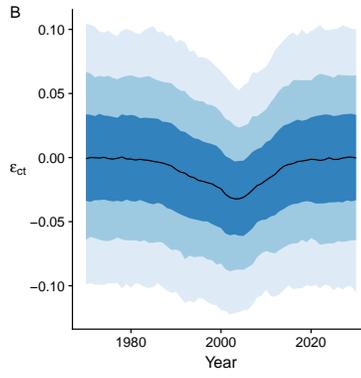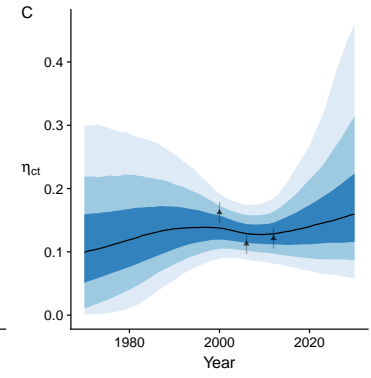

**Botswana**

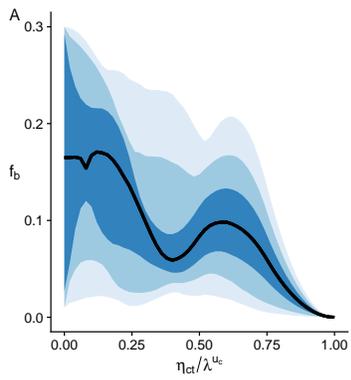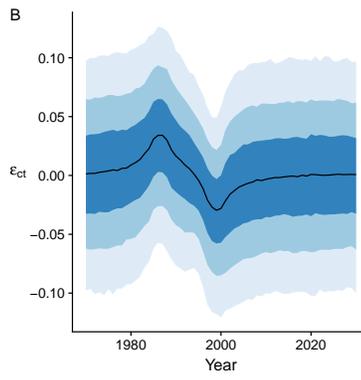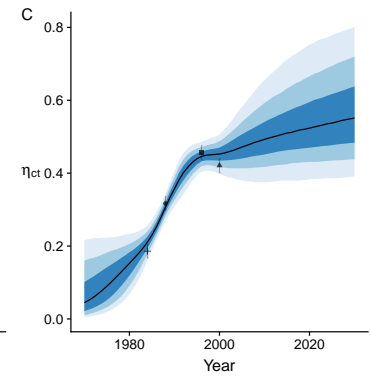

**Brazil**

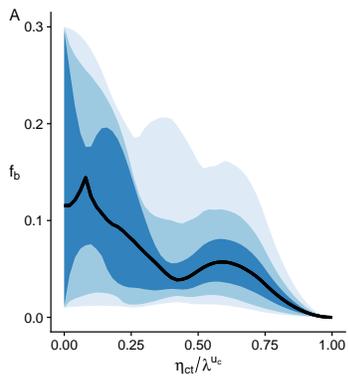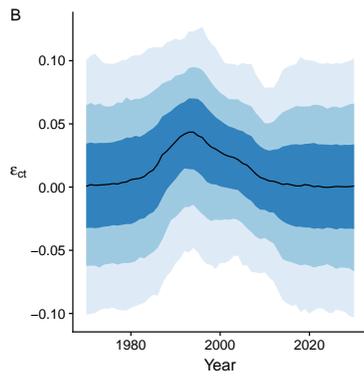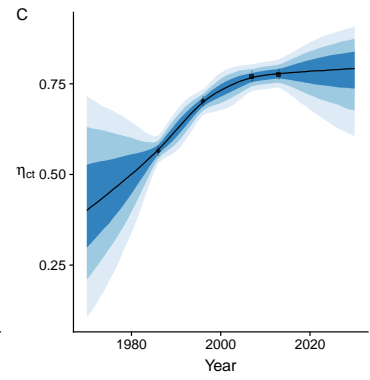

**Bulgaria**

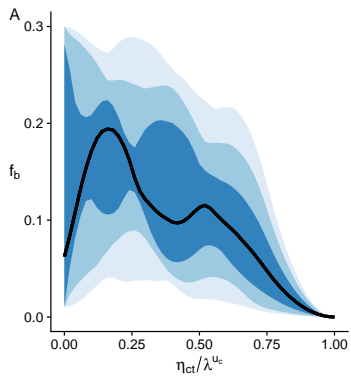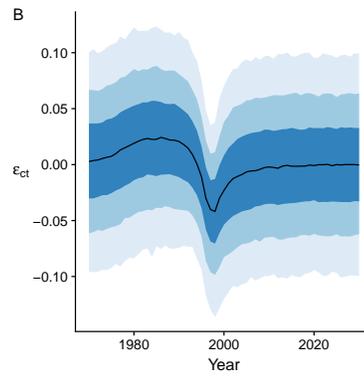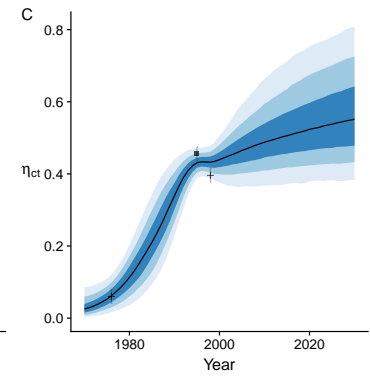

**Burkina Faso**

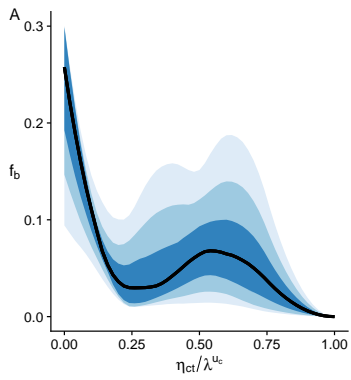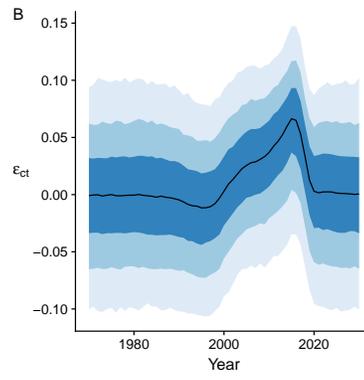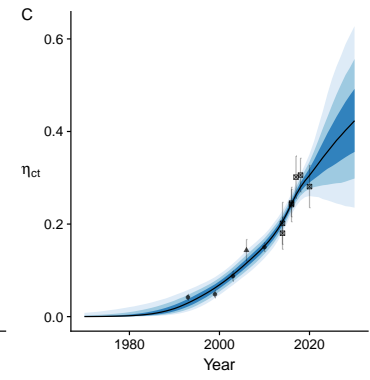

**Burundi**

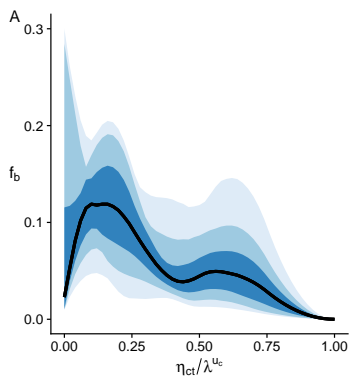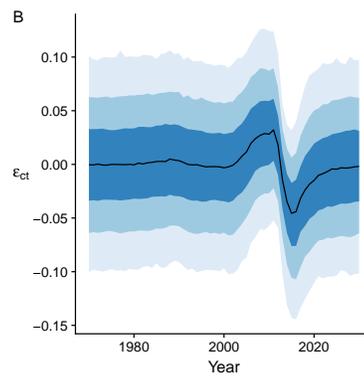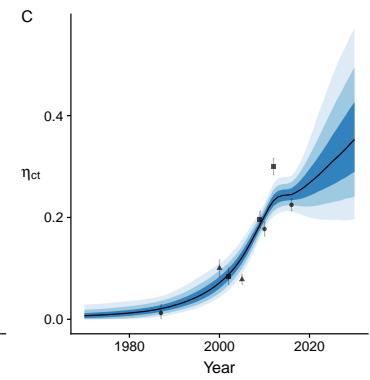

**Cabo Verde**

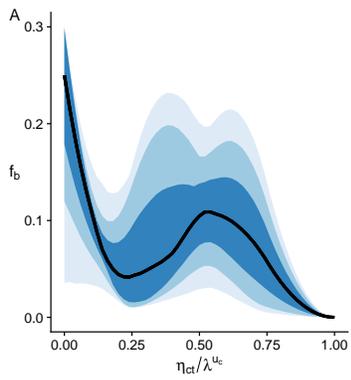
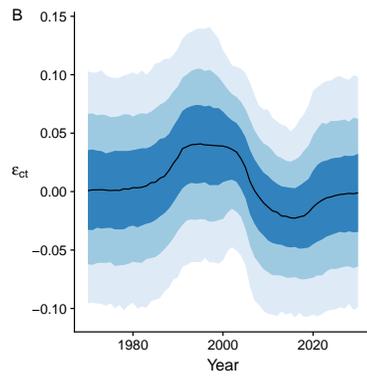
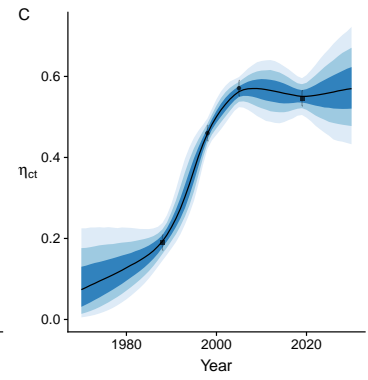

**Cambodia**

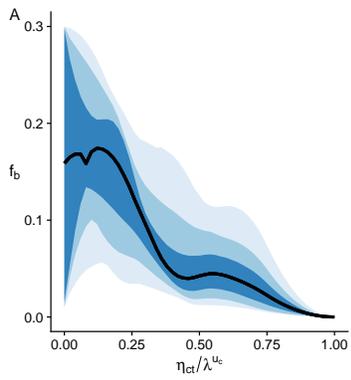
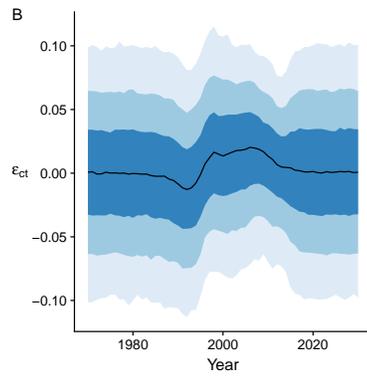
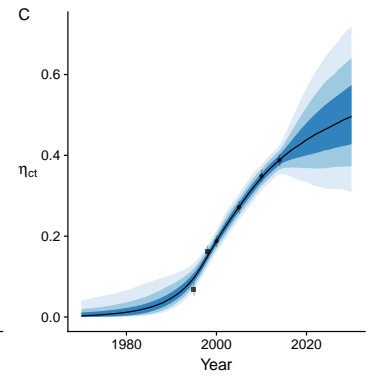

**Cameroon**

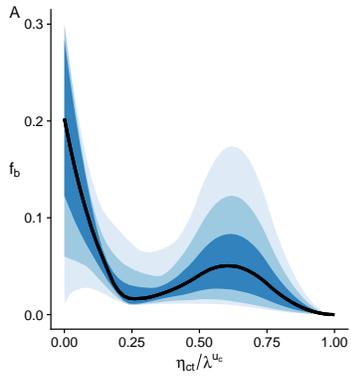
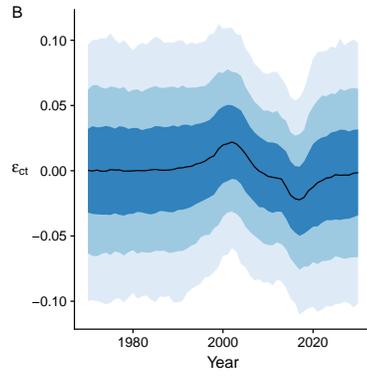
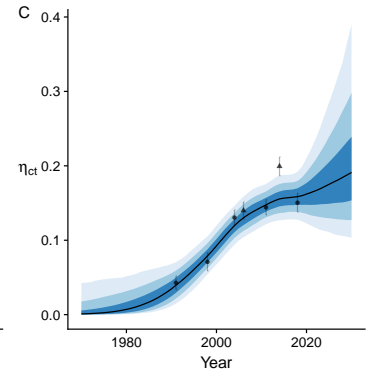

**Canada**

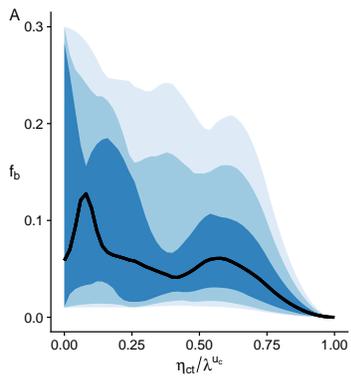
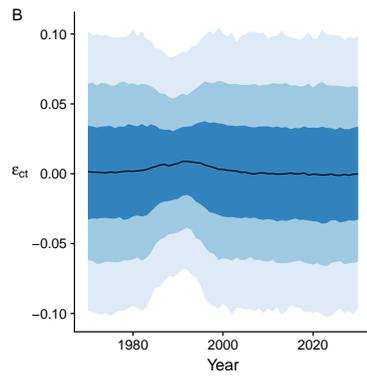
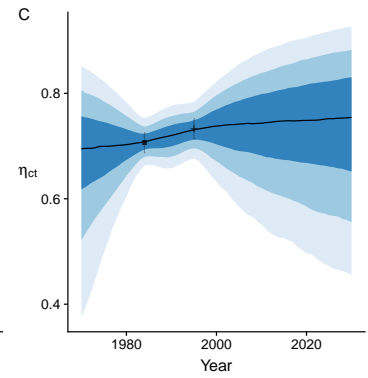

**Central African Republic**

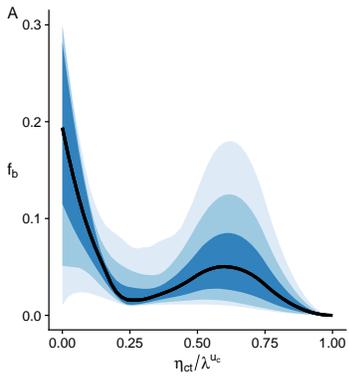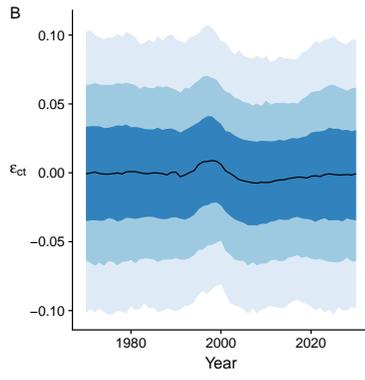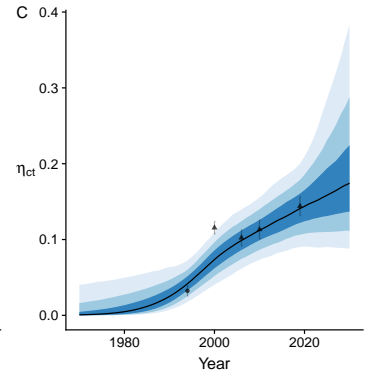

**Chad**

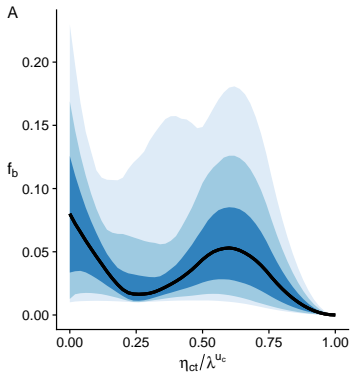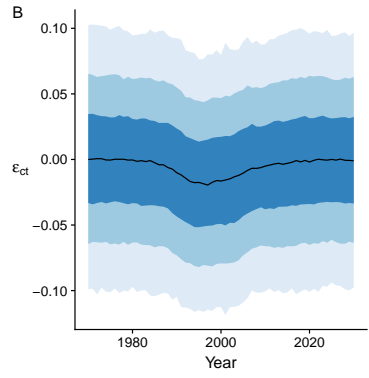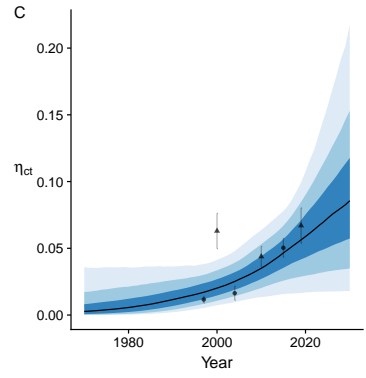

**Chile**

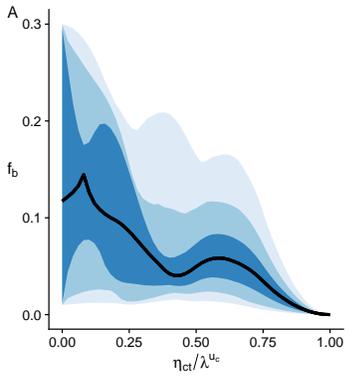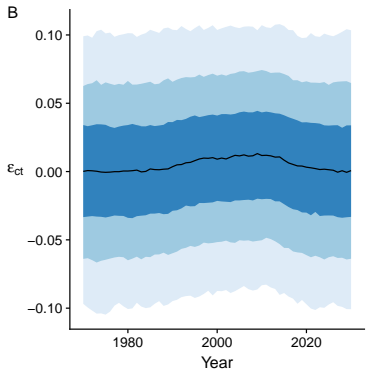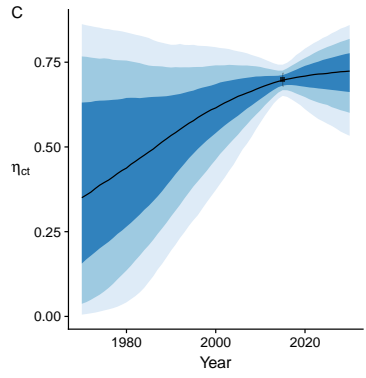

**China**

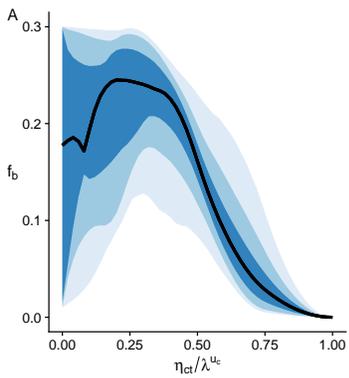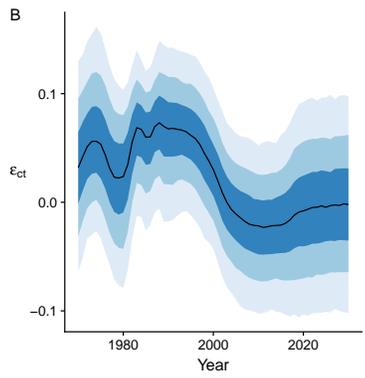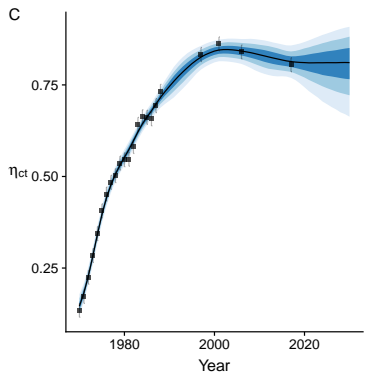

**China, Hong Kong Special Administrative Region**

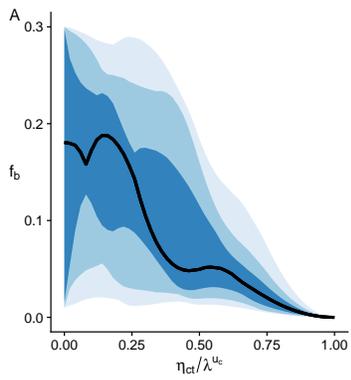 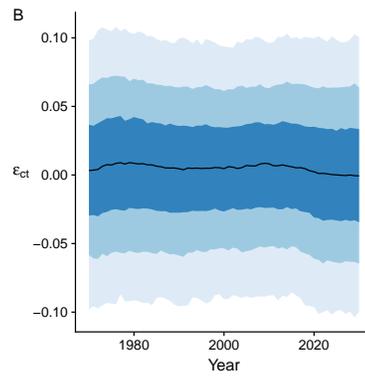 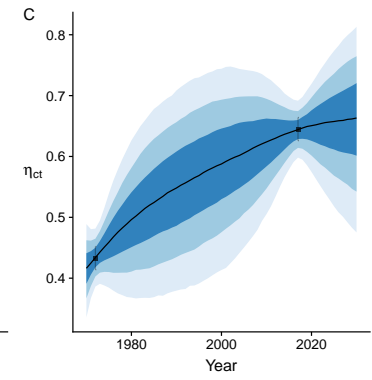

**Colombia**

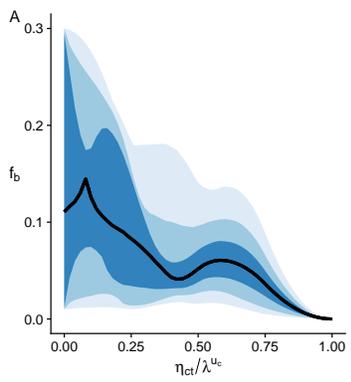 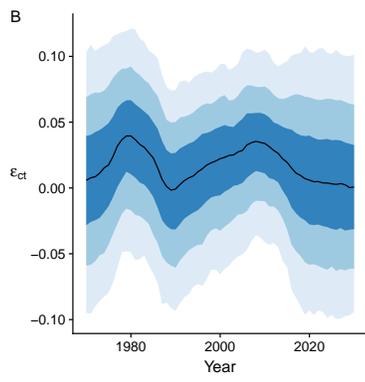 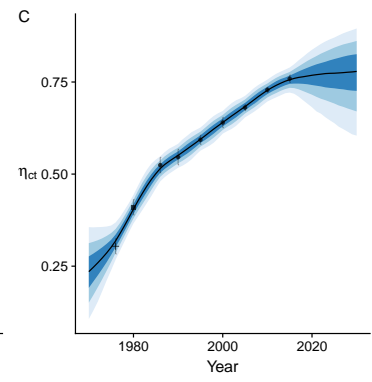

**Comoros**

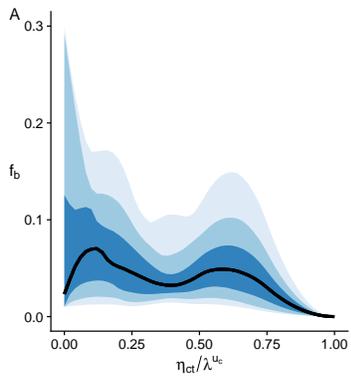 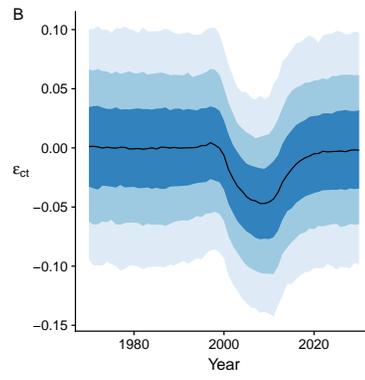 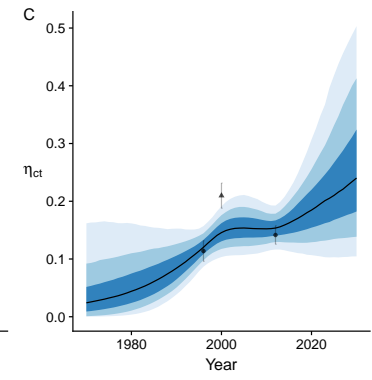

**Congo**

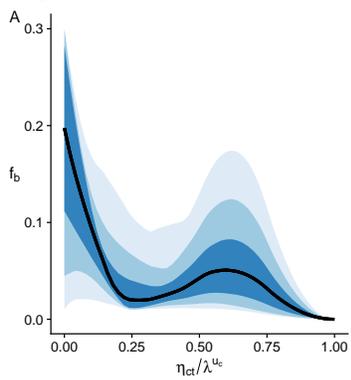 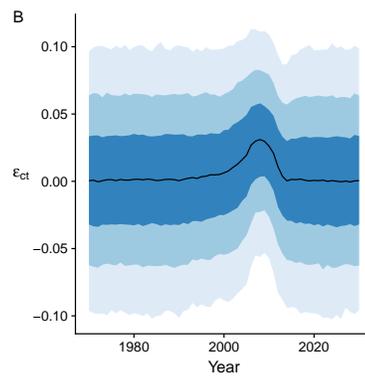 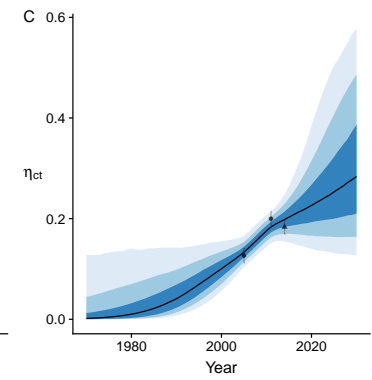

**Costa Rica**

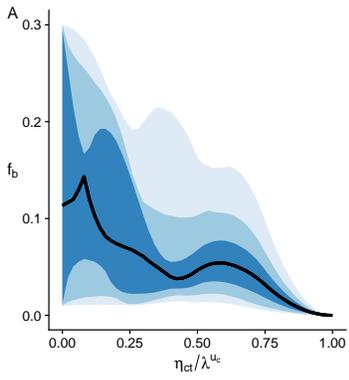 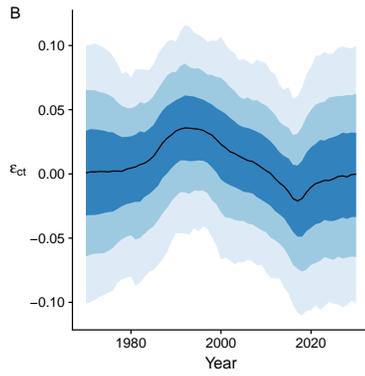 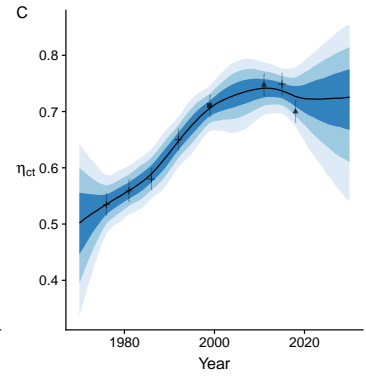

**Côte d'Ivoire**

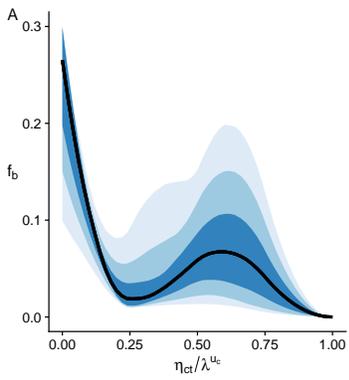 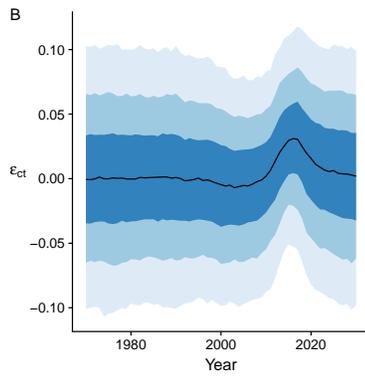 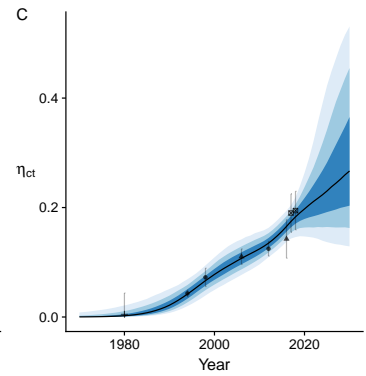

**Cuba**

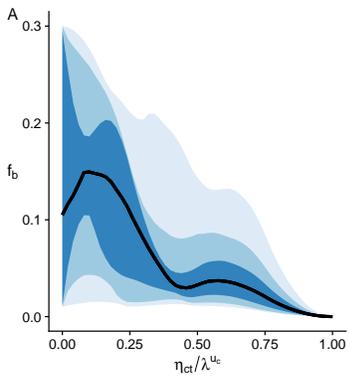 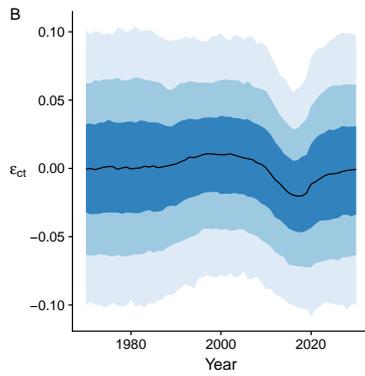 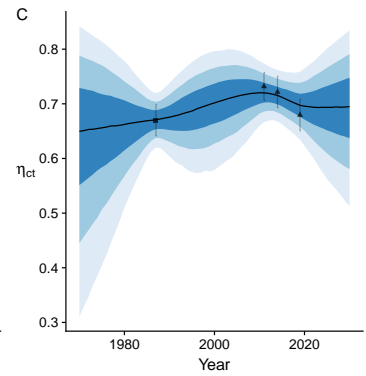

**Czechia**

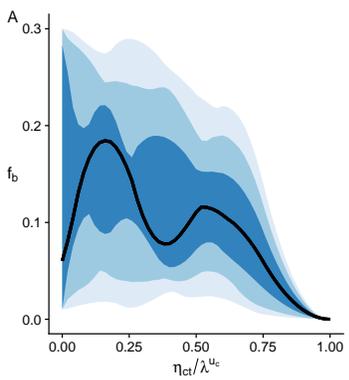 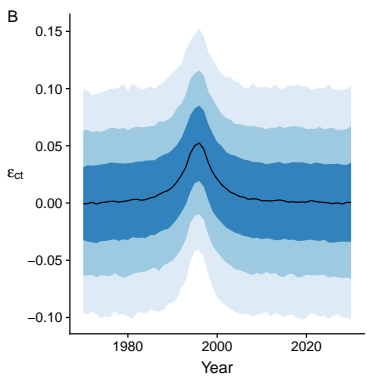 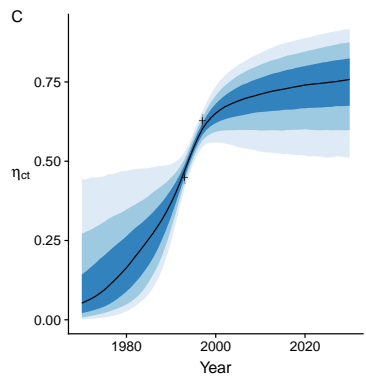

**Democratic People's Republic of Korea**

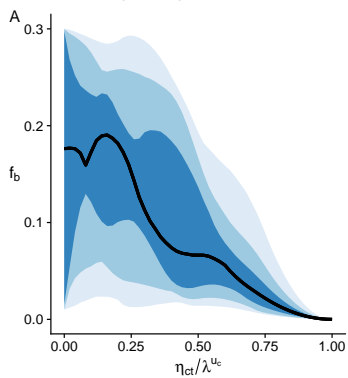 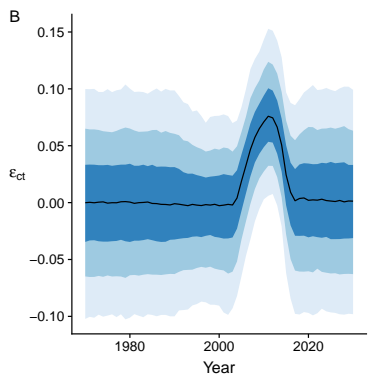 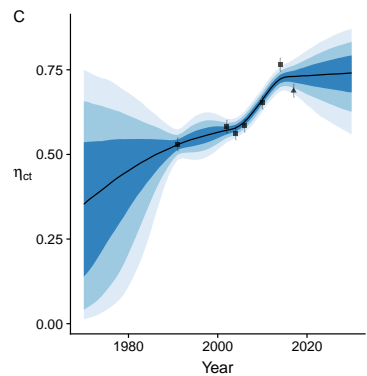

**Democratic Republic of the Congo**

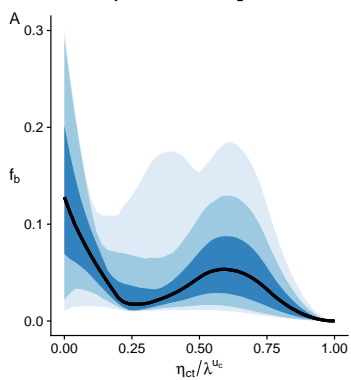 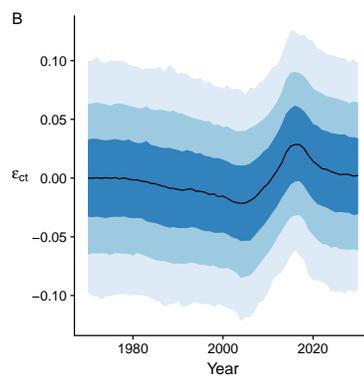 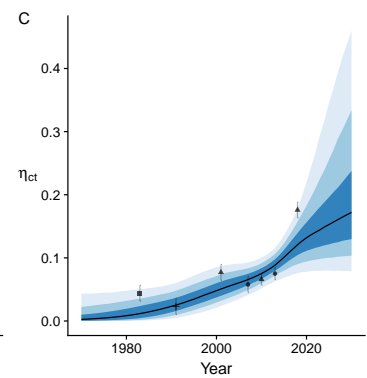

**Denmark**

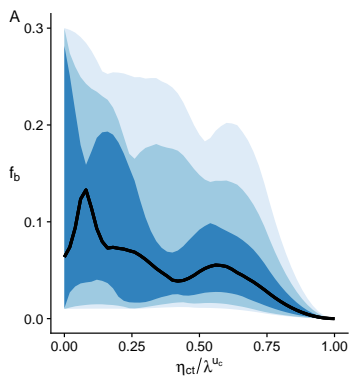 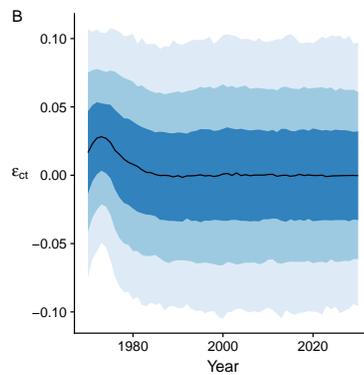 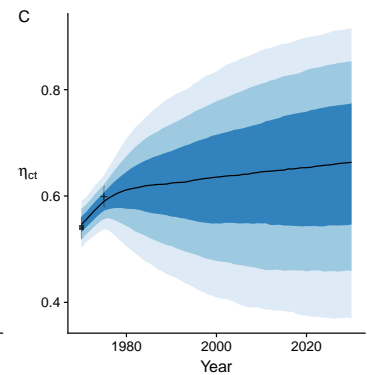

**Djibouti**

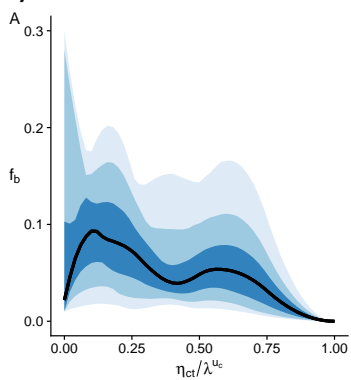 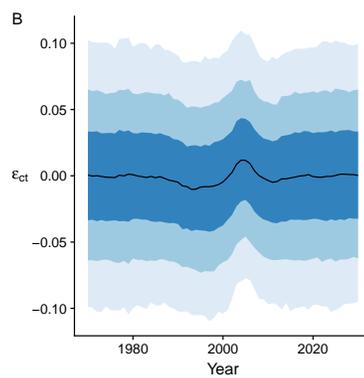 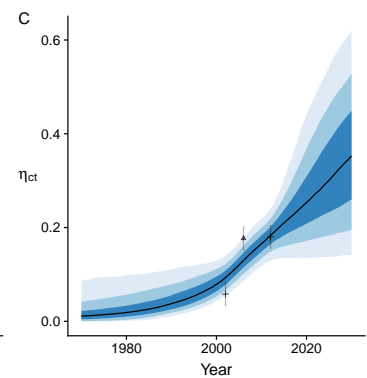

**Dominica**

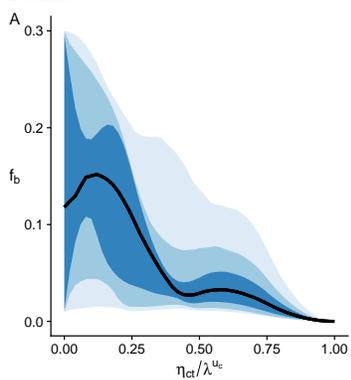
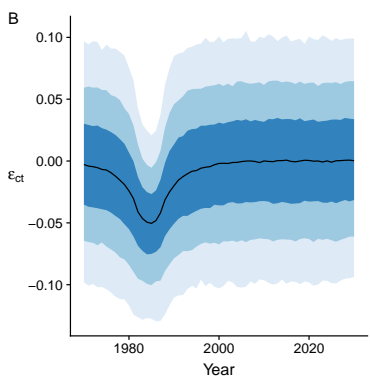
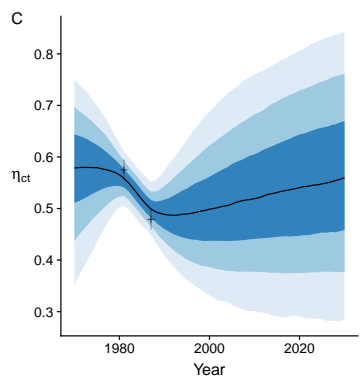

**Dominican Republic**

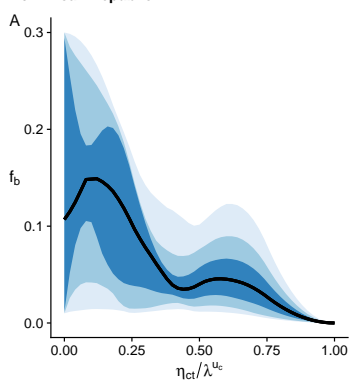
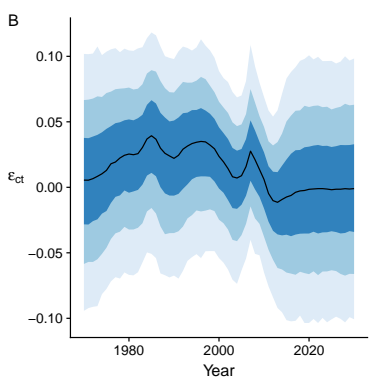
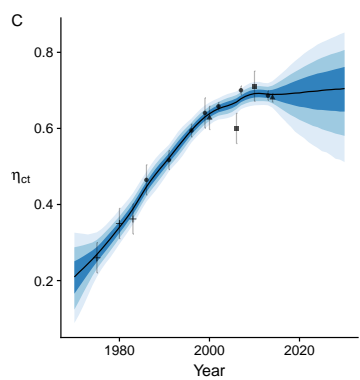

**Ecuador**

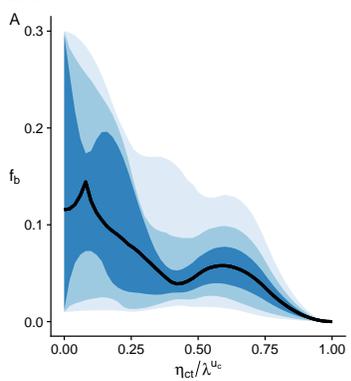
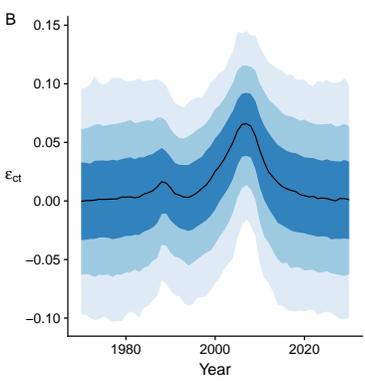
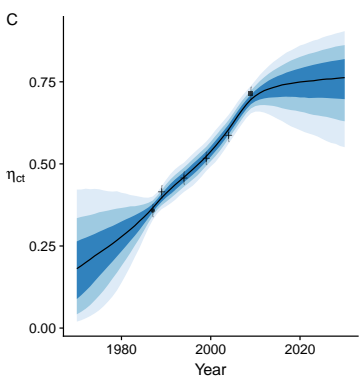

**Egypt**

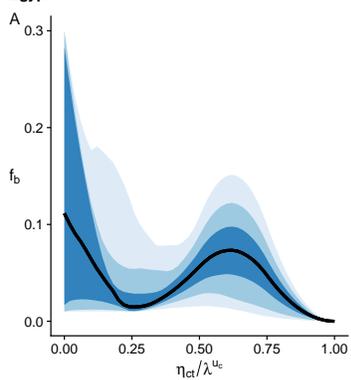
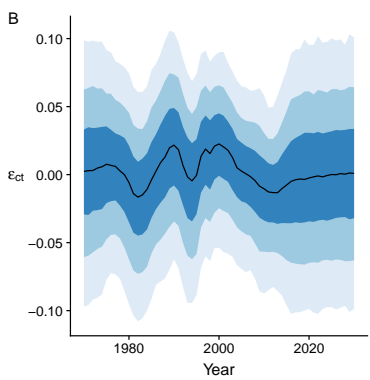
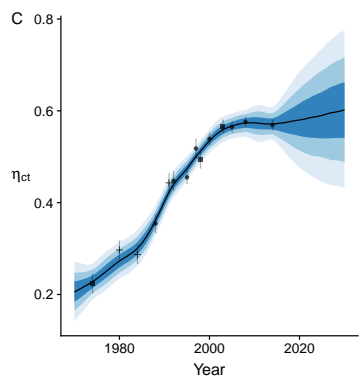

**El Salvador**

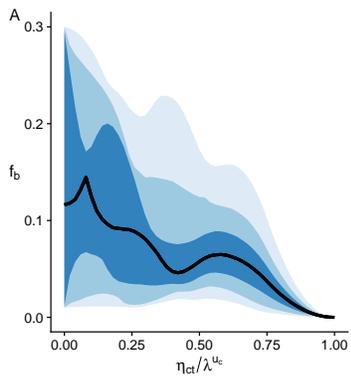
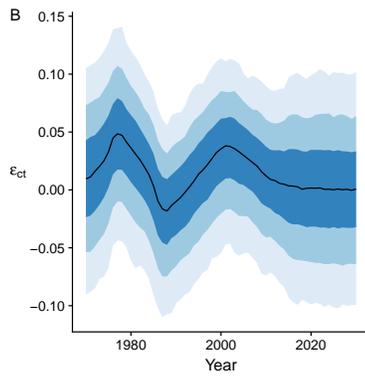
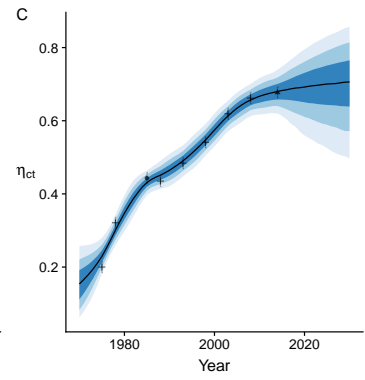

**Equatorial Guinea**

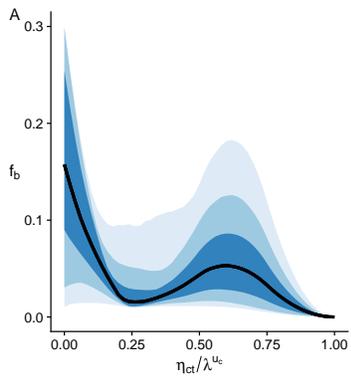
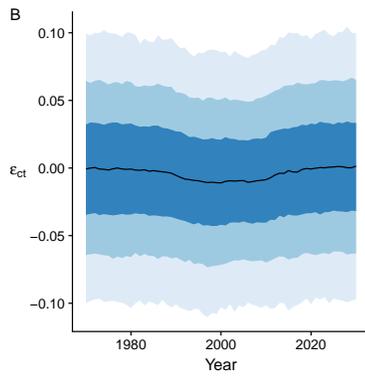
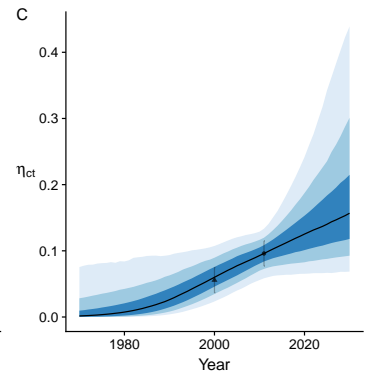

**Eritrea**

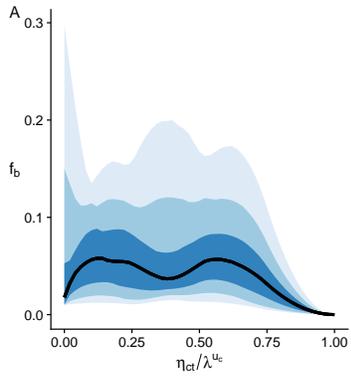
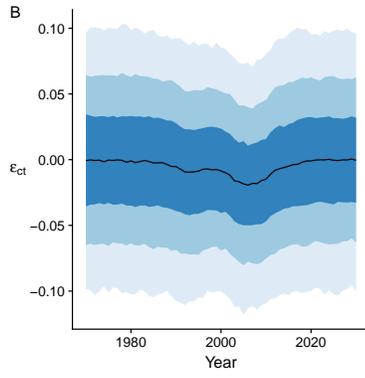
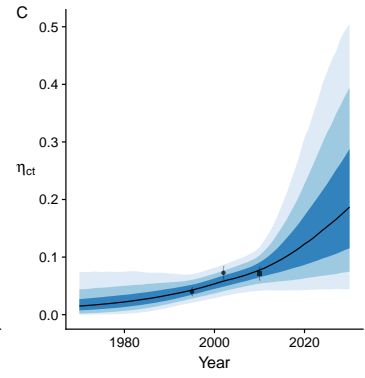

**Estonia**

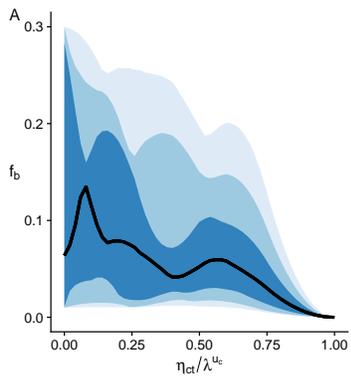
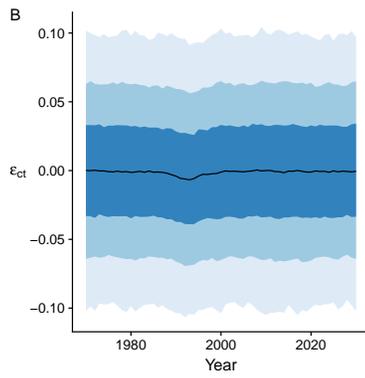
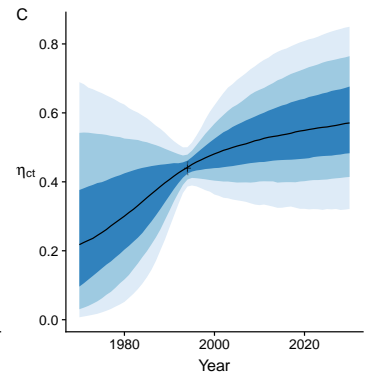

**Swaziland**

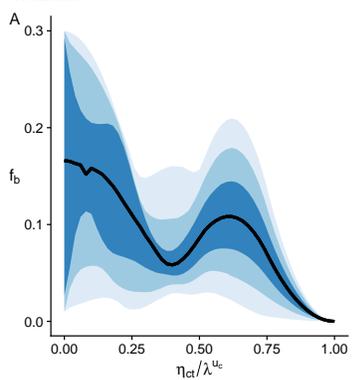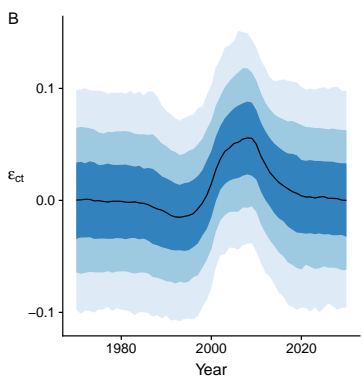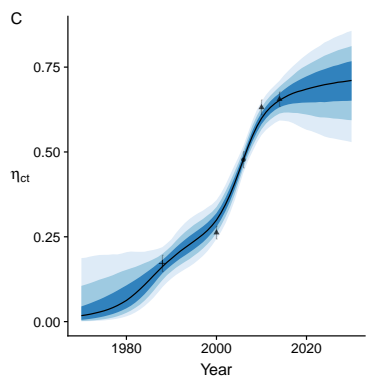

**Ethiopia**

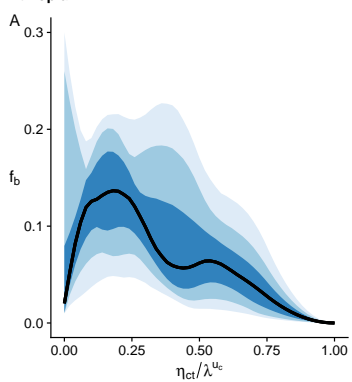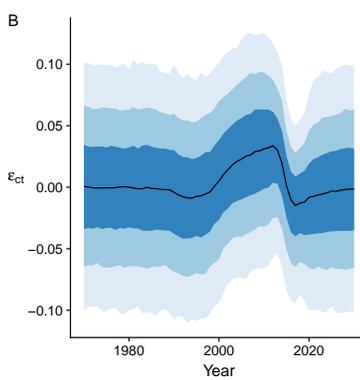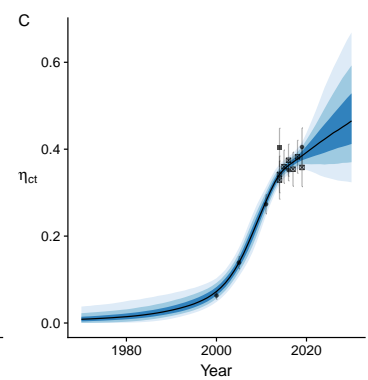

**Fiji**

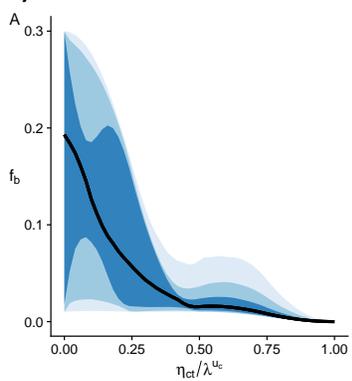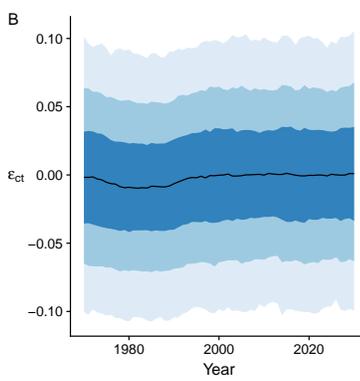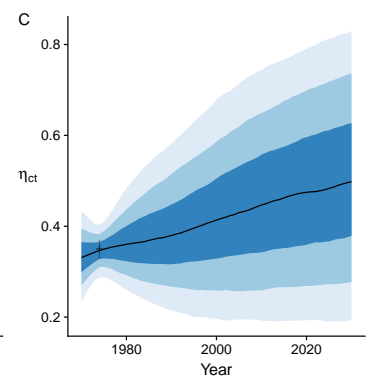

**Finland**

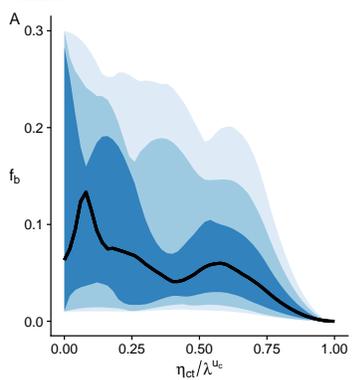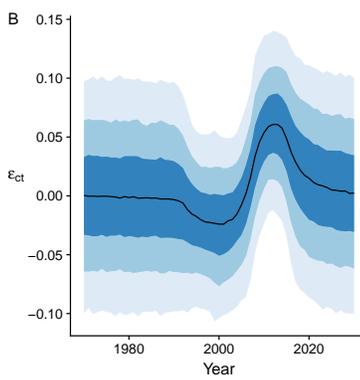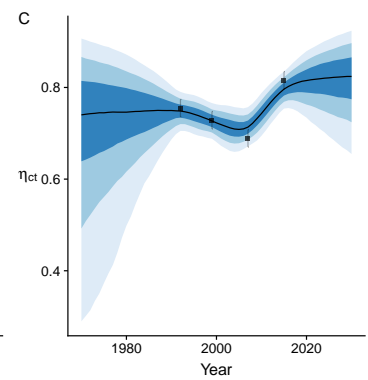

**France**

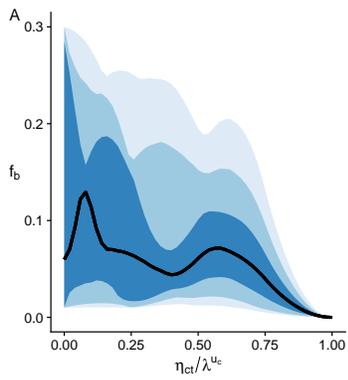 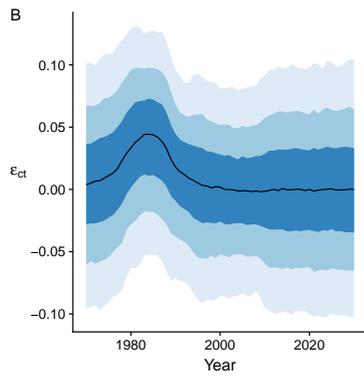 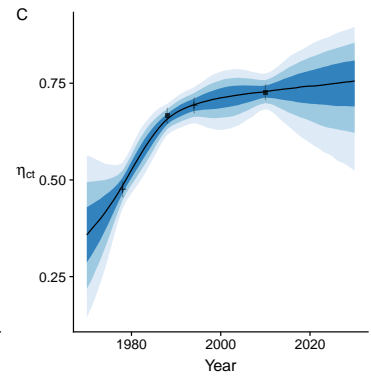

**Gabon**

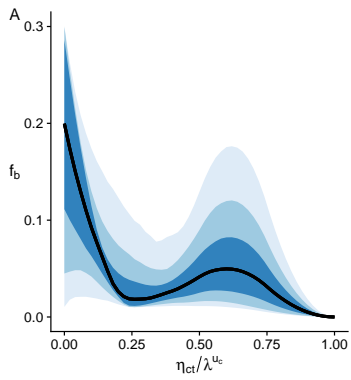 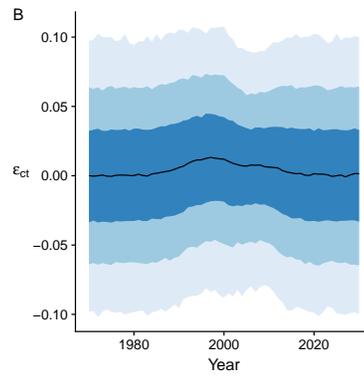 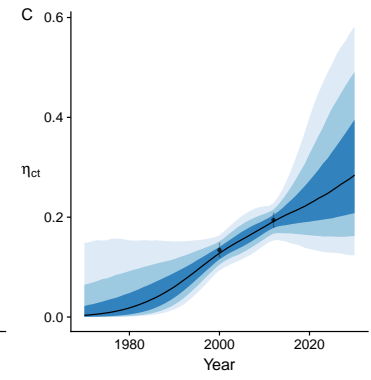

**Gambia**

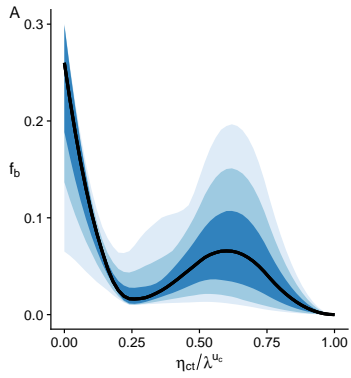 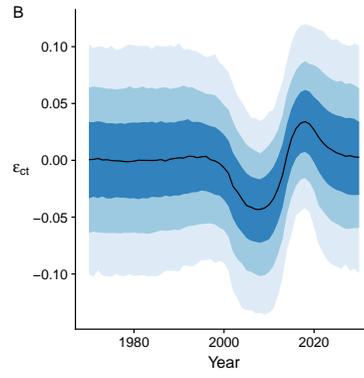 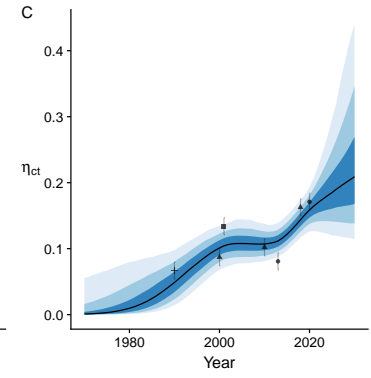

**Georgia**

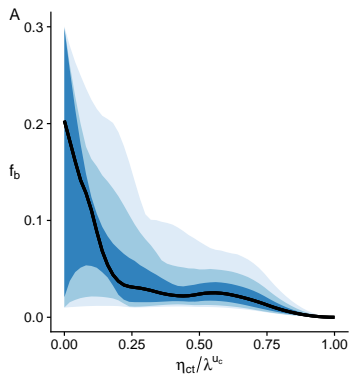 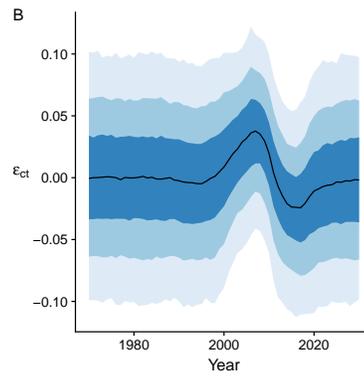 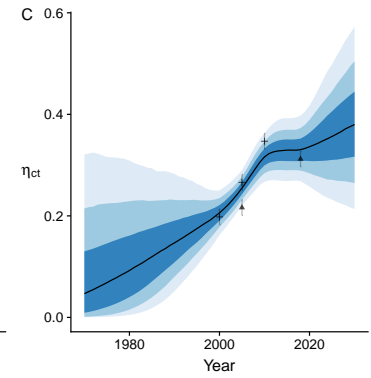

**Germany**

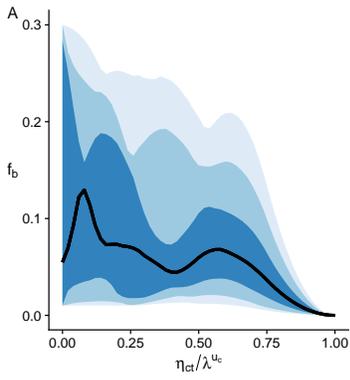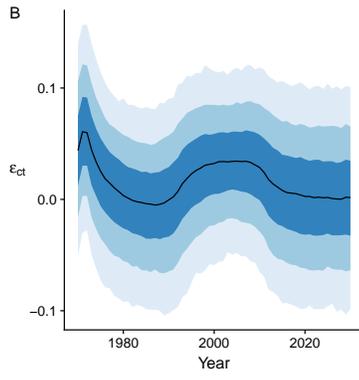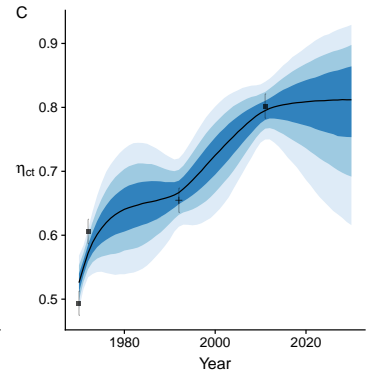

**Ghana**

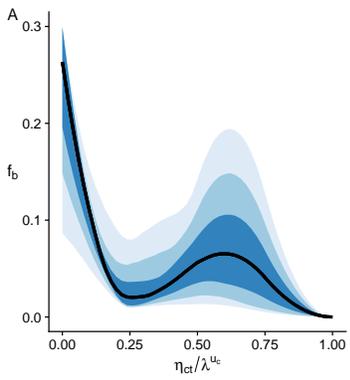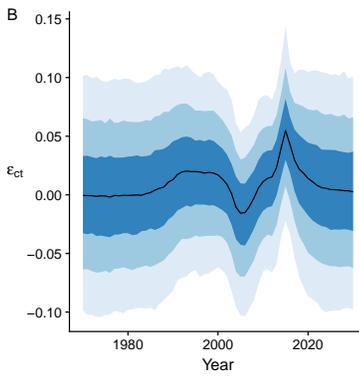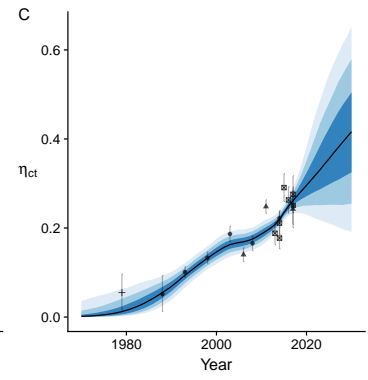

**Grenada**

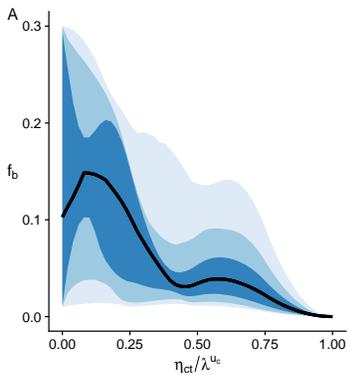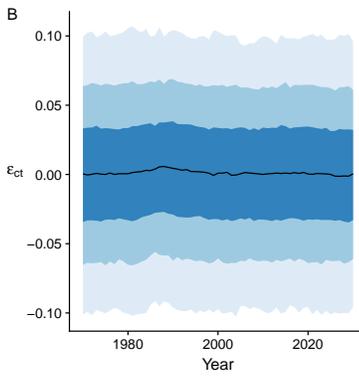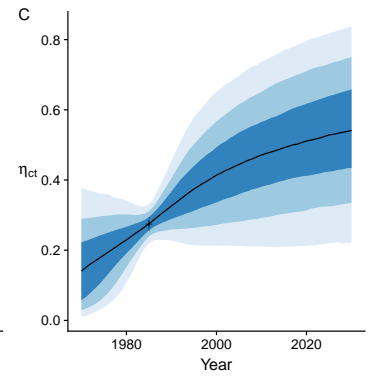

**Guatemala**

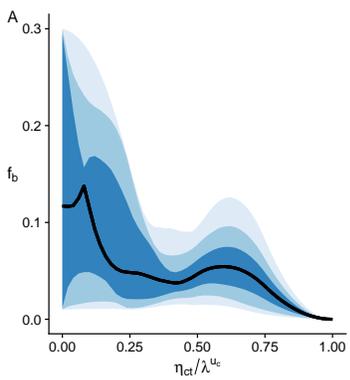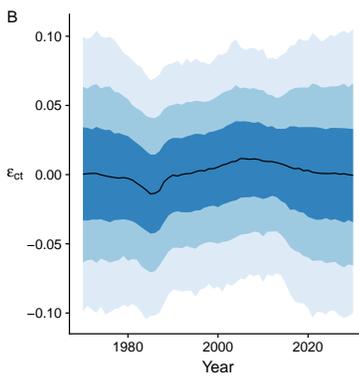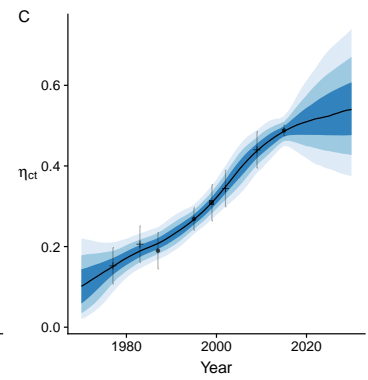

**Guinea**

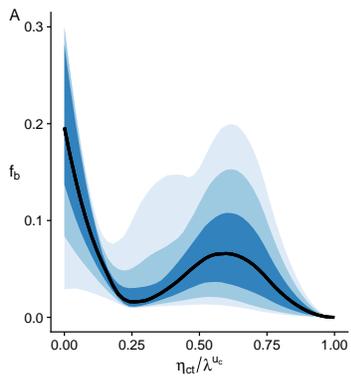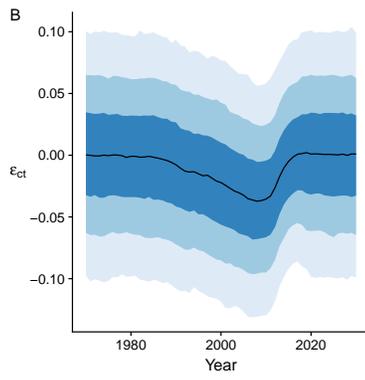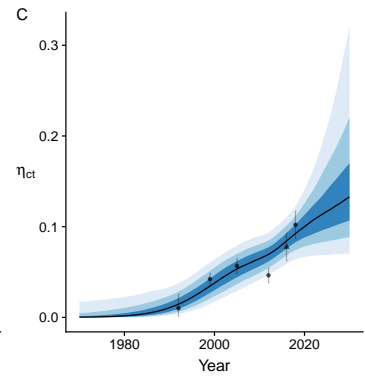

**Guinea–Bissau**

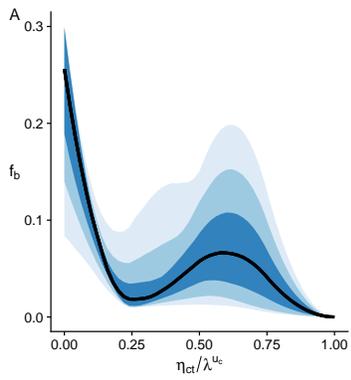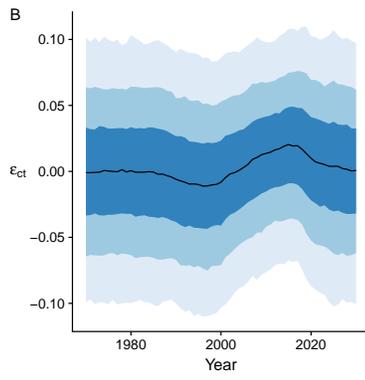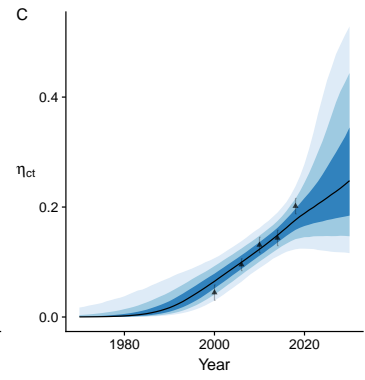

**Guyana**

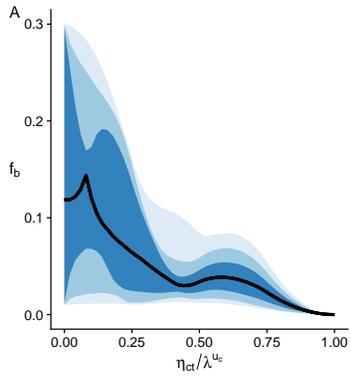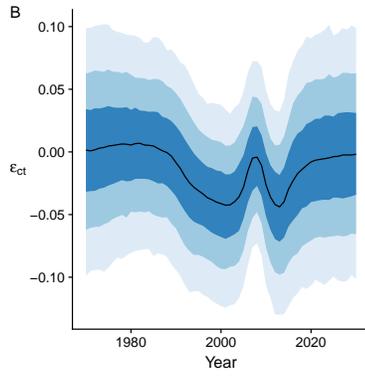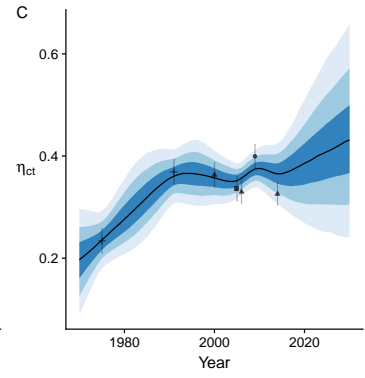

**Haiti**

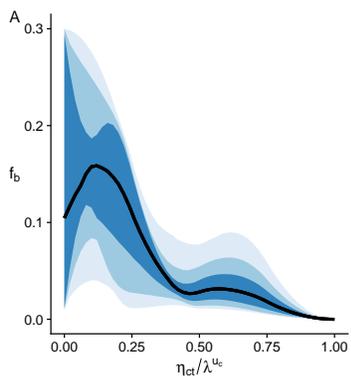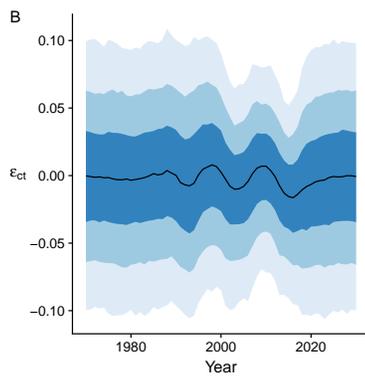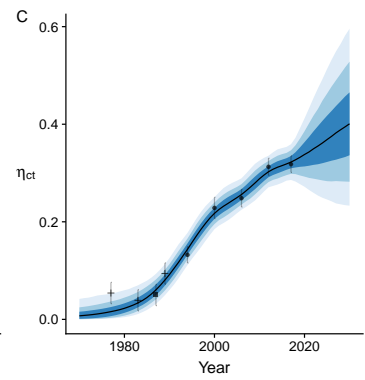

**Honduras**

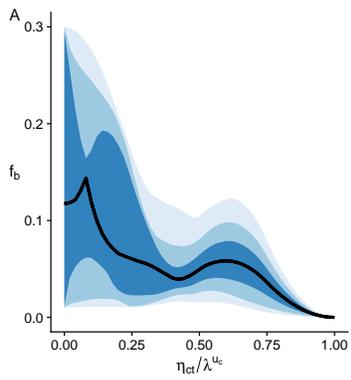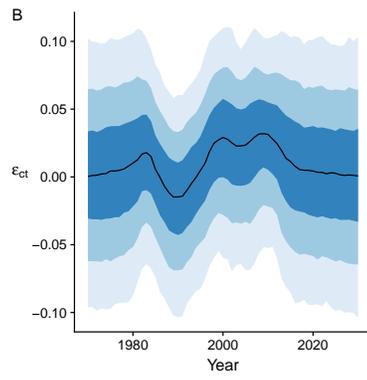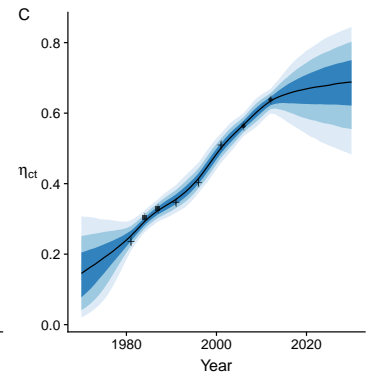

**Hungary**

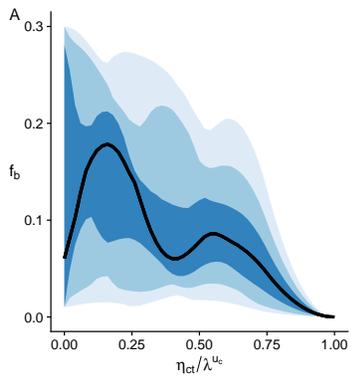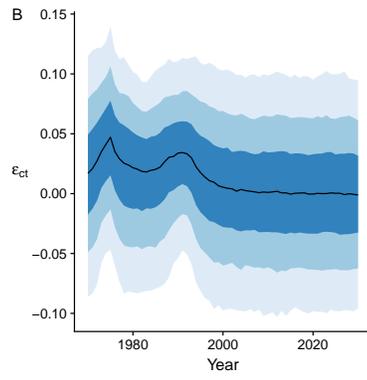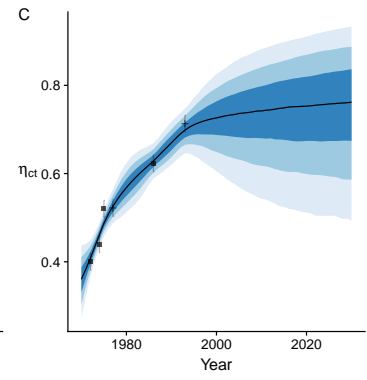

**India**

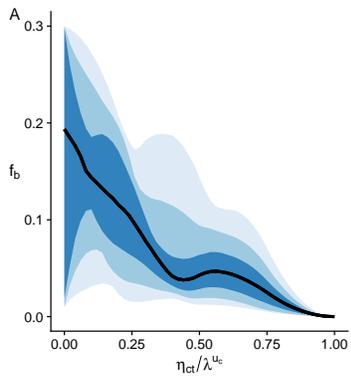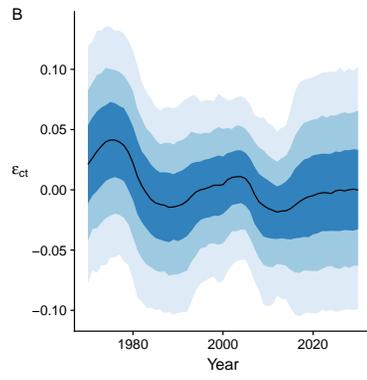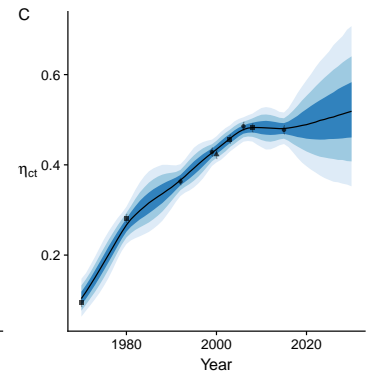

**Indonesia**

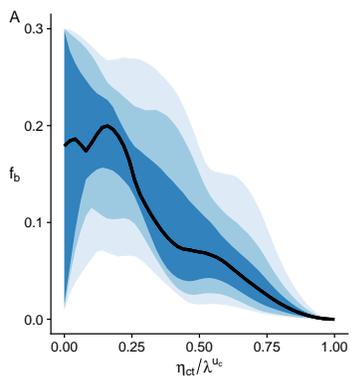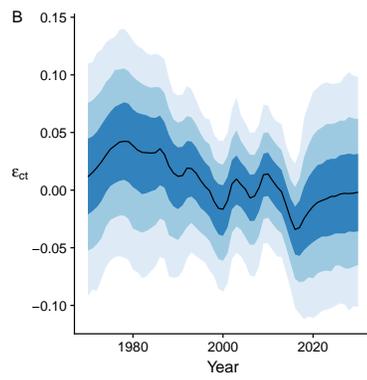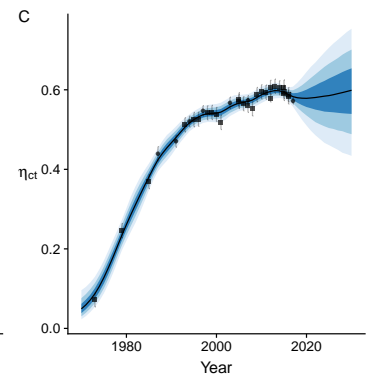

**Iran (Islamic Republic of)**

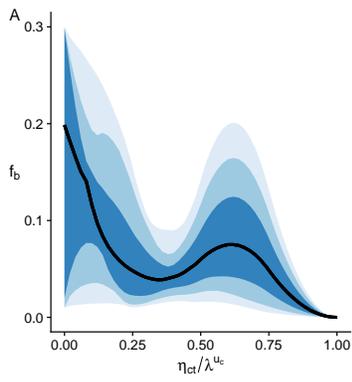 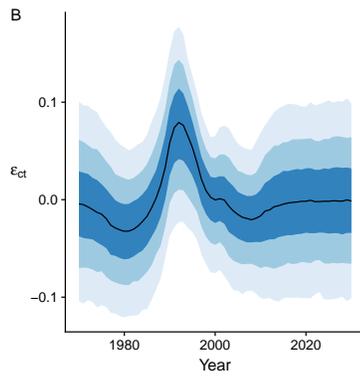 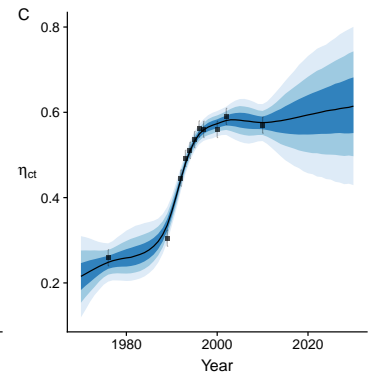

**Iraq**

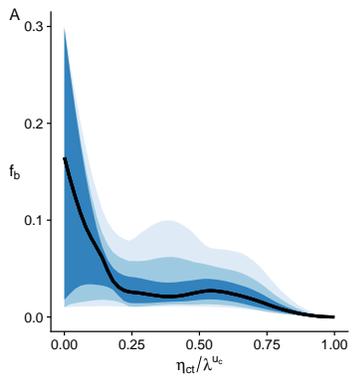 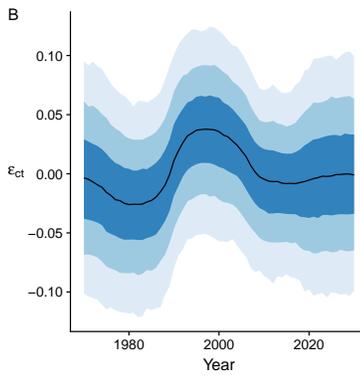 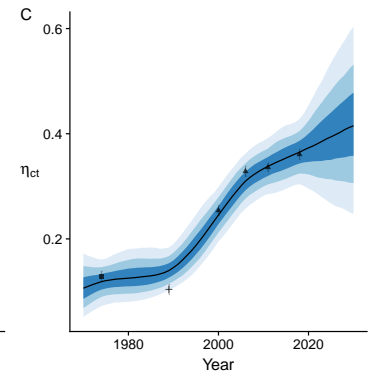

**Ireland**

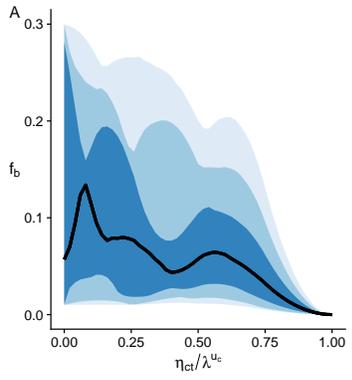 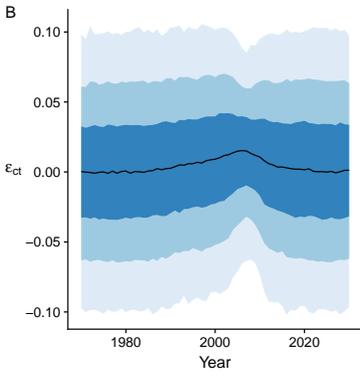 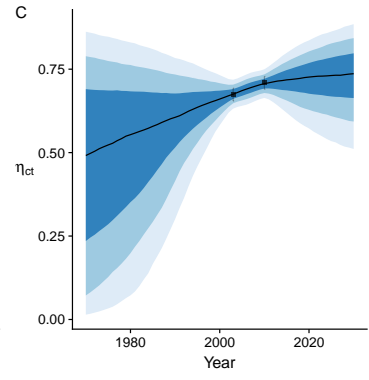

**Israel**

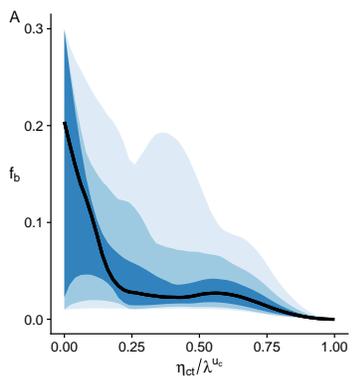 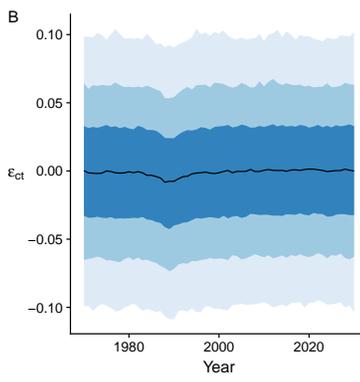 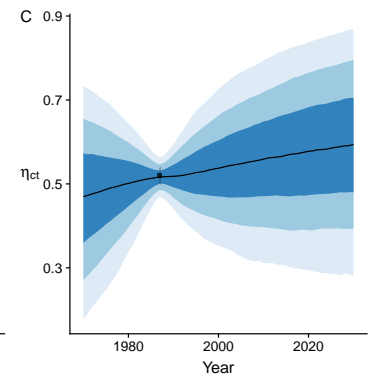

**Italy**

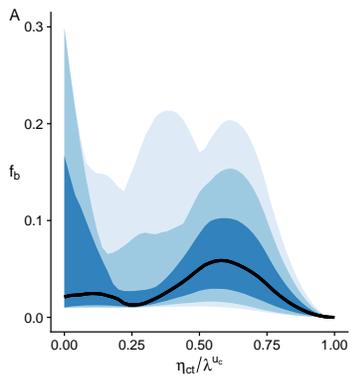
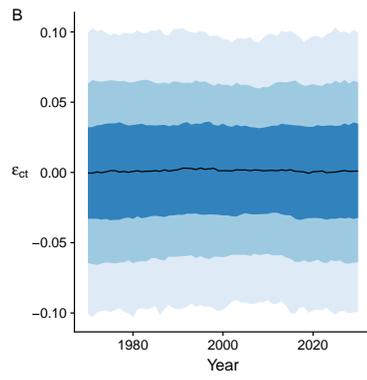
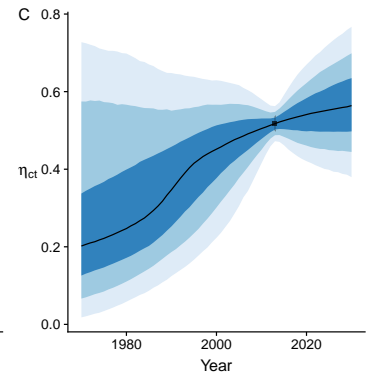

**Jamaica**

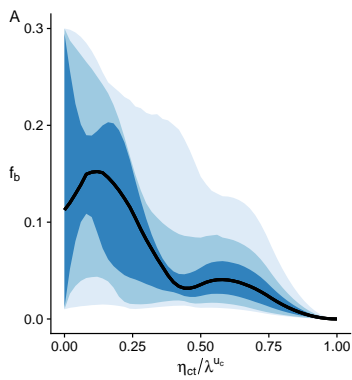
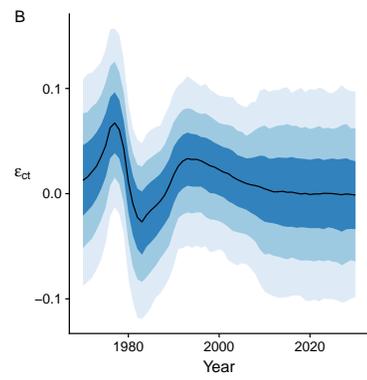
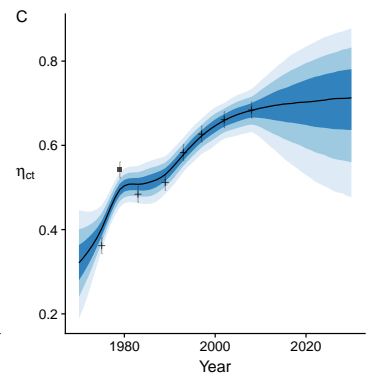

**Japan**

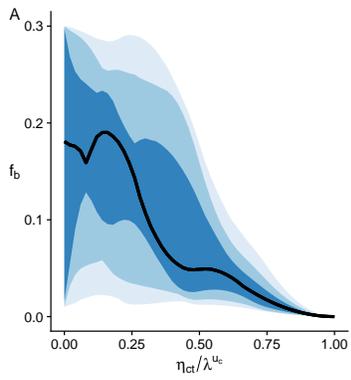
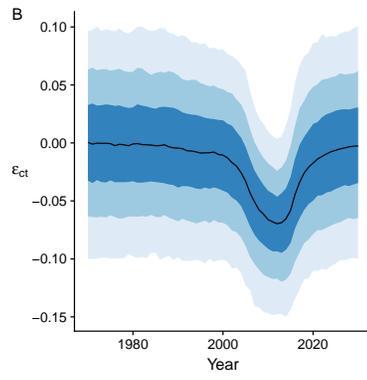
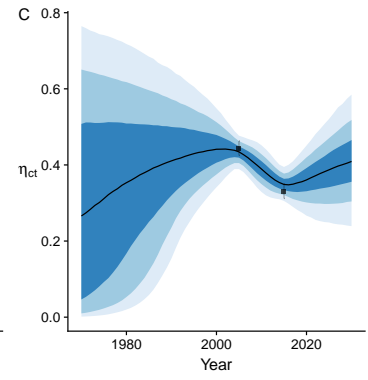

**Jordan**

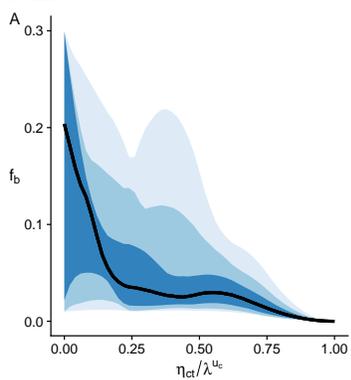
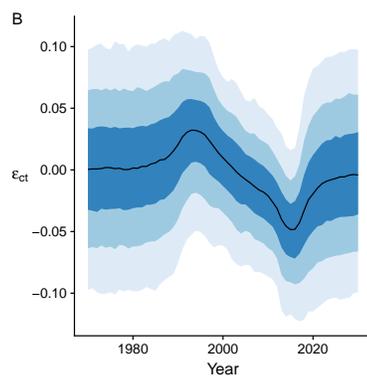
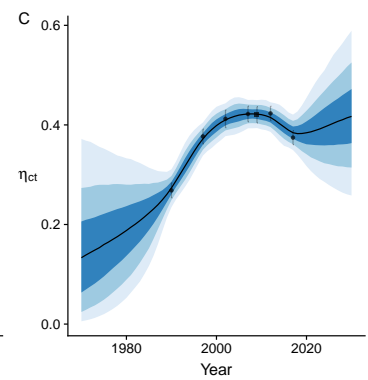

**Kazakhstan**

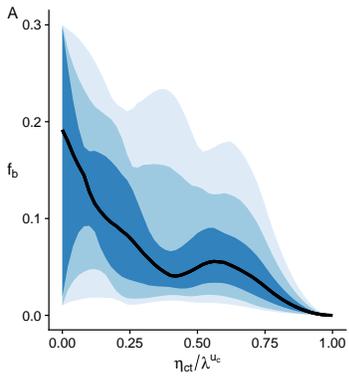 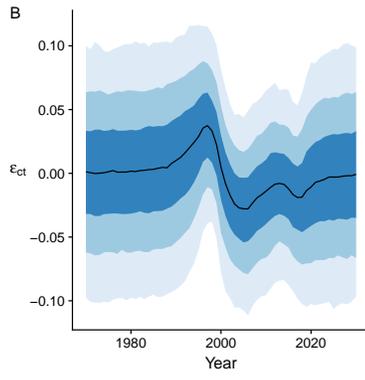 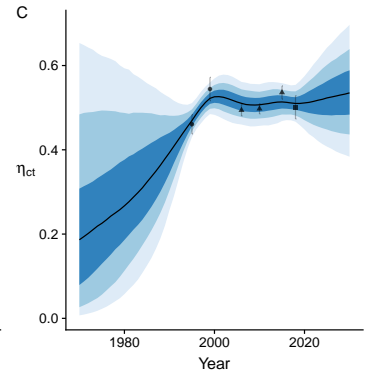

**Kenya**

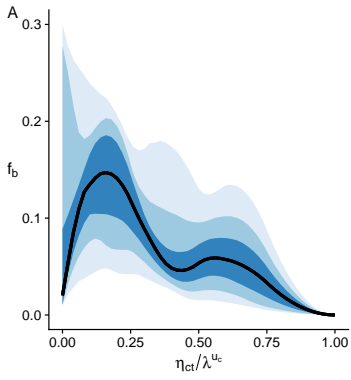 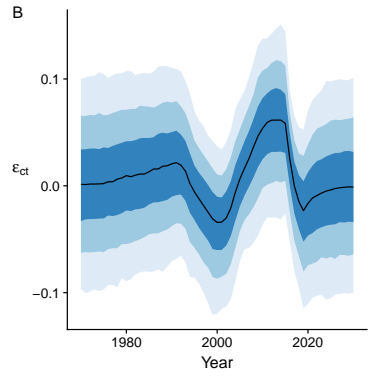 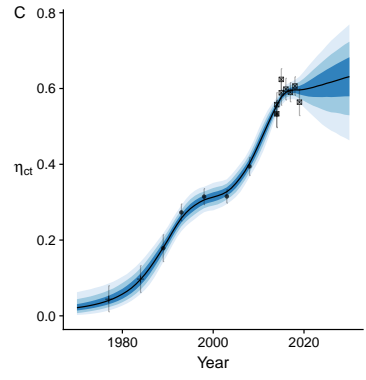

**Kiribati**

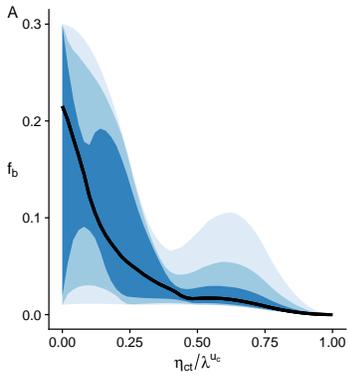 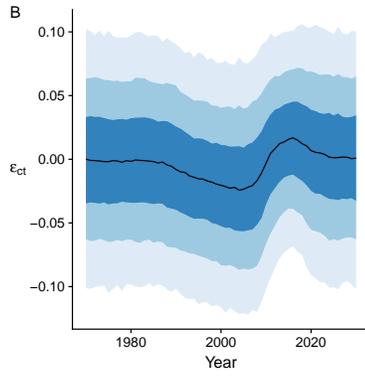 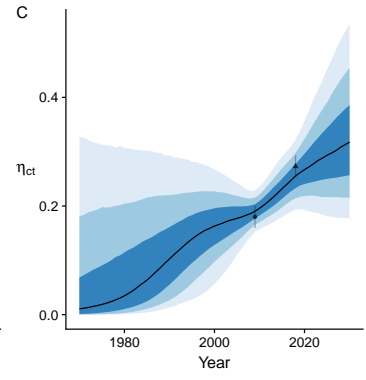

**Kuwait**

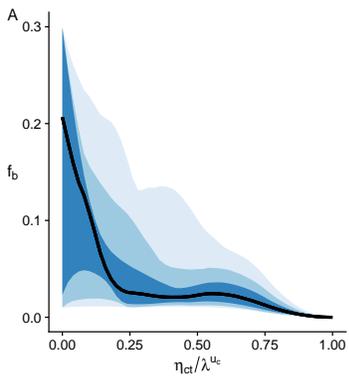 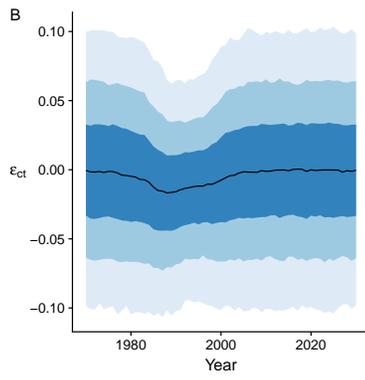 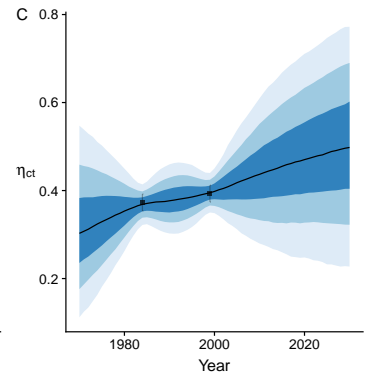

## Kyrgyzstan

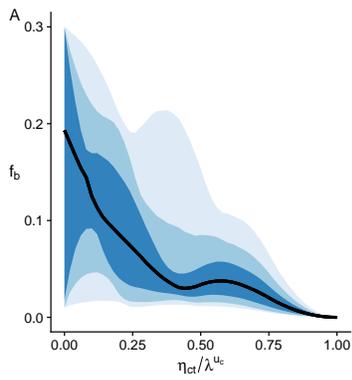 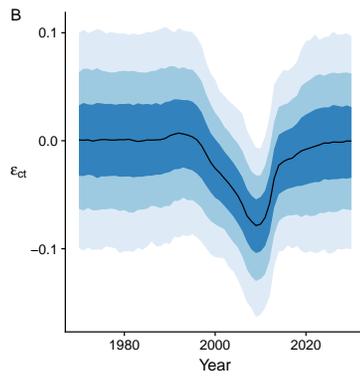 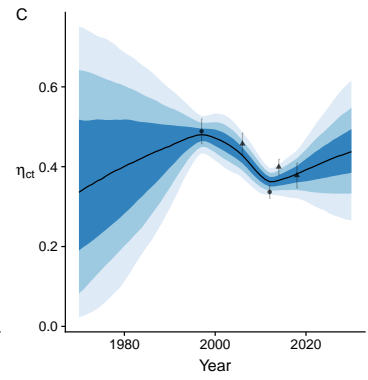

## Lao People's Democratic Republic

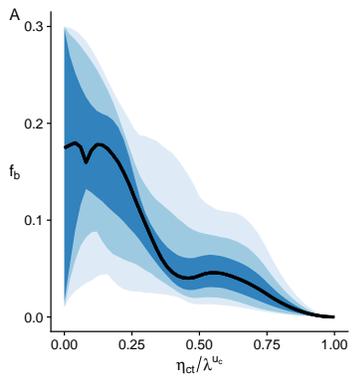 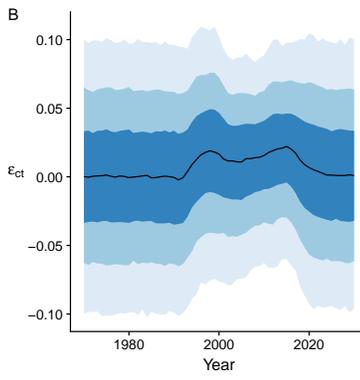 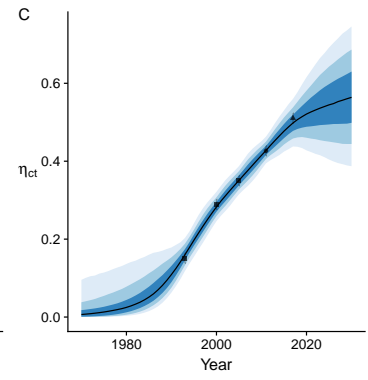

## Latvia

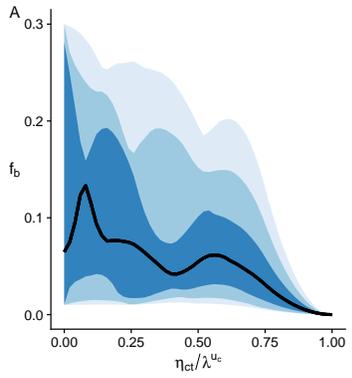 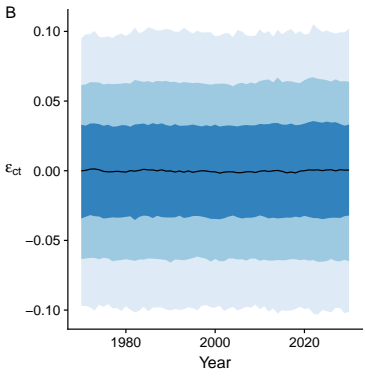 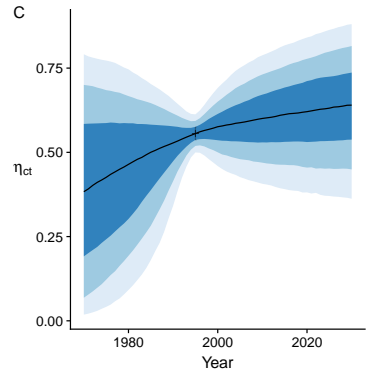

## Lebanon

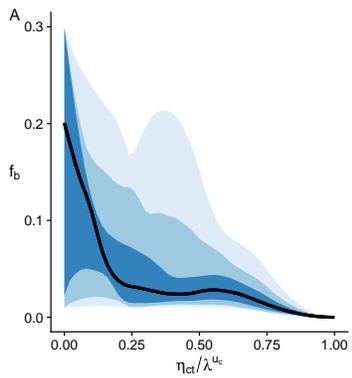 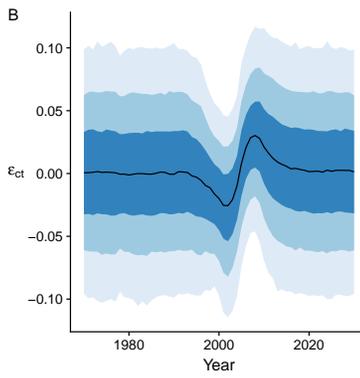 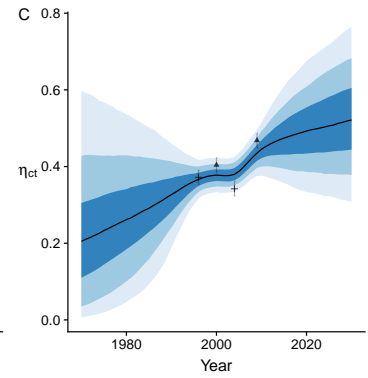

**Lesotho**

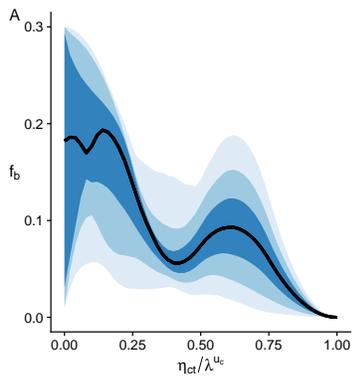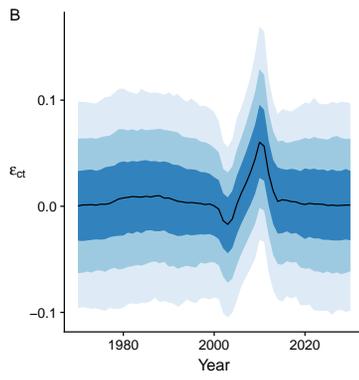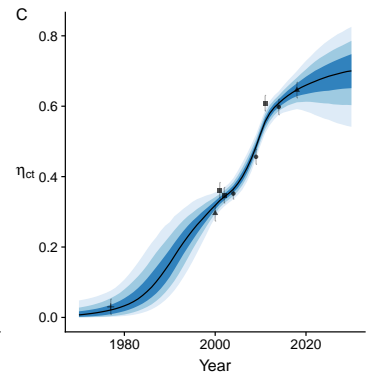

**Liberia**

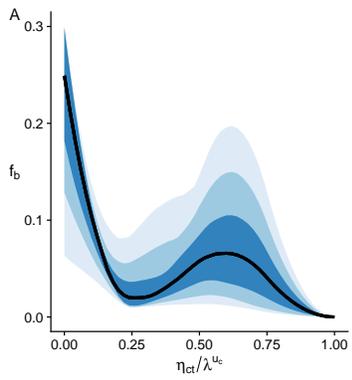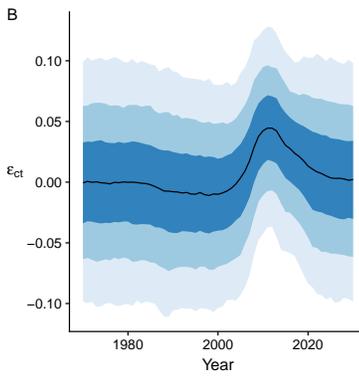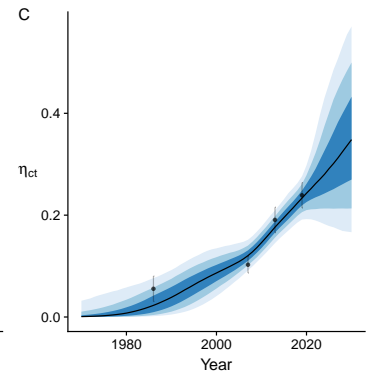

**Libya**

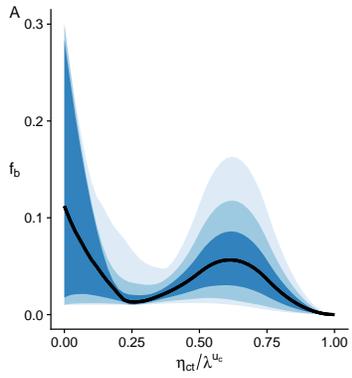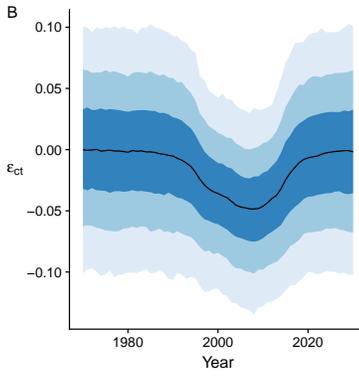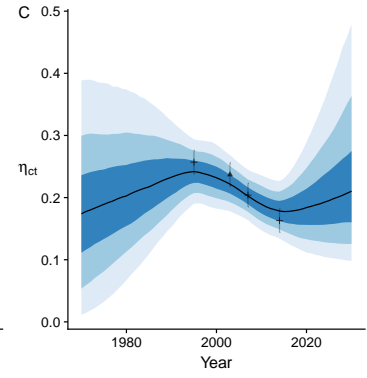

**Lithuania**

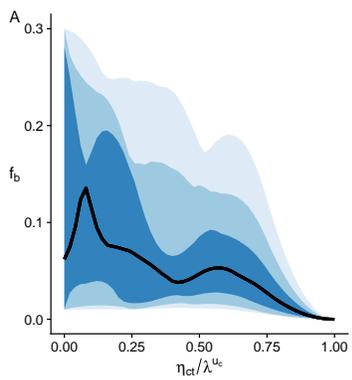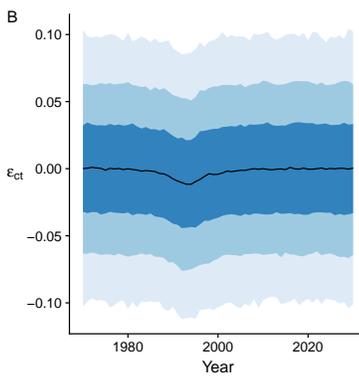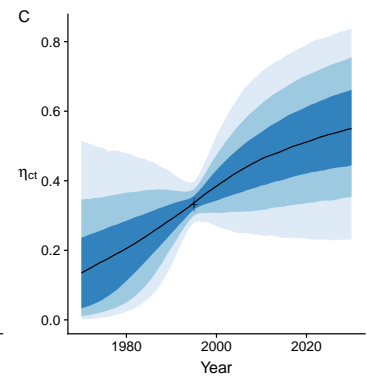

**Madagascar**

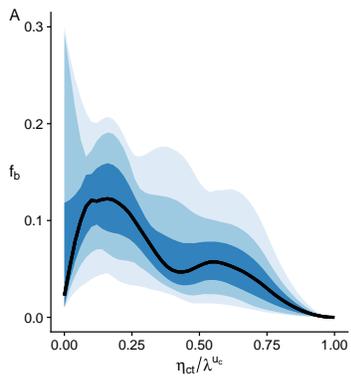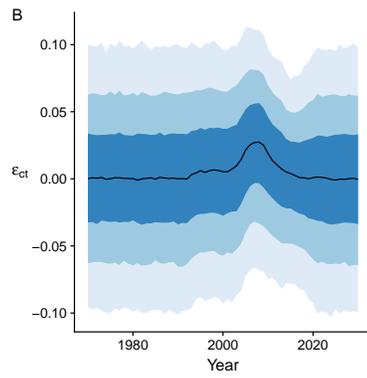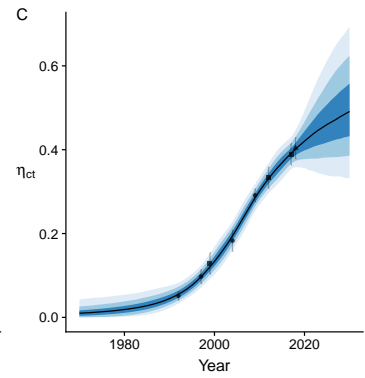

**Malawi**

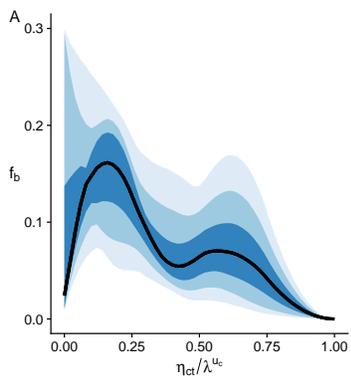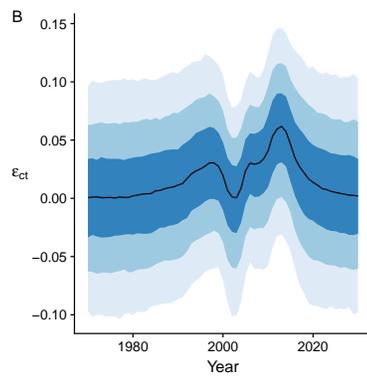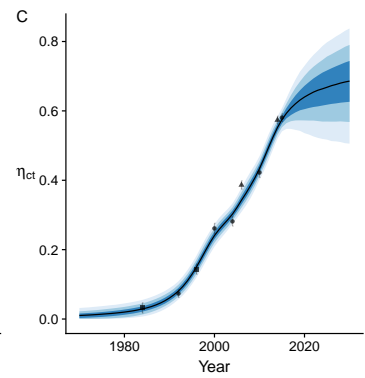

**Malaysia**

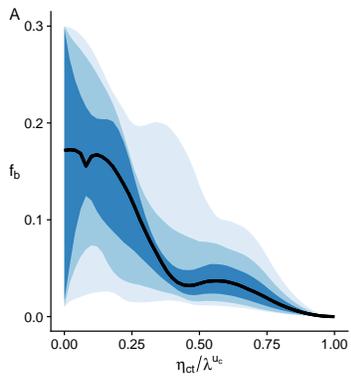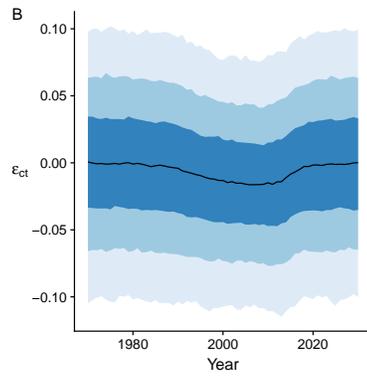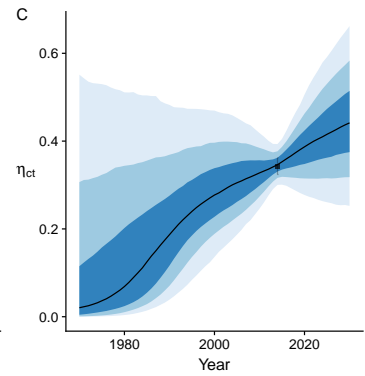

**Maldives**

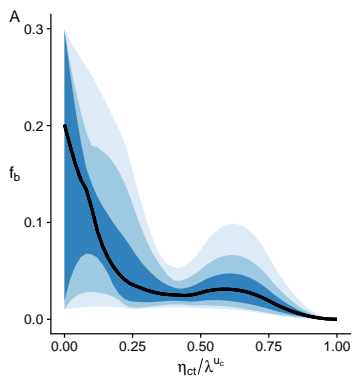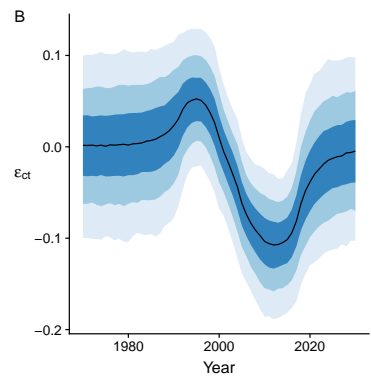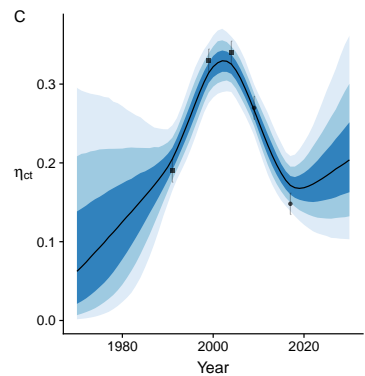

**Mali**

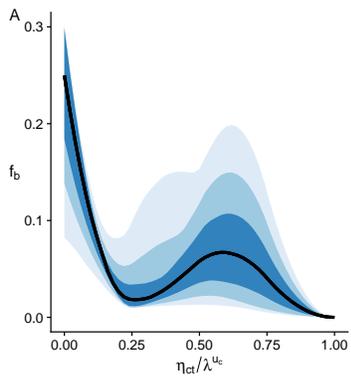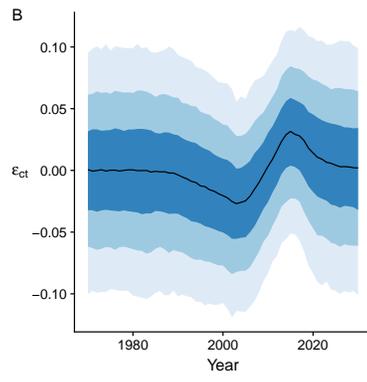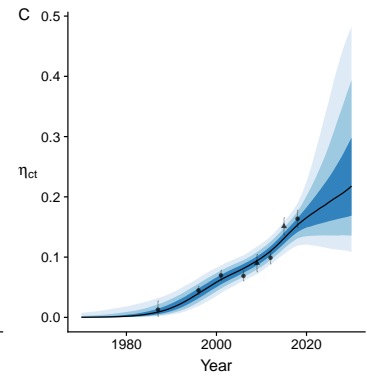

**Marshall Islands**

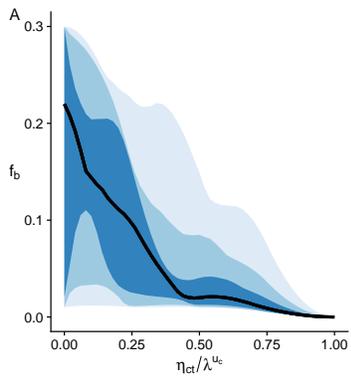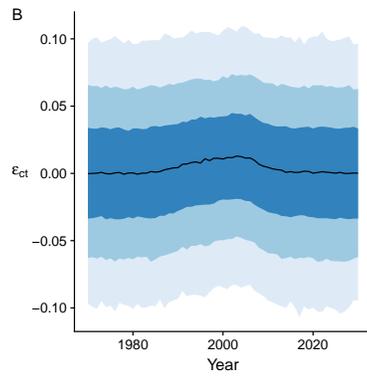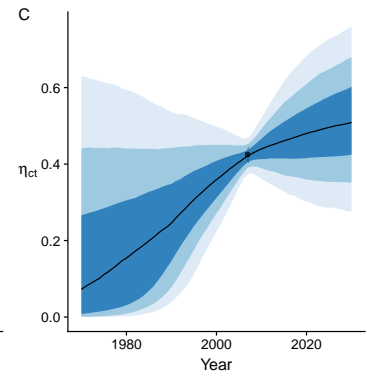

**Mauritania**

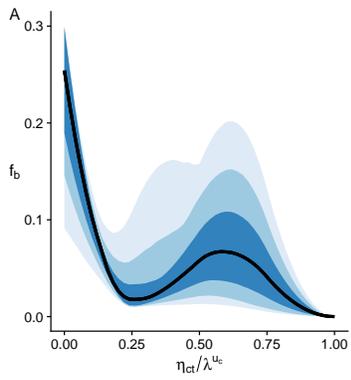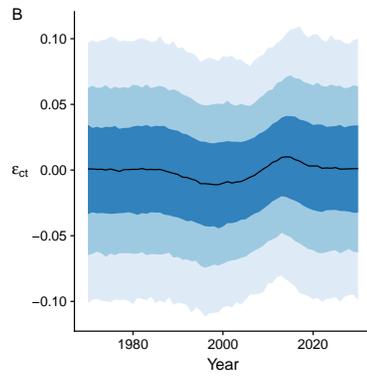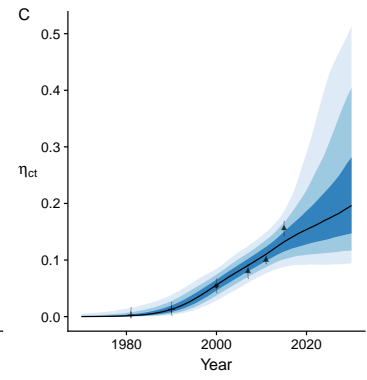

**Mauritius**

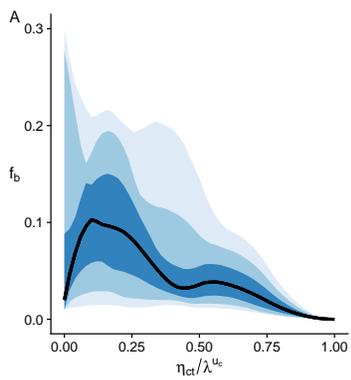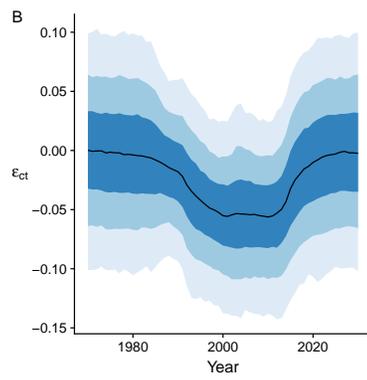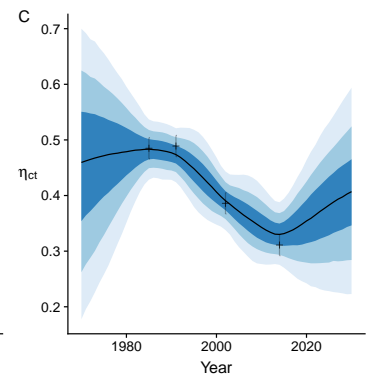

**Mexico**

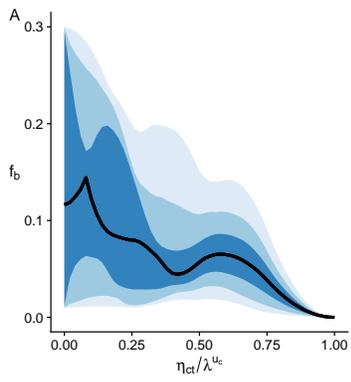 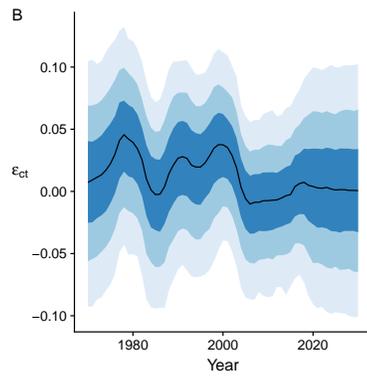 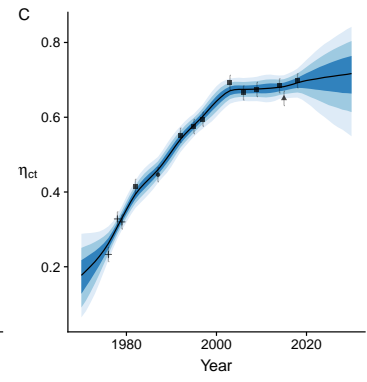

**Mongolia**

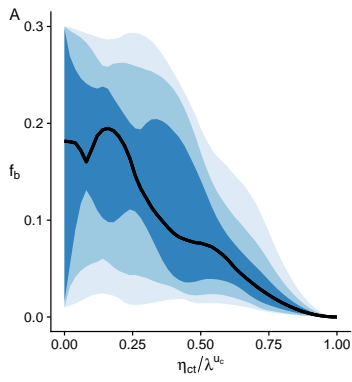 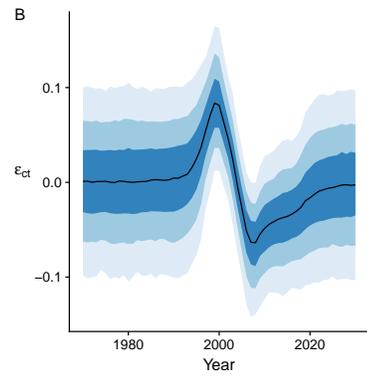 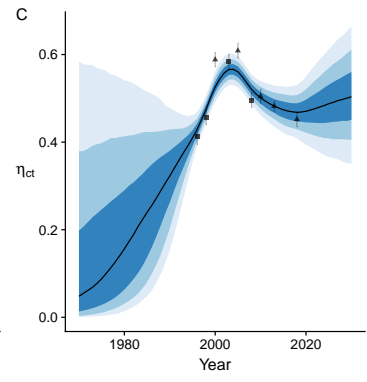

**Montenegro**

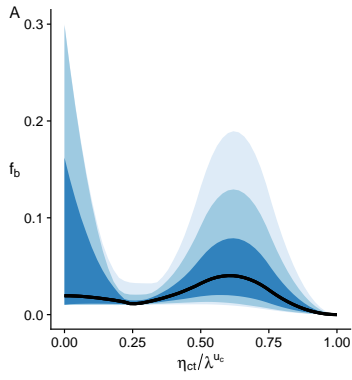 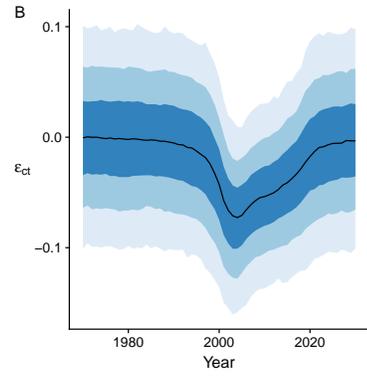 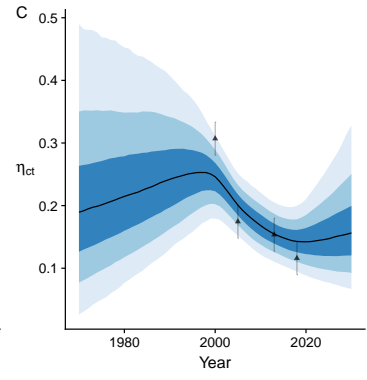

**Morocco**

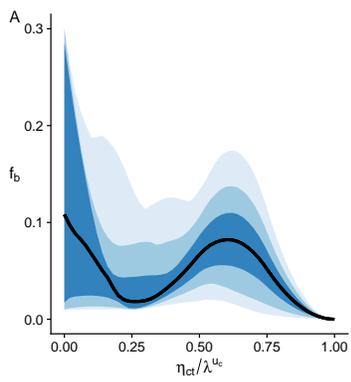 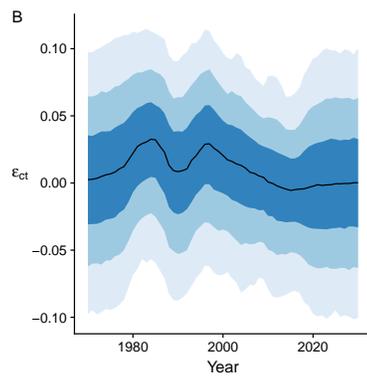 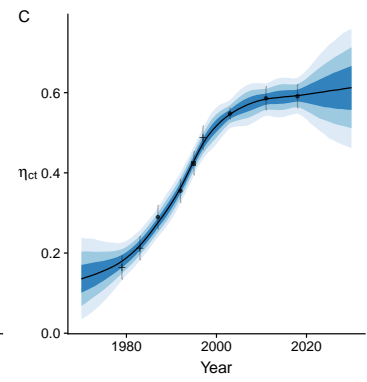

**Mozambique**

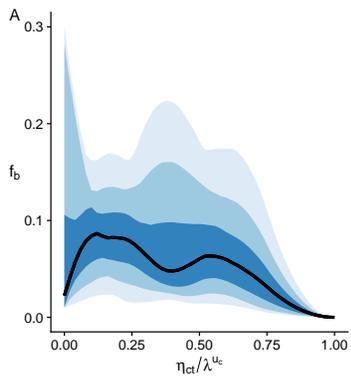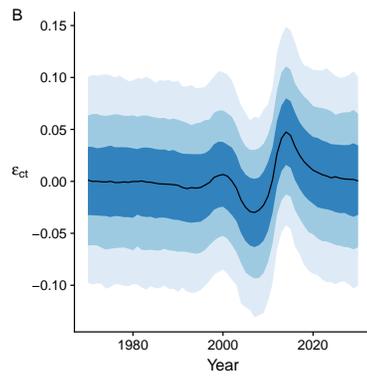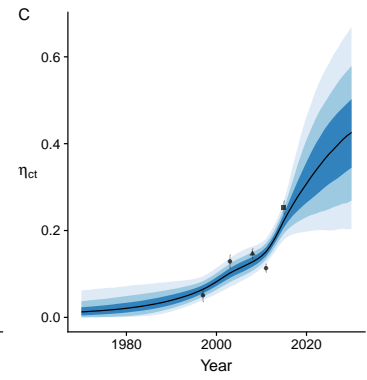

**Myanmar**

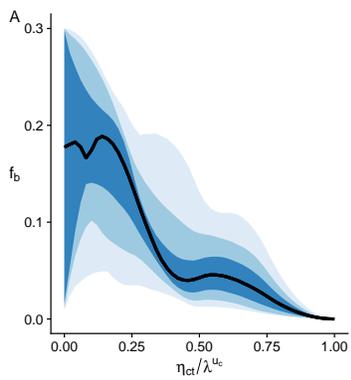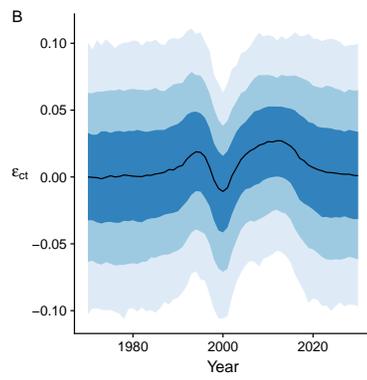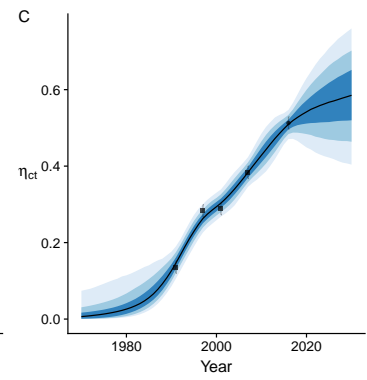

**Namibia**

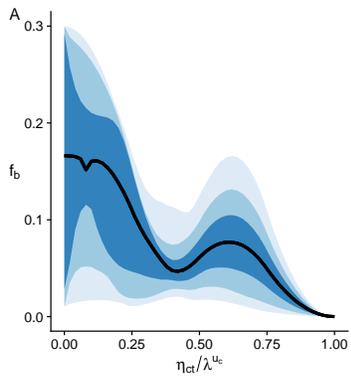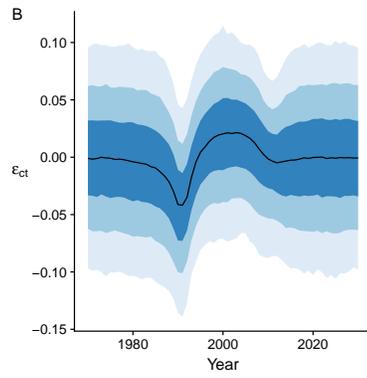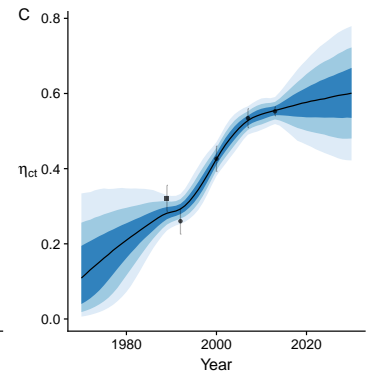

**Nauru**

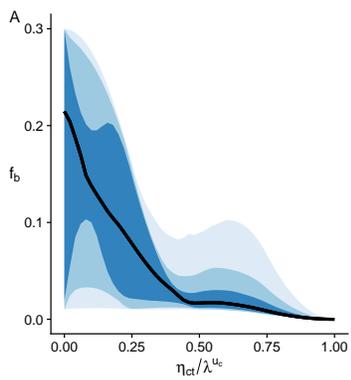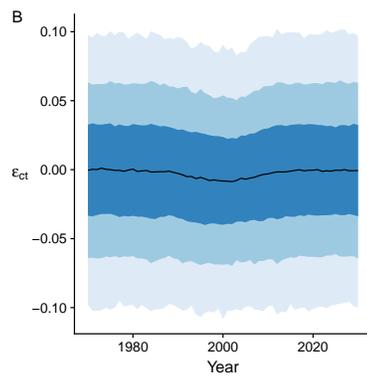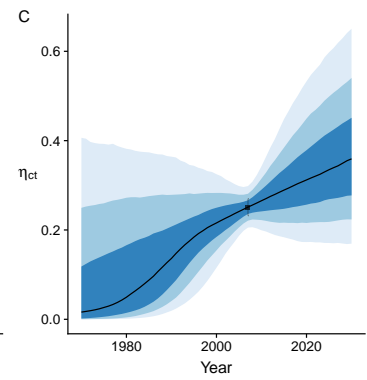

**Nepal**

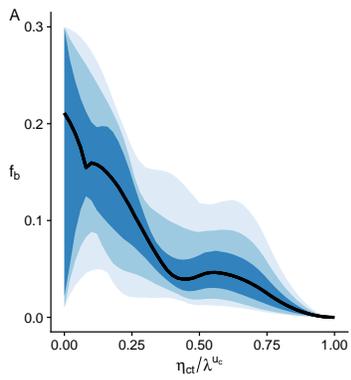 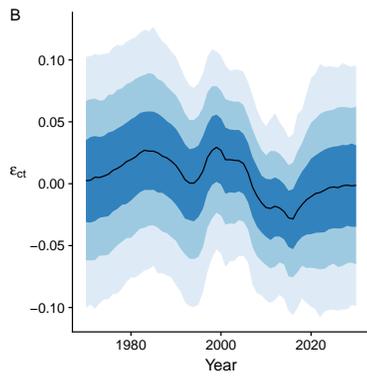 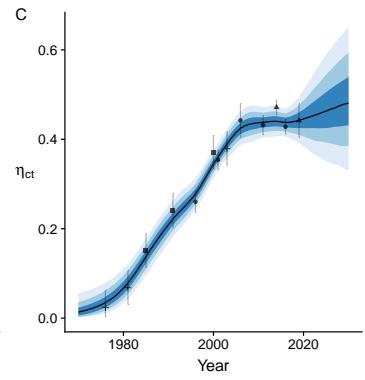

**Netherlands**

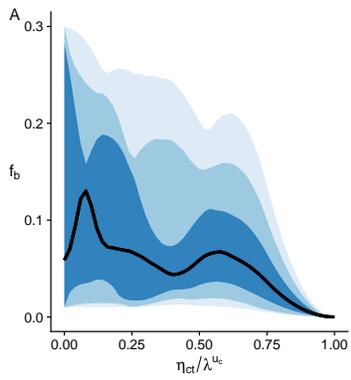 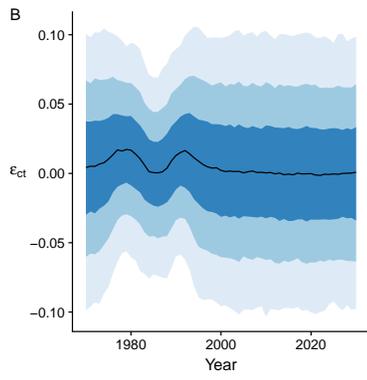 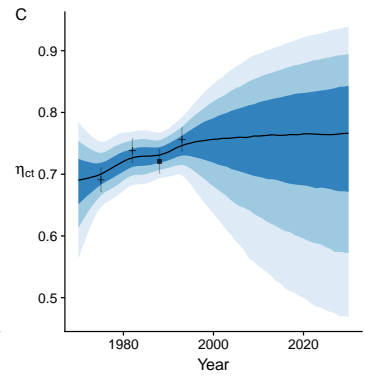

**New Zealand**

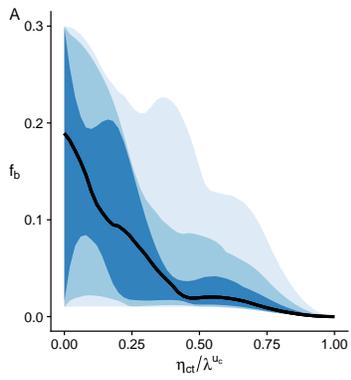 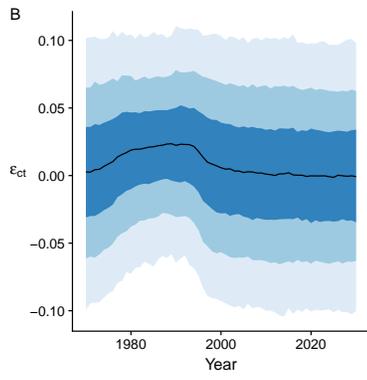 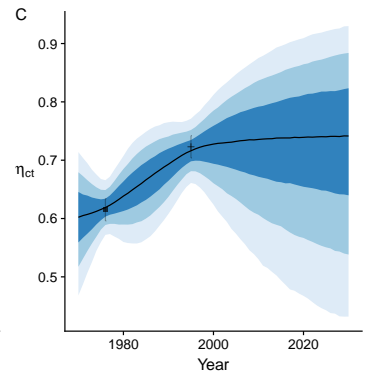

**Nicaragua**

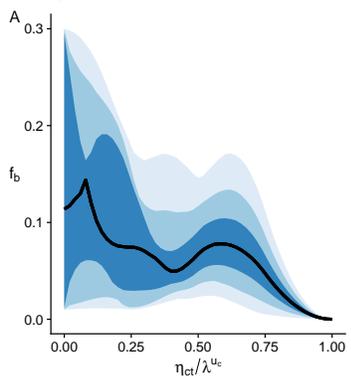 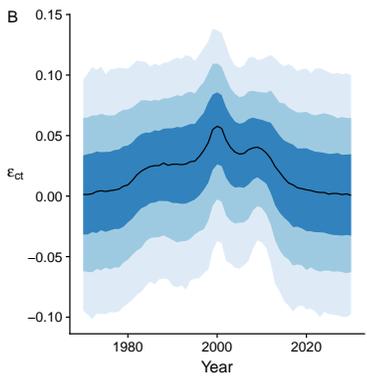 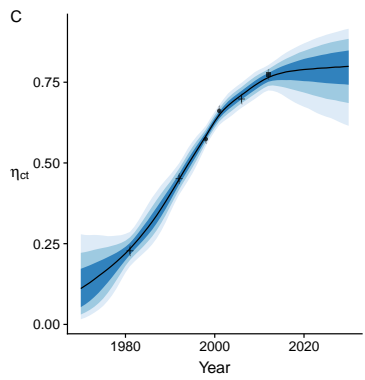

**Niger**

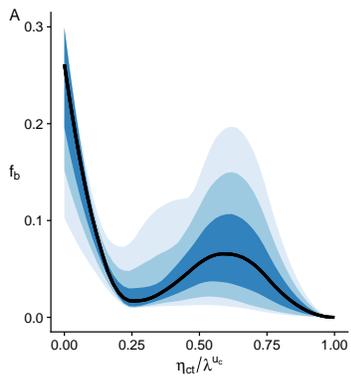
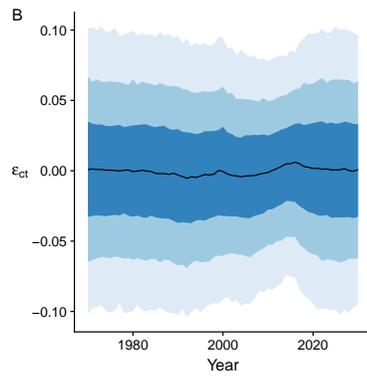
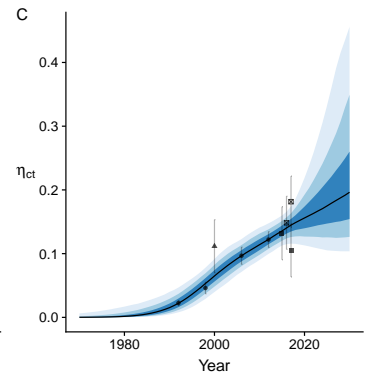

**Nigeria**

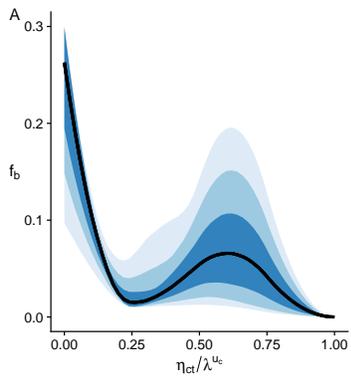
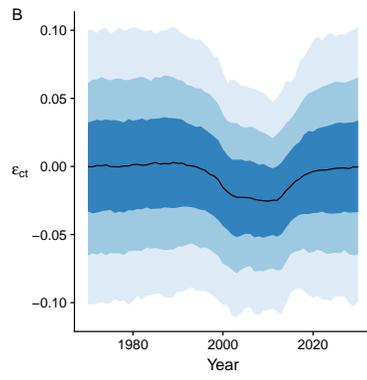
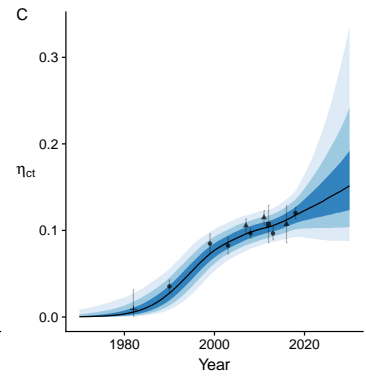

**The former Yugoslav Republic of Macedonia**

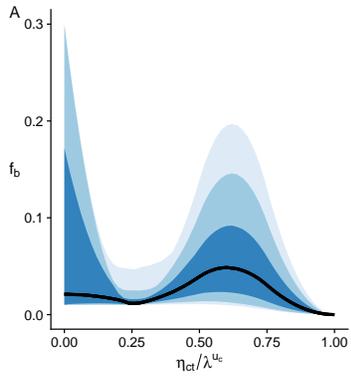
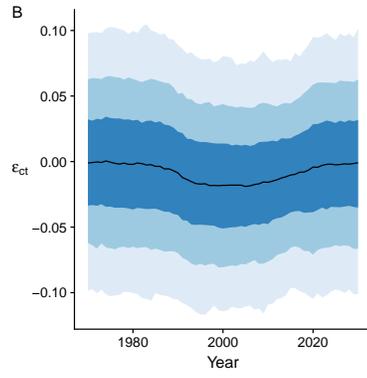
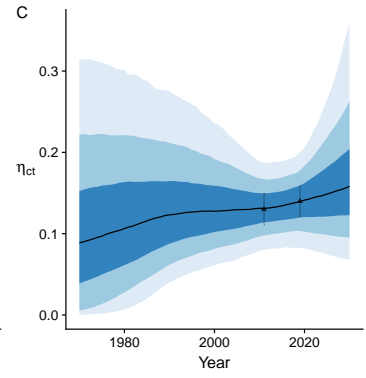

**Norway**

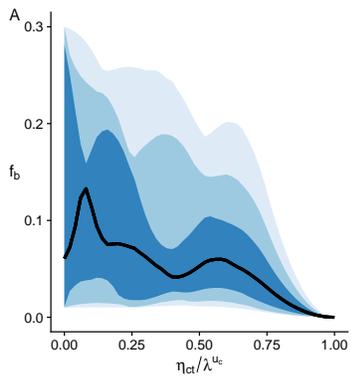
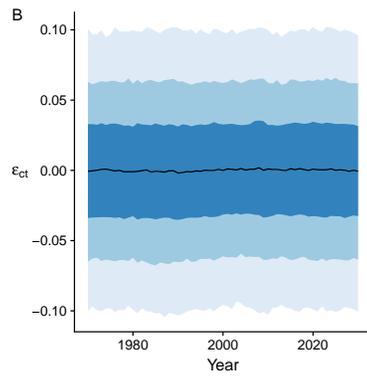
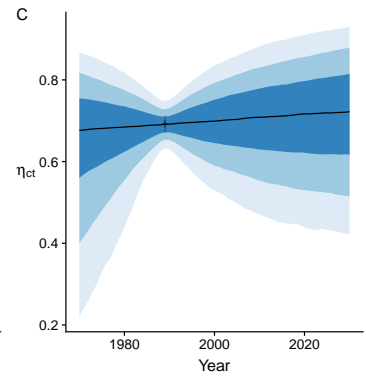

**Oman**

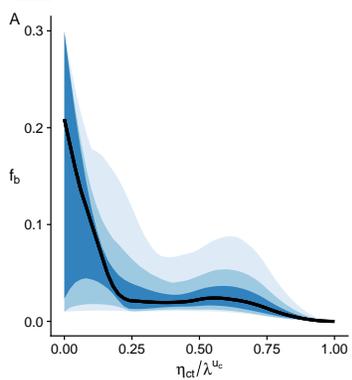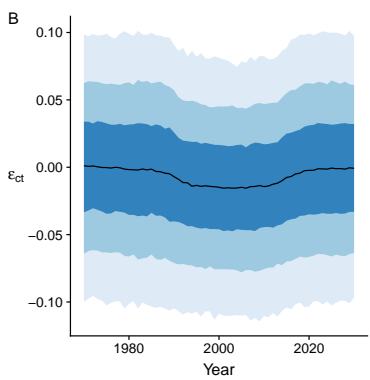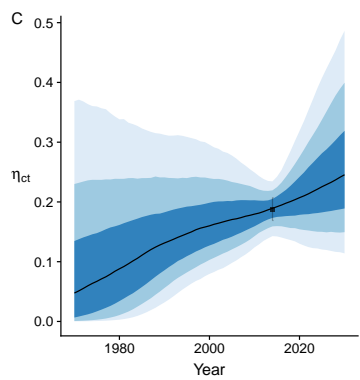

**Pakistan**

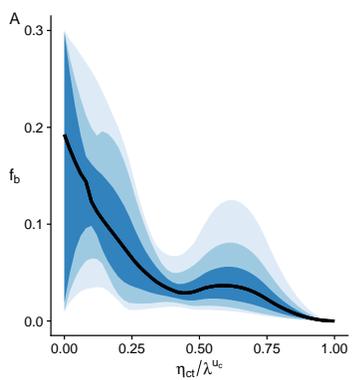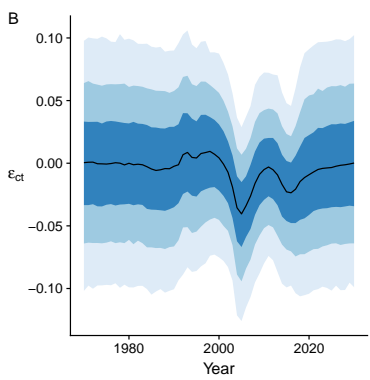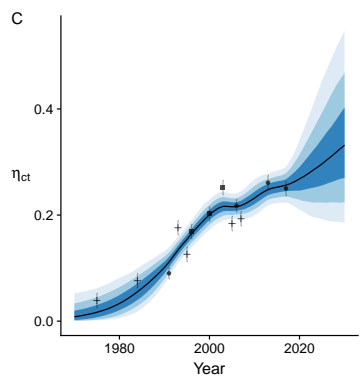

**Panama**

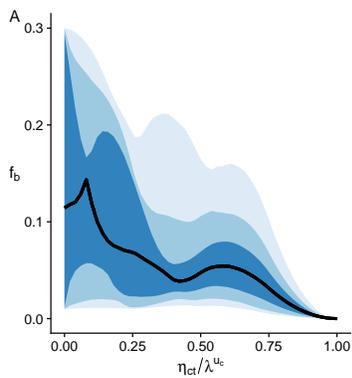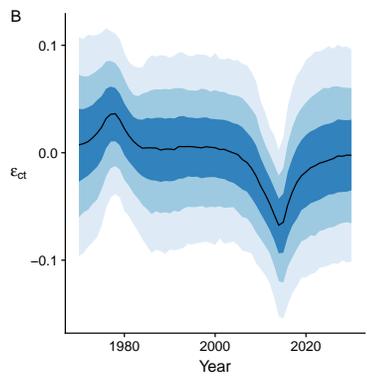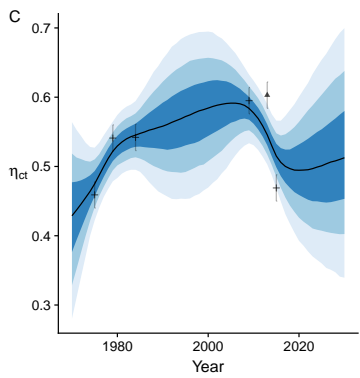

**Papua New Guinea**

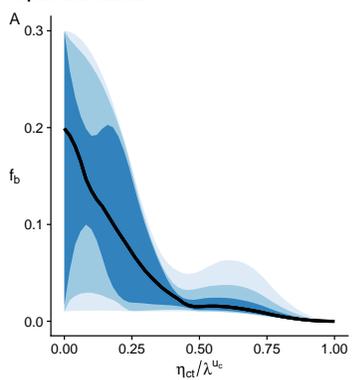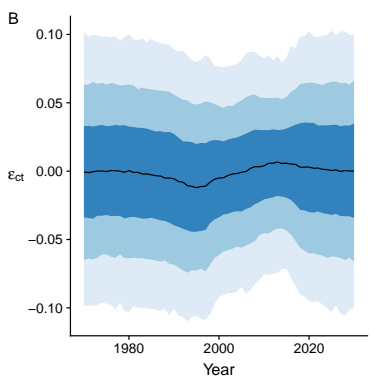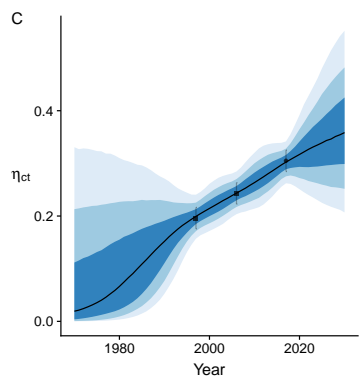

**Paraguay**

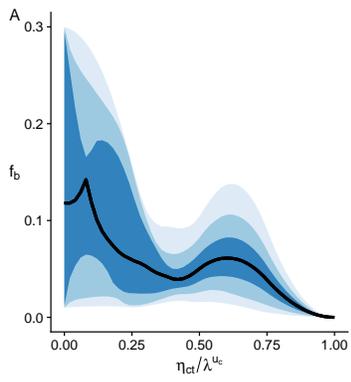 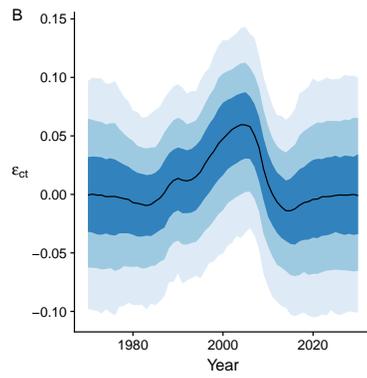 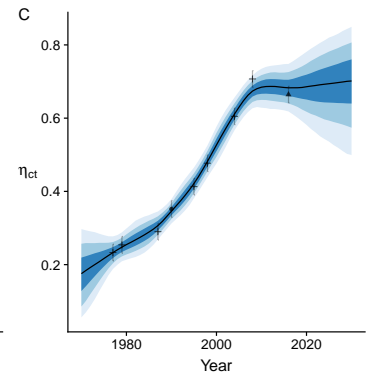

**Peru**

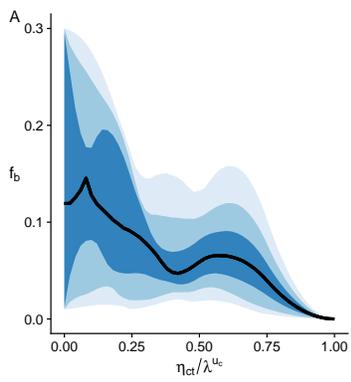 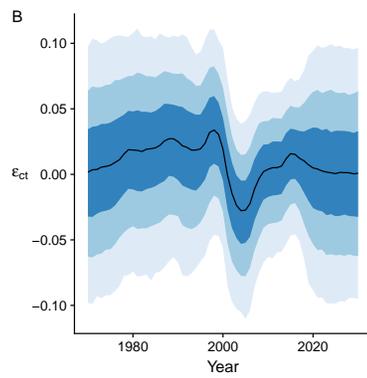 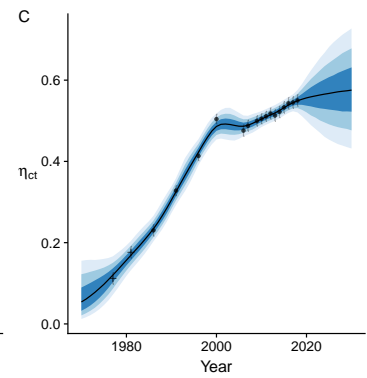

**Philippines**

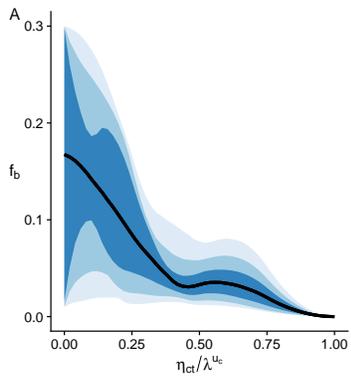 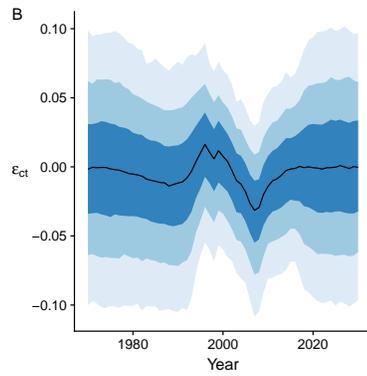 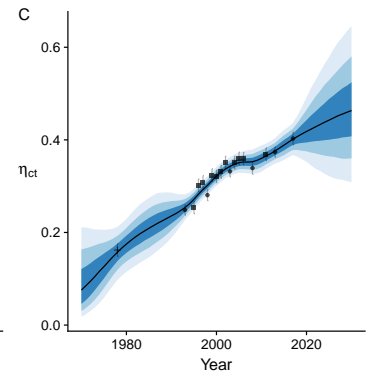

**Poland**

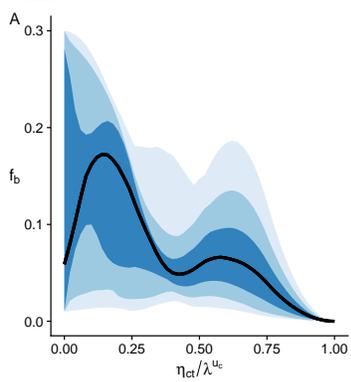 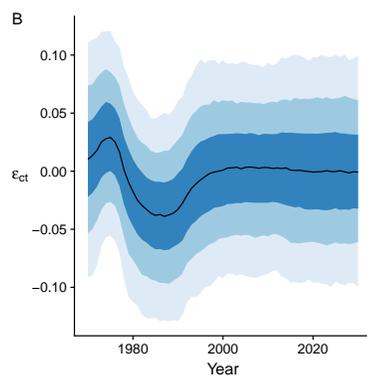 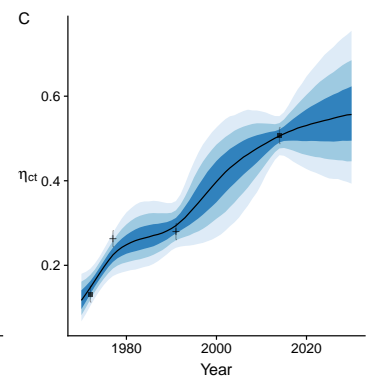

**Puerto Rico**

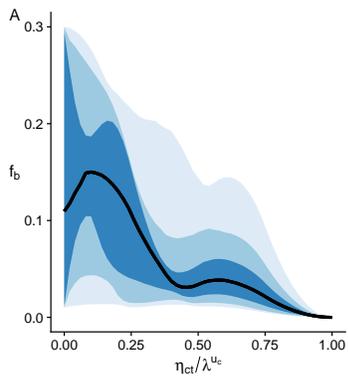
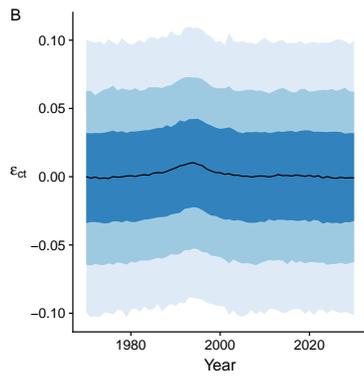
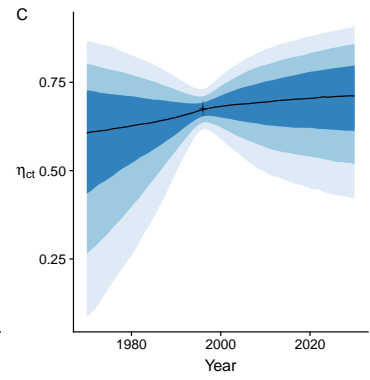

**Qatar**

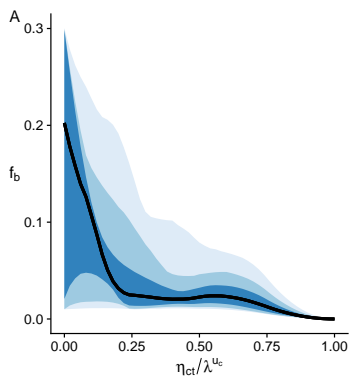
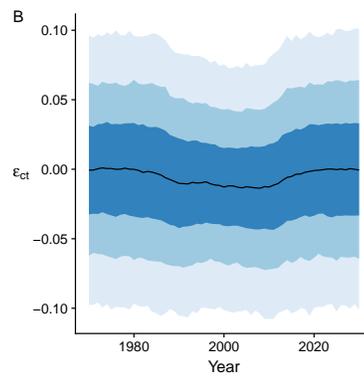
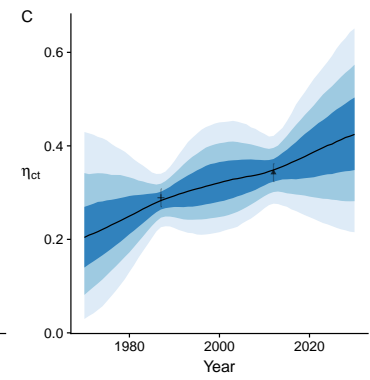

**Republic of Korea**

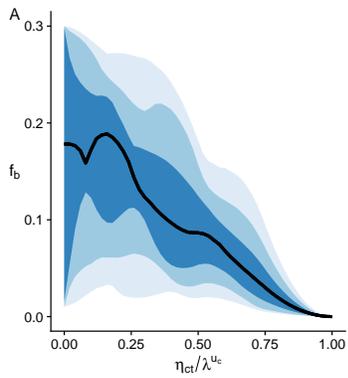
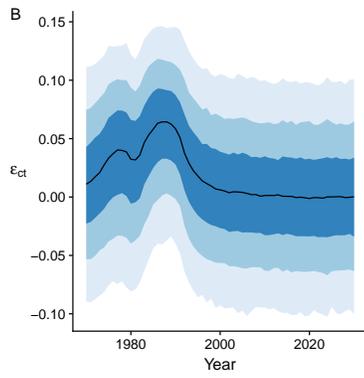
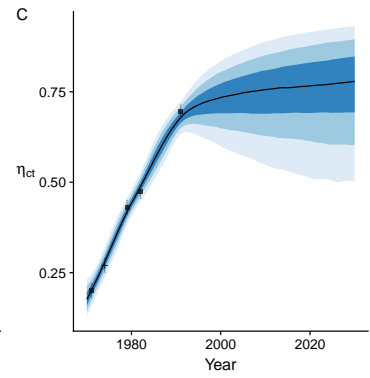

**Republic of Moldova**

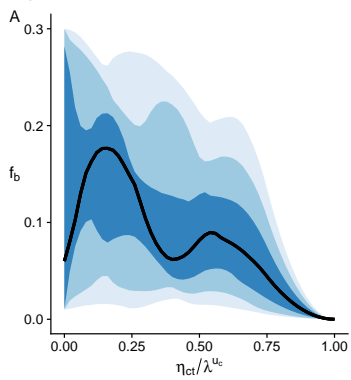
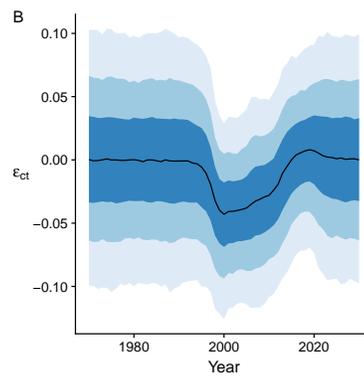
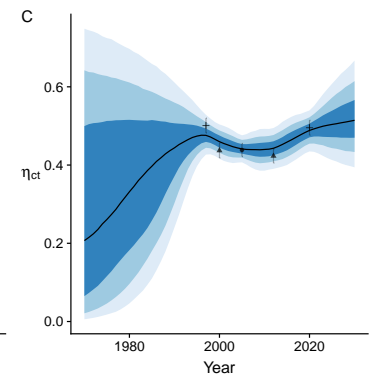

**Réunion**

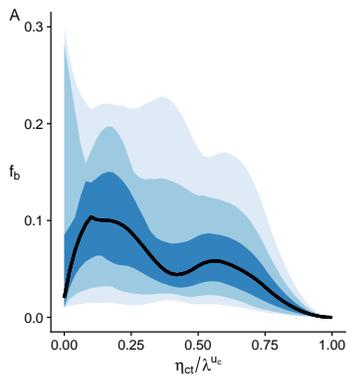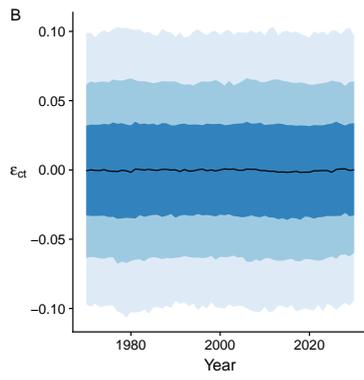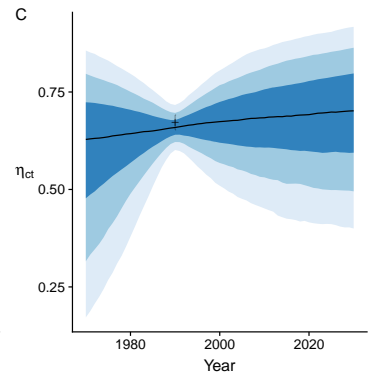

**Romania**

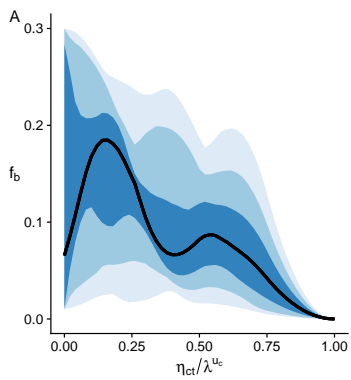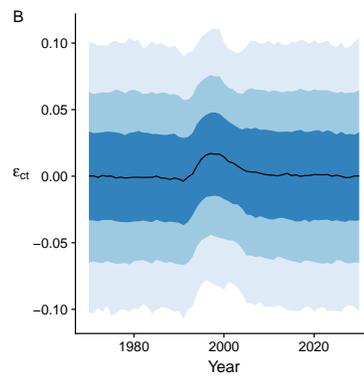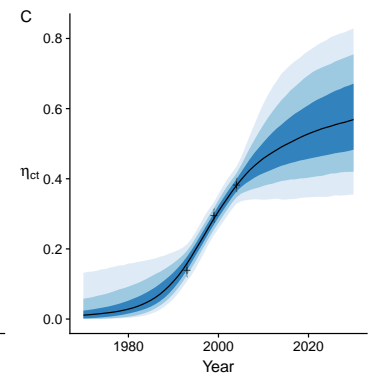

**Russian Federation**

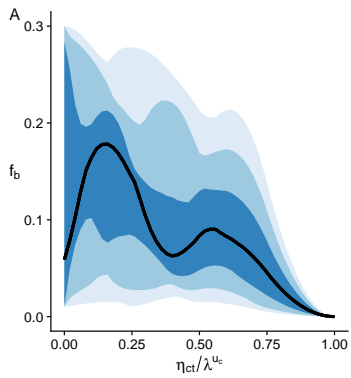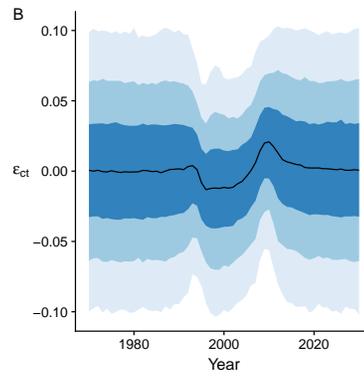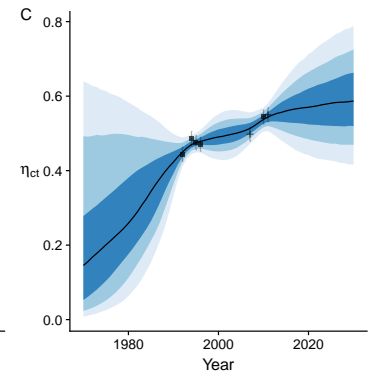

**Rwanda**

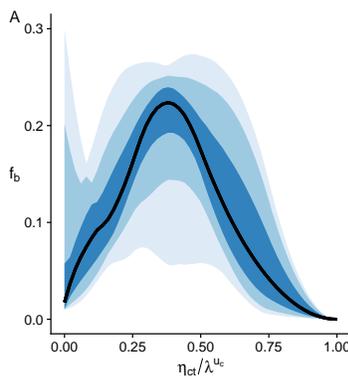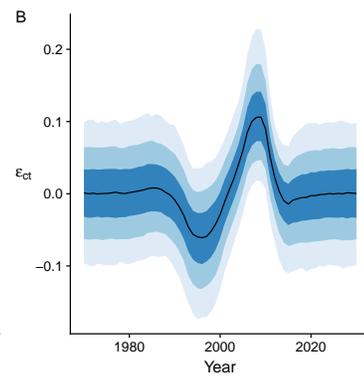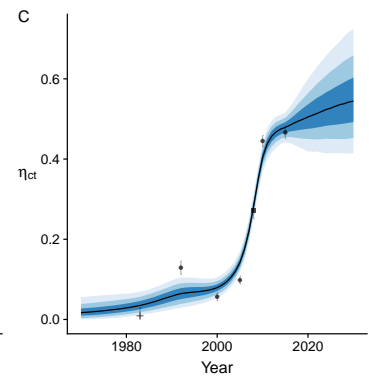

**Saint Lucia**

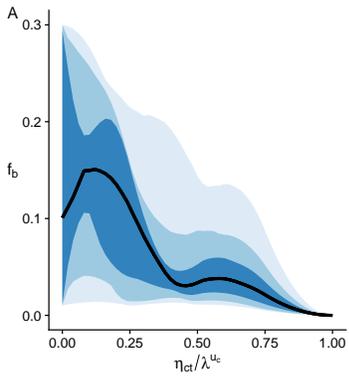
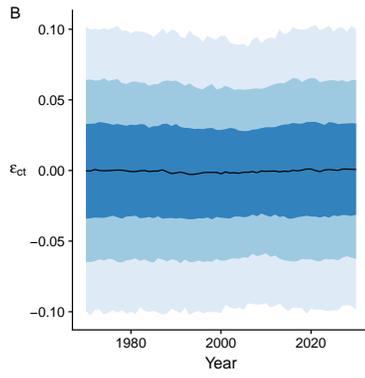
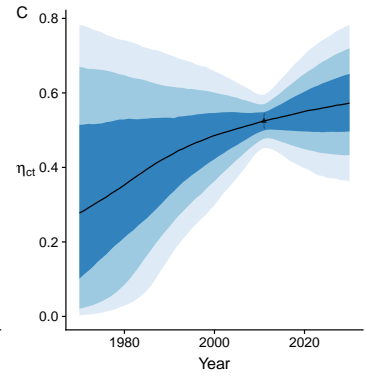

**Saint Vincent and the Grenadines**

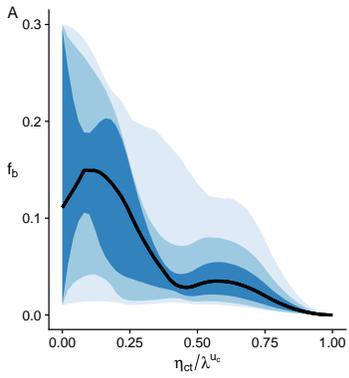
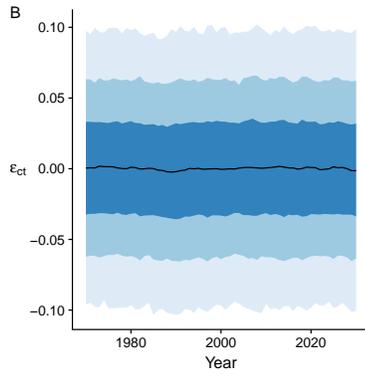
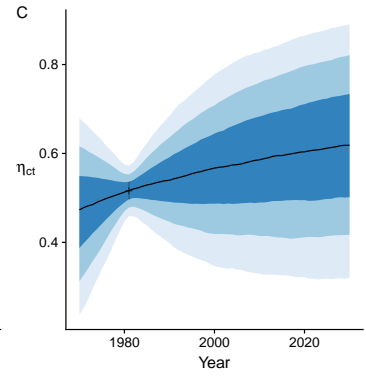

**Samoa**

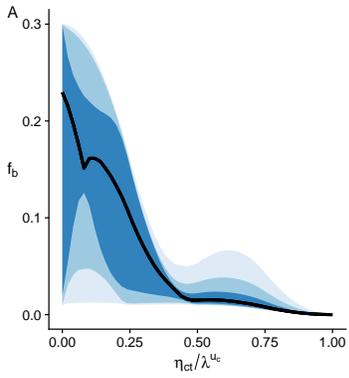
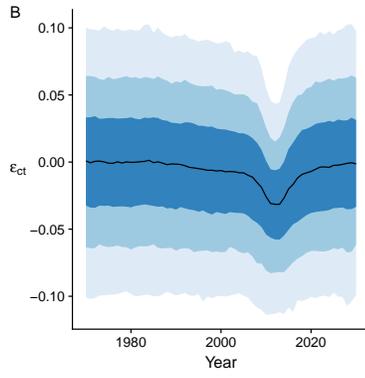
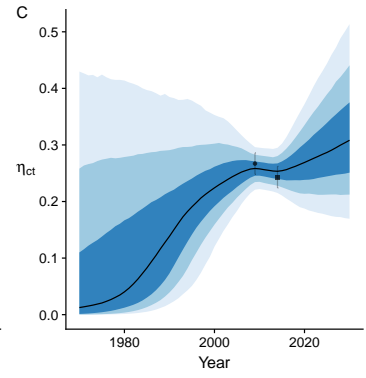

**Sao Tome and Principe**

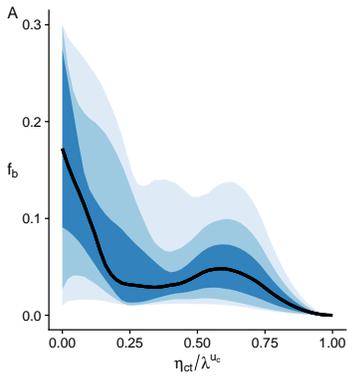
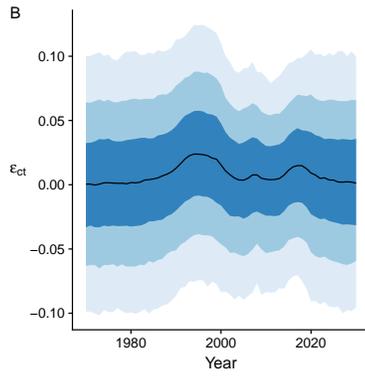
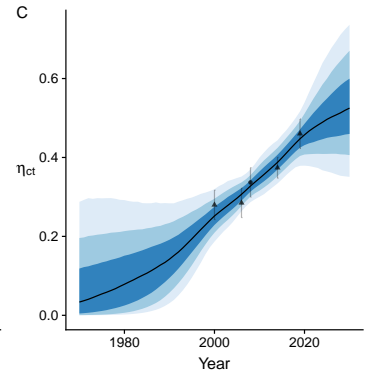

**Senegal**

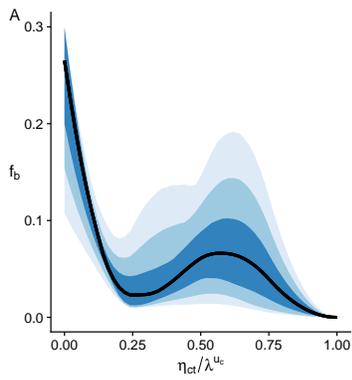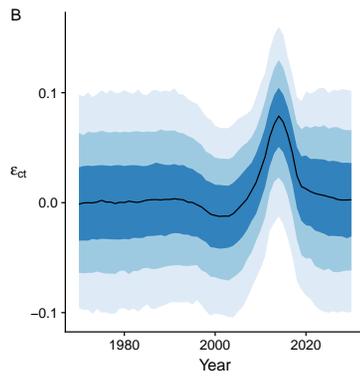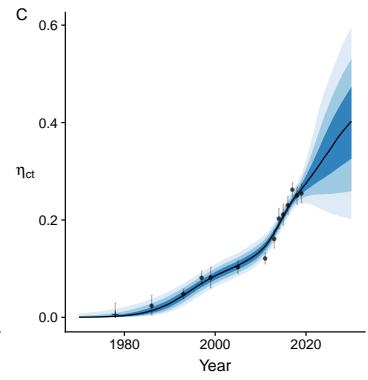

**Serbia**

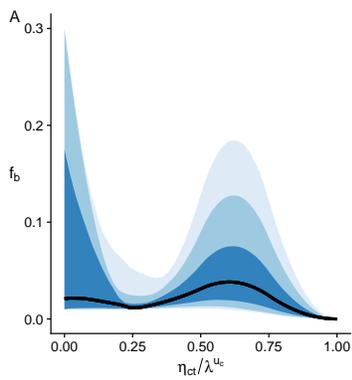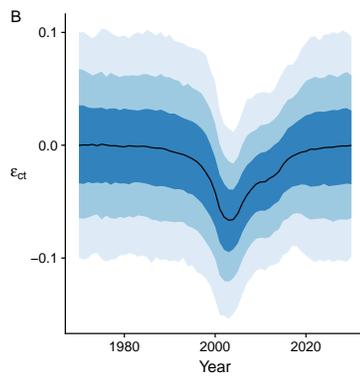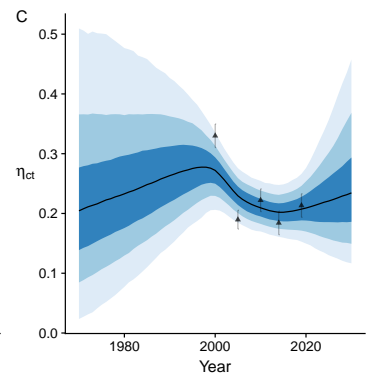

**Sierra Leone**

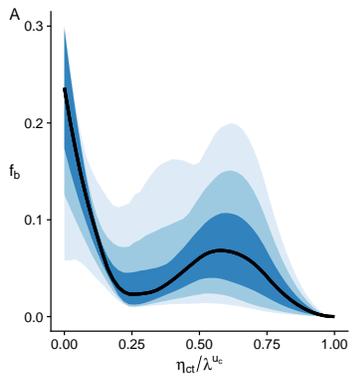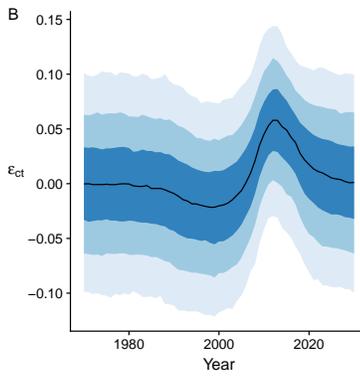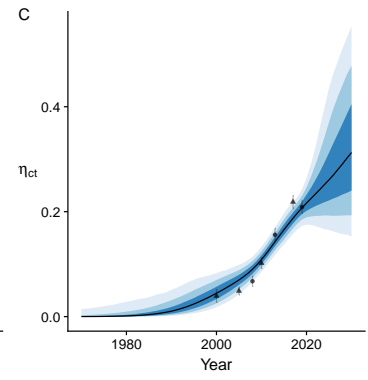

**Singapore**

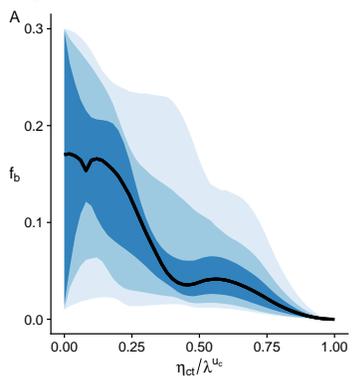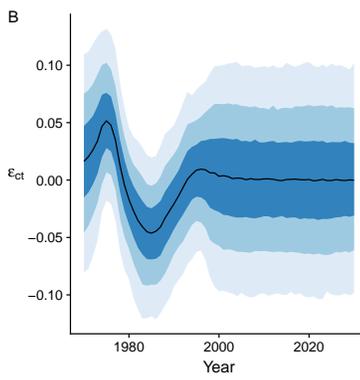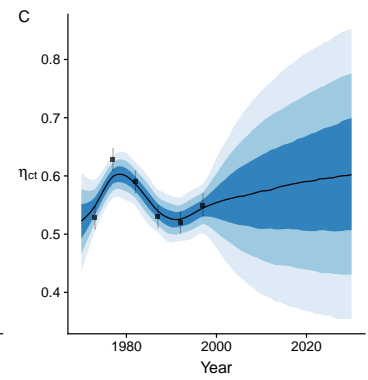

**Slovenia**

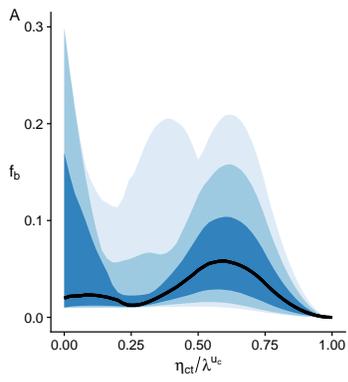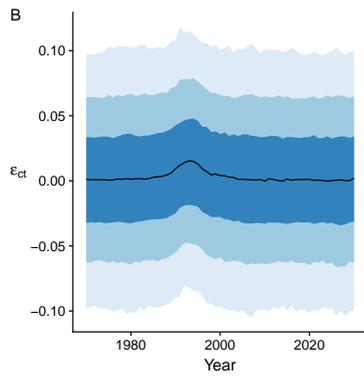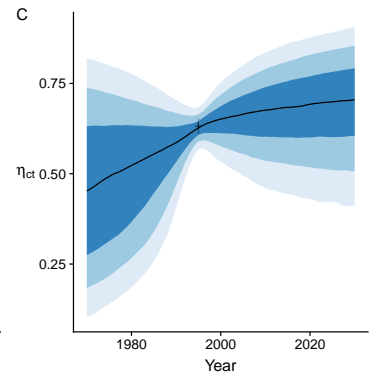

**Solomon Islands**

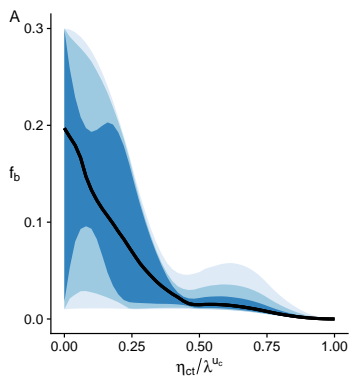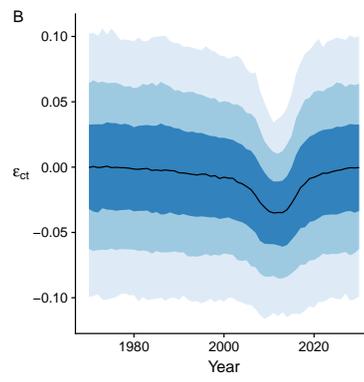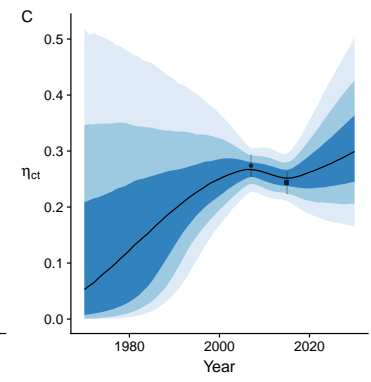

**Somalia**

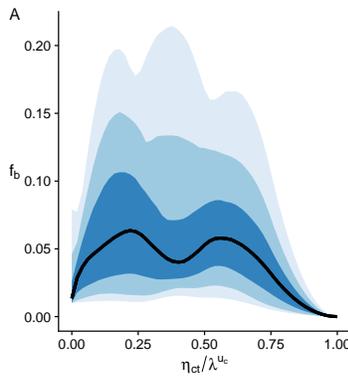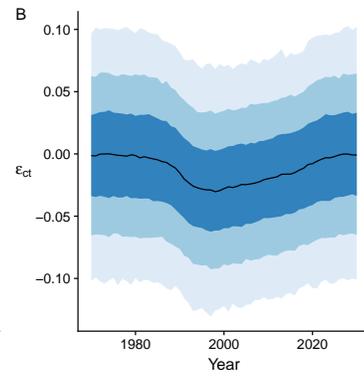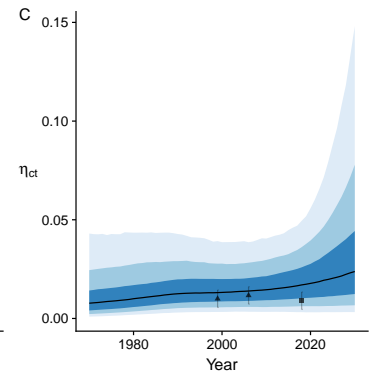

**South Africa**

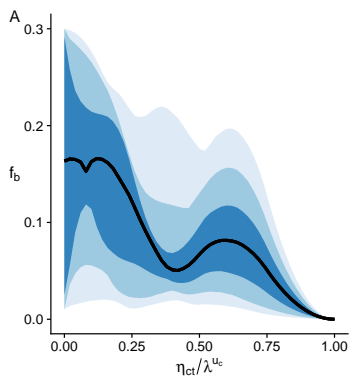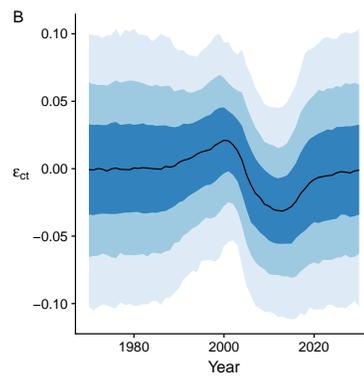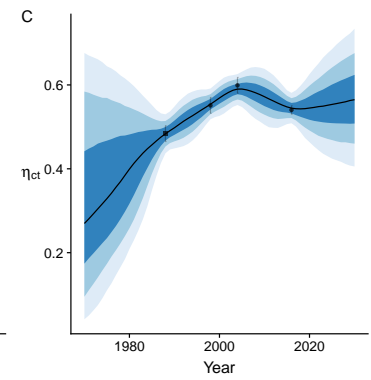

**South Sudan**

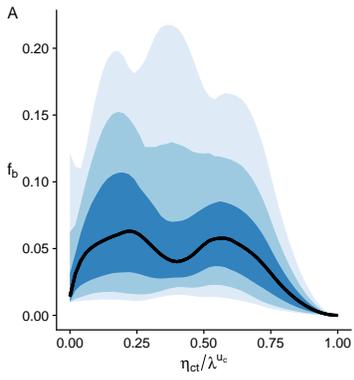
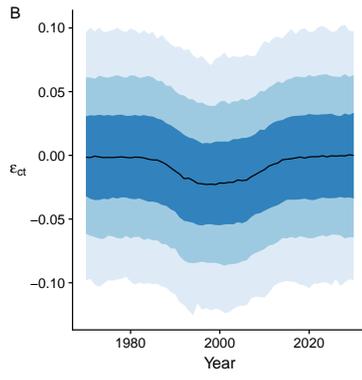
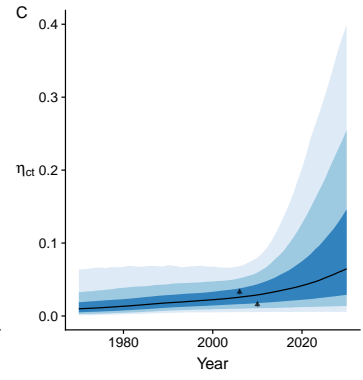

**Spain**

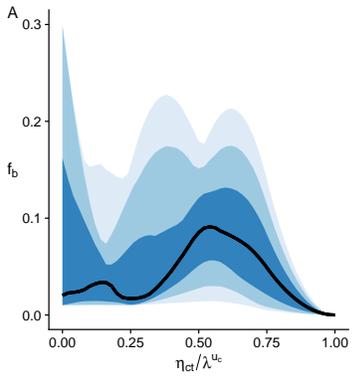
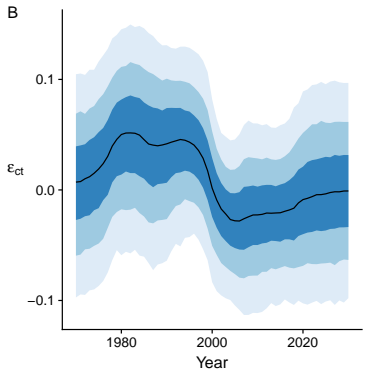
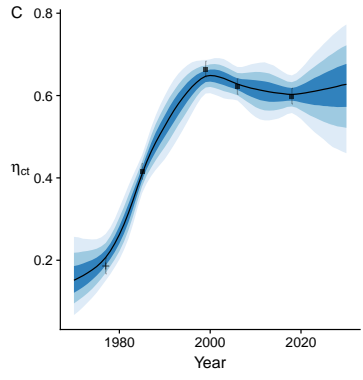

**Sri Lanka**

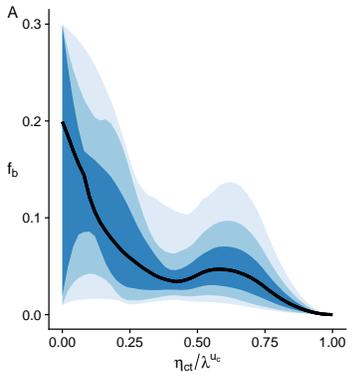
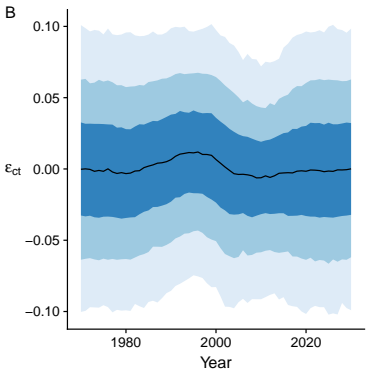
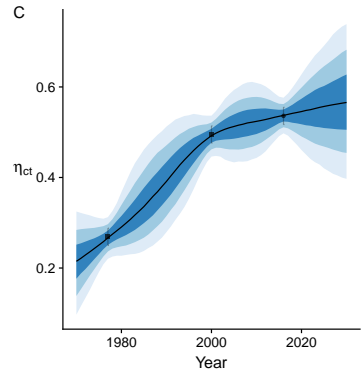

**State of Palestine**

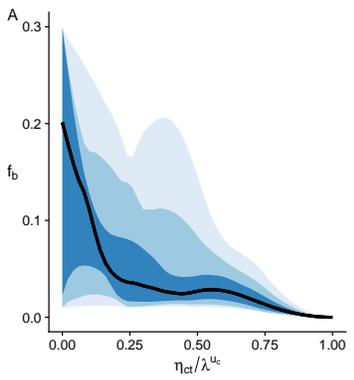
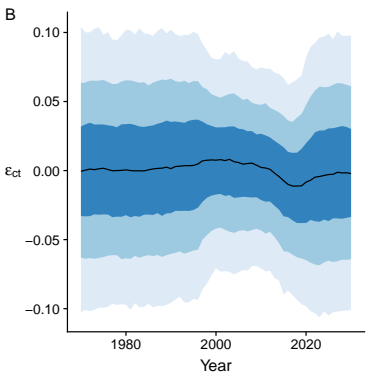
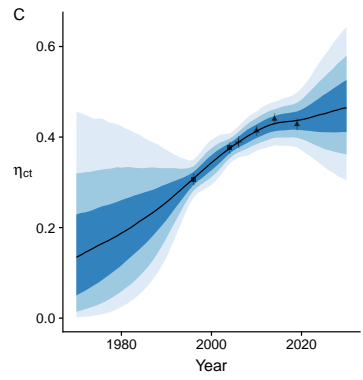

**Sudan**

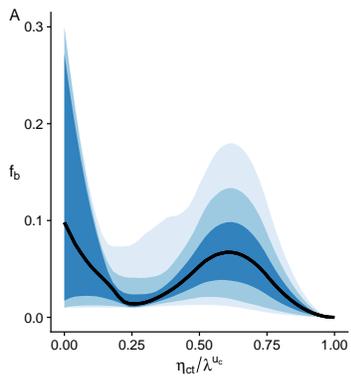 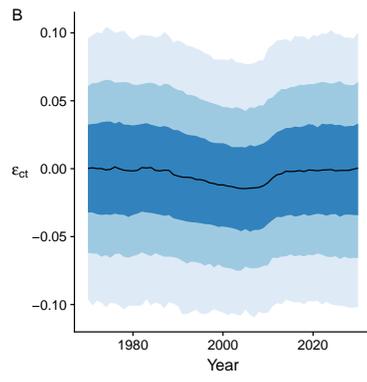 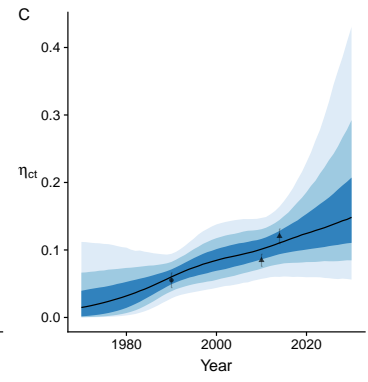

**Suriname**

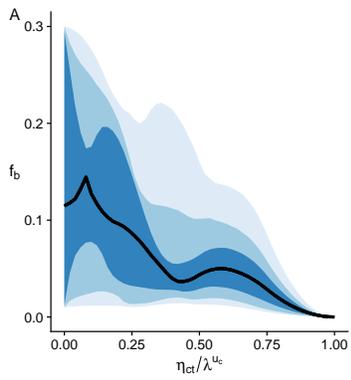 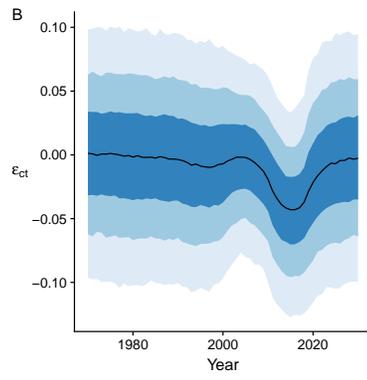 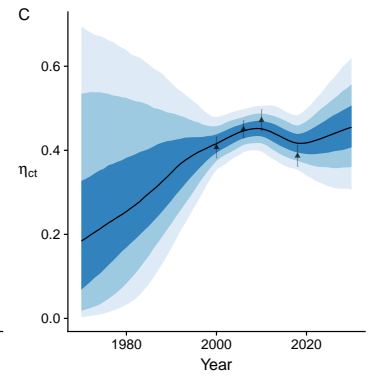

**Switzerland**

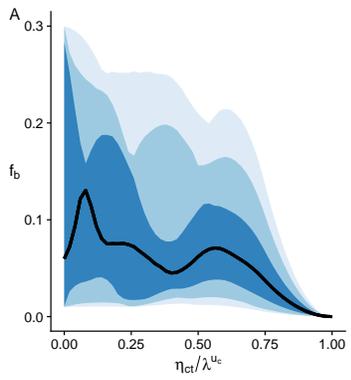 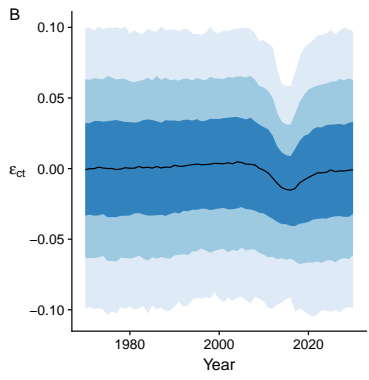 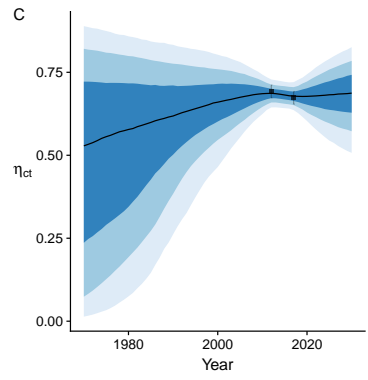

**Syrian Arab Republic**

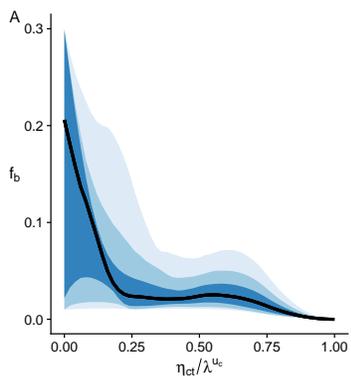 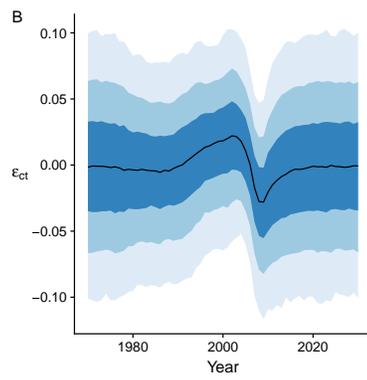 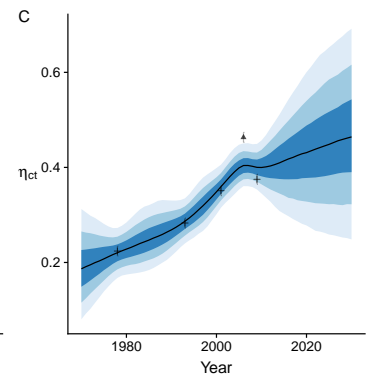

**Tajikistan**

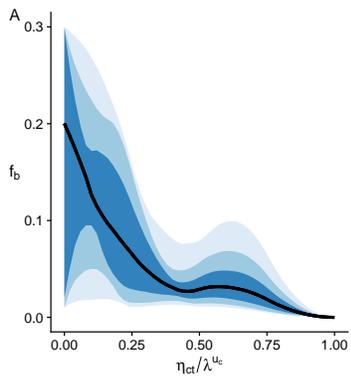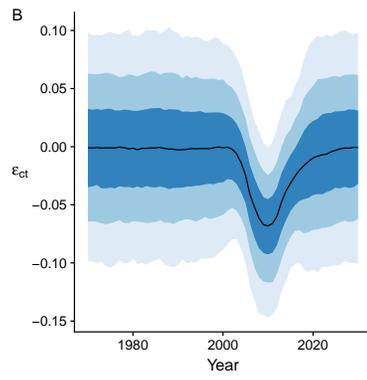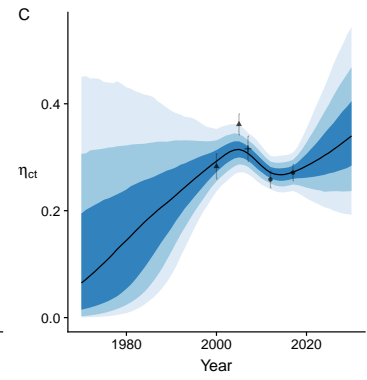

**Thailand**

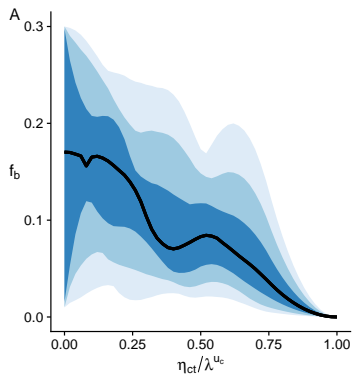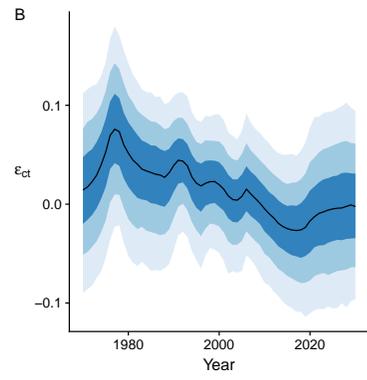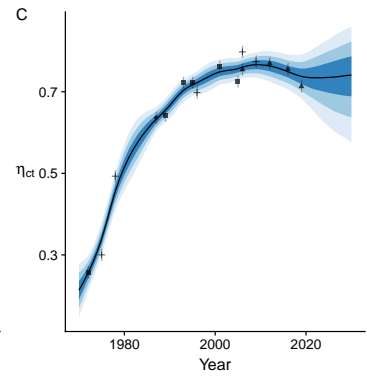

**Timor-Leste**

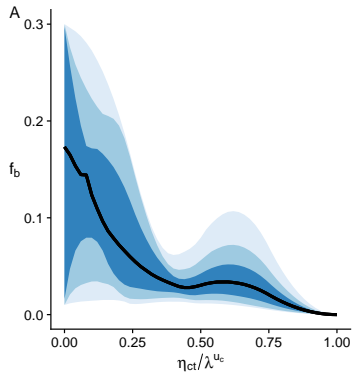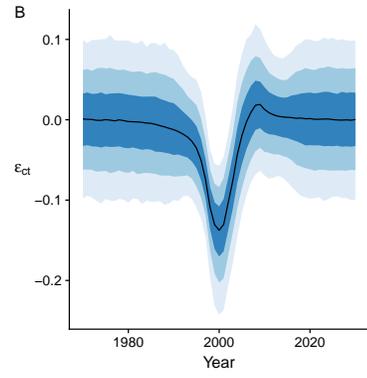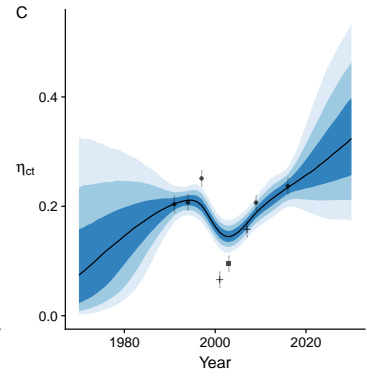

**Togo**

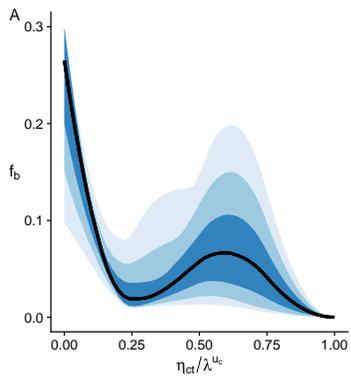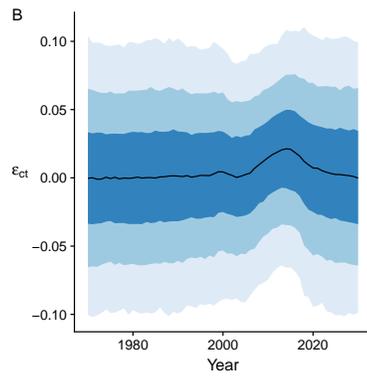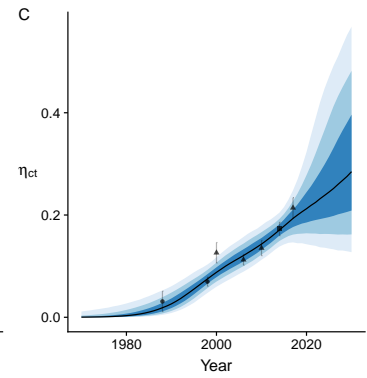

**Tonga**

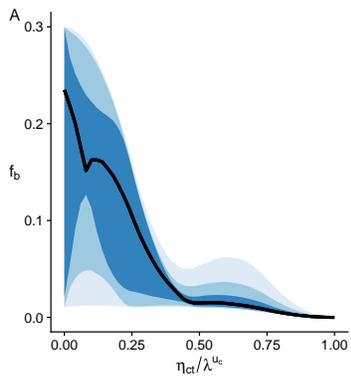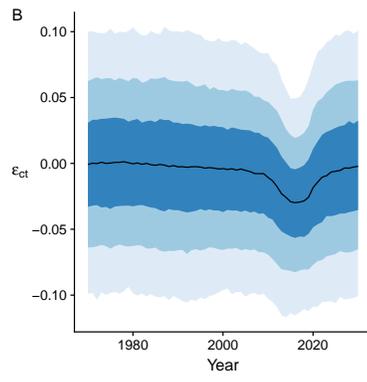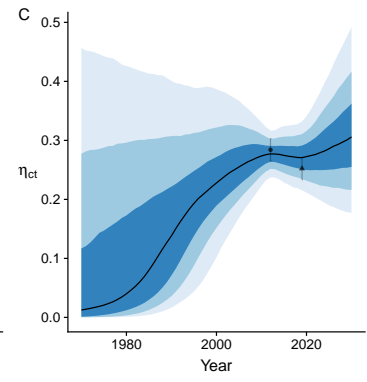

**Trinidad and Tobago**

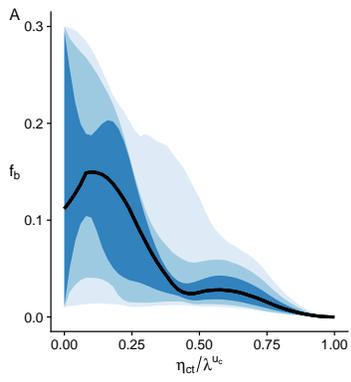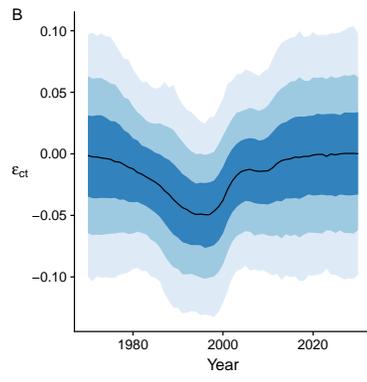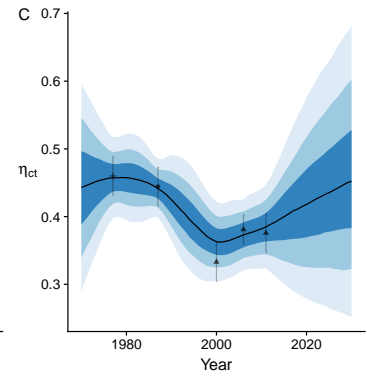

**Tunisia**

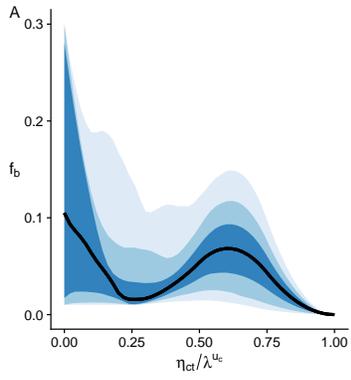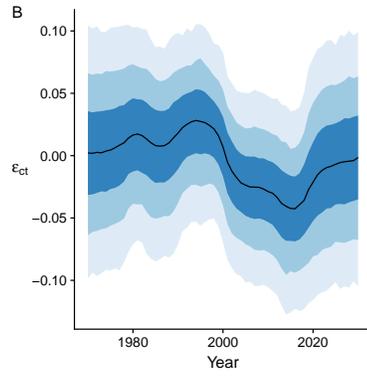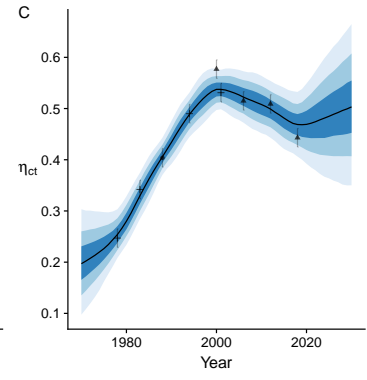

**Turkey**

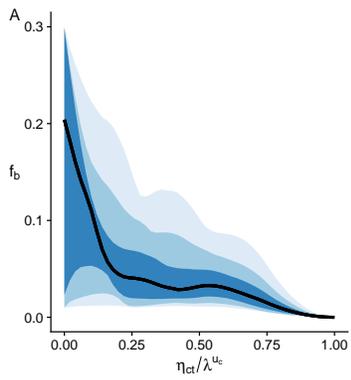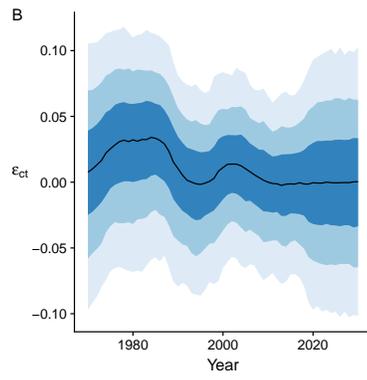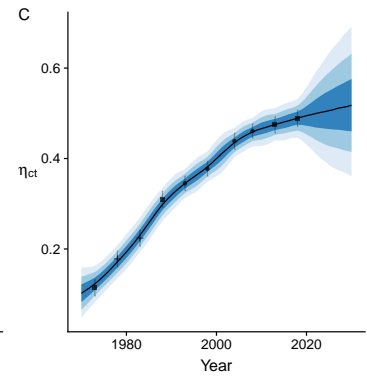

**Turkmenistan**

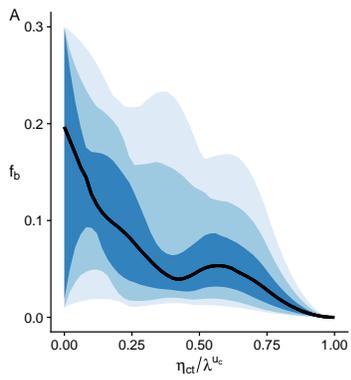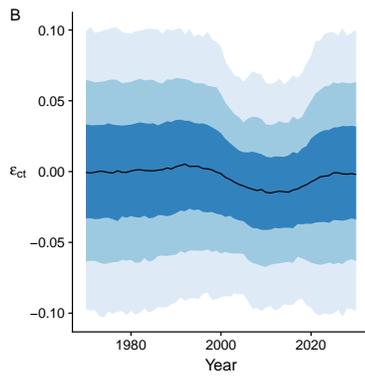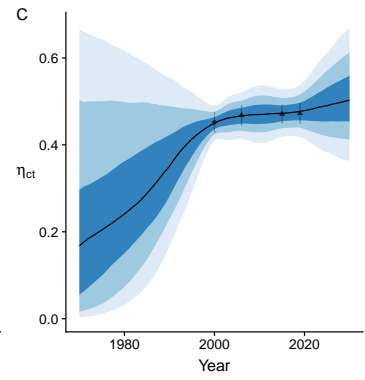

**Tuvalu**

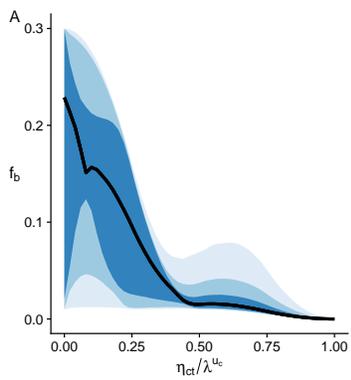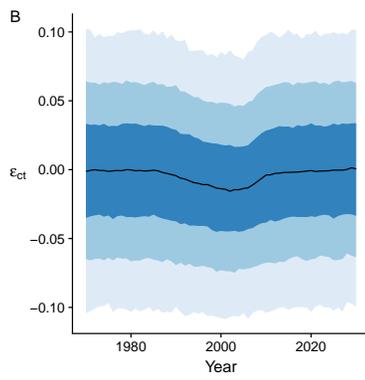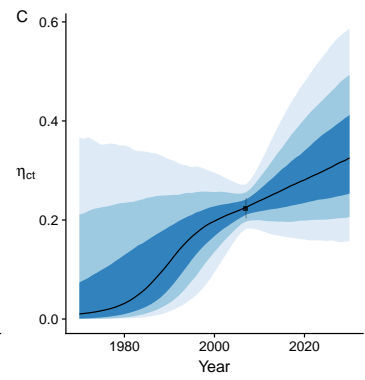

**Uganda**

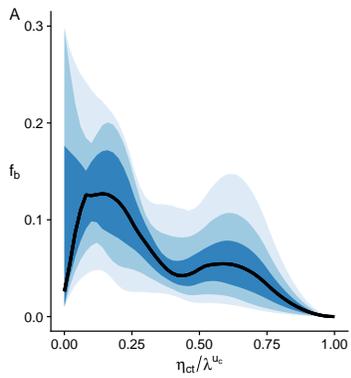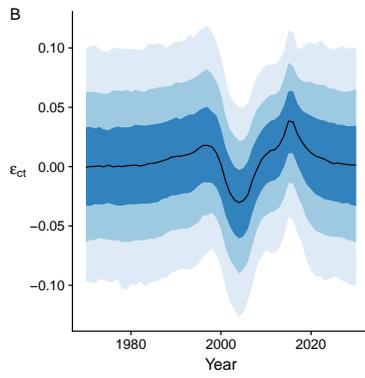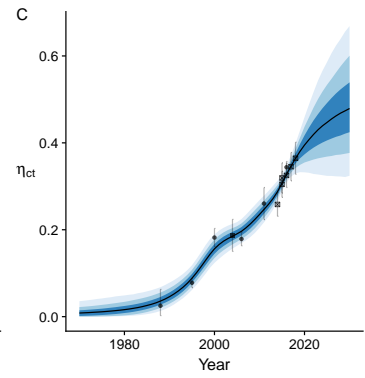

**Ukraine**

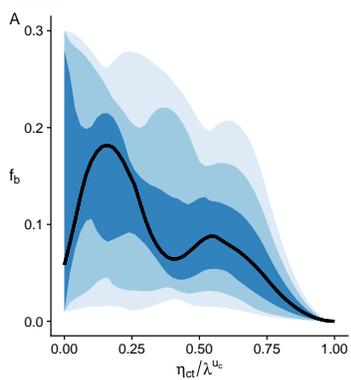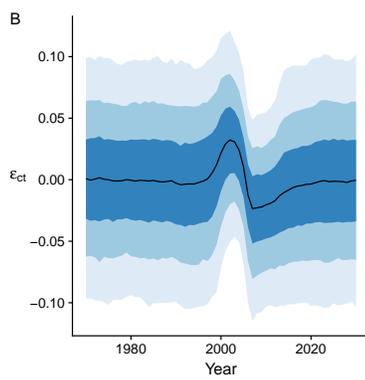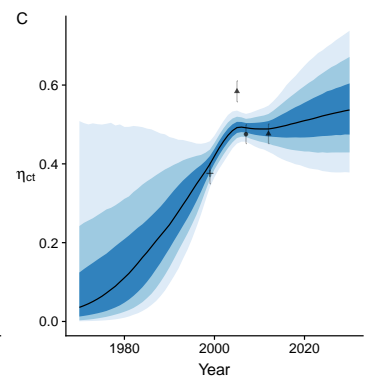

**United Republic of Tanzania**

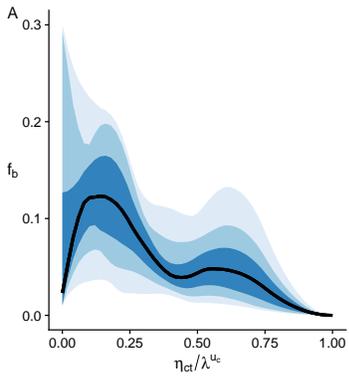
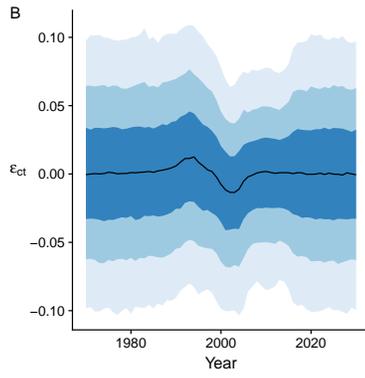
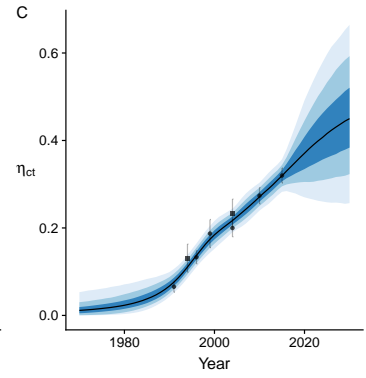

**United States of America**

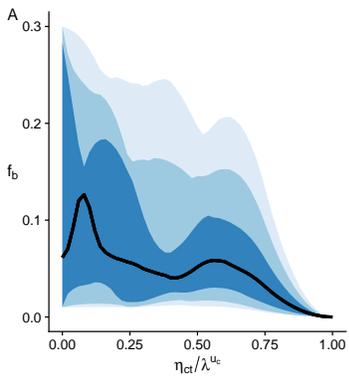
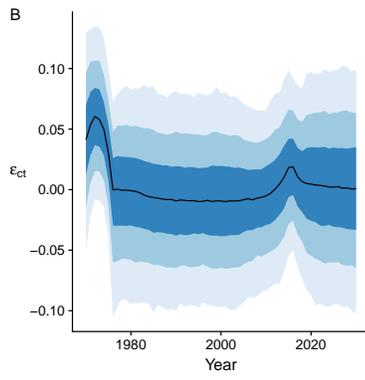
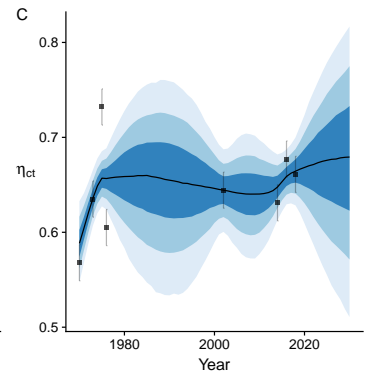

**Uruguay**

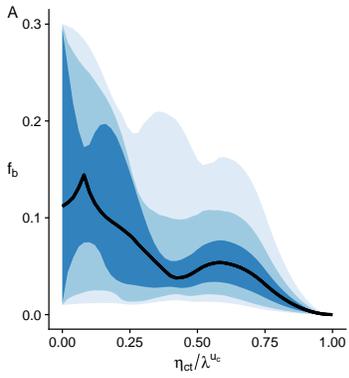
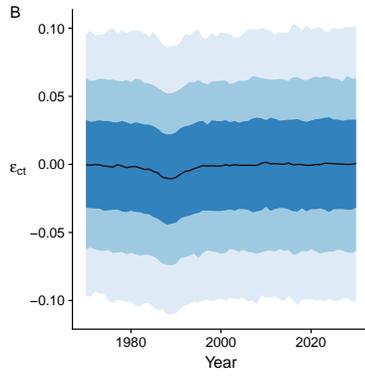
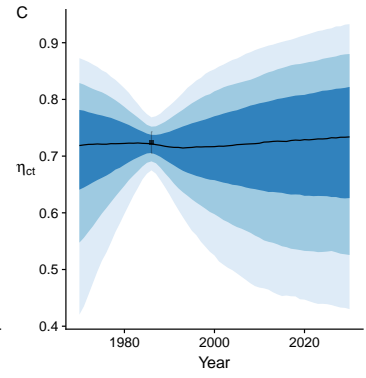

**Uzbekistan**

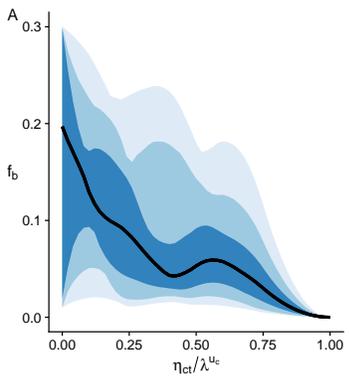
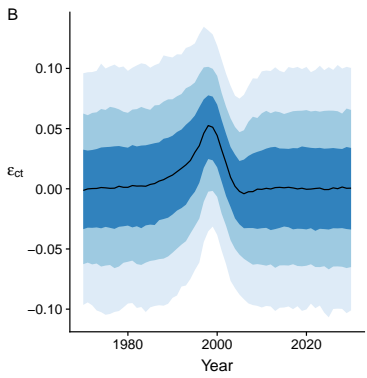
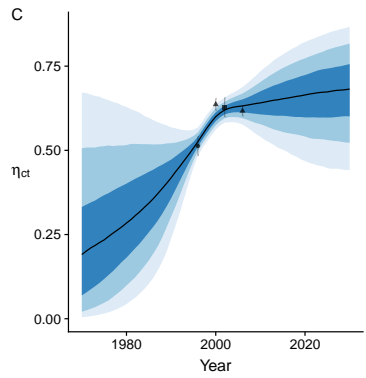

**Vanuatu**

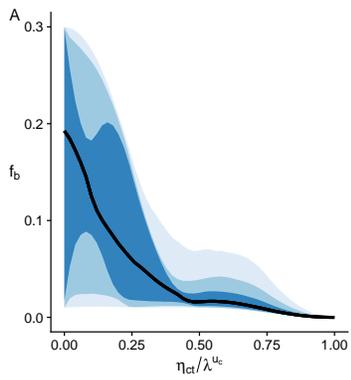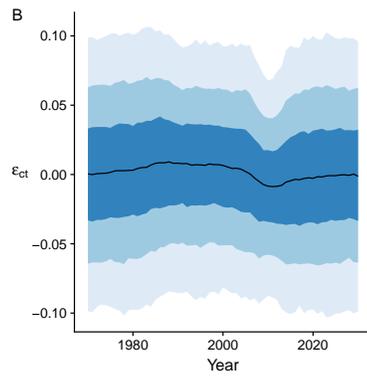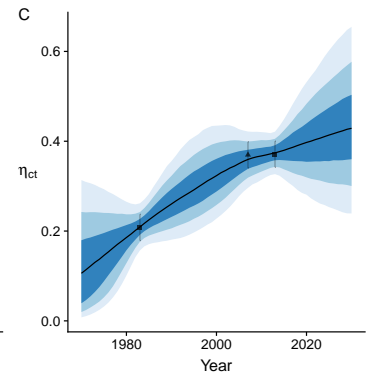

**Venezuela (Bolivarian Republic of)**

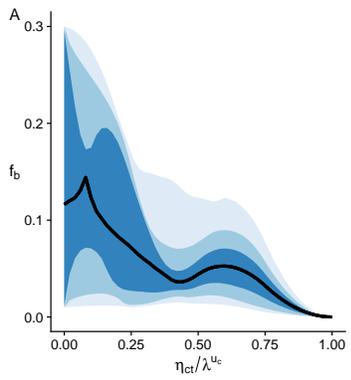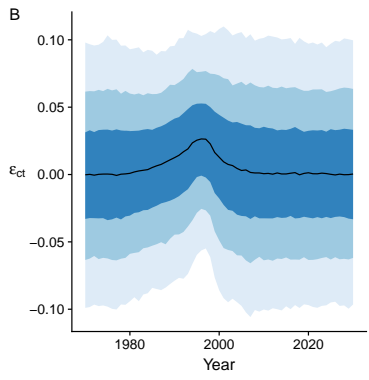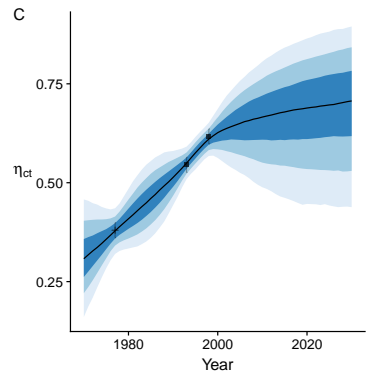

**Viet Nam**

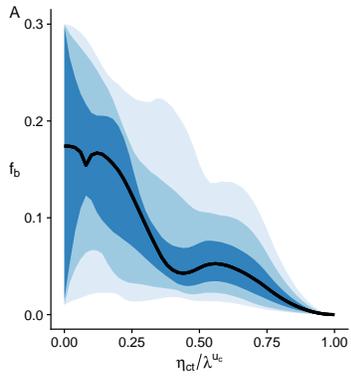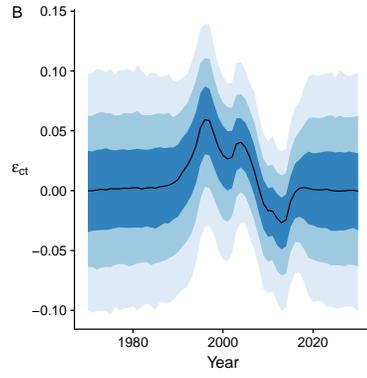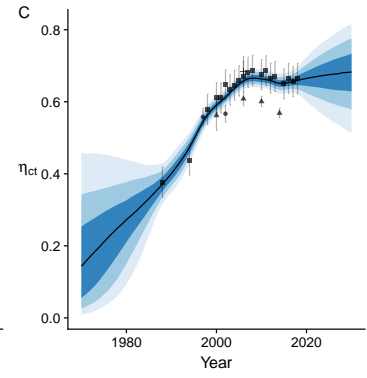

**Yemen**

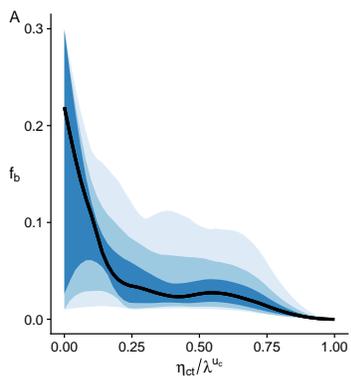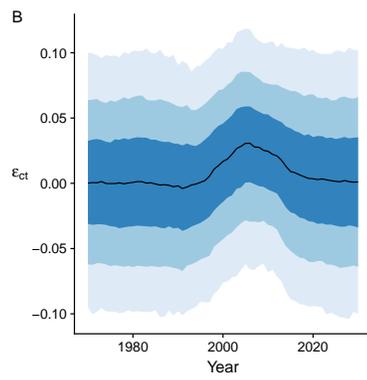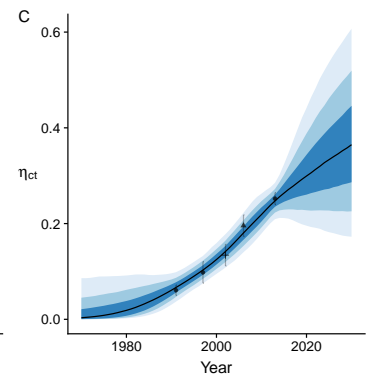

**Zambia**

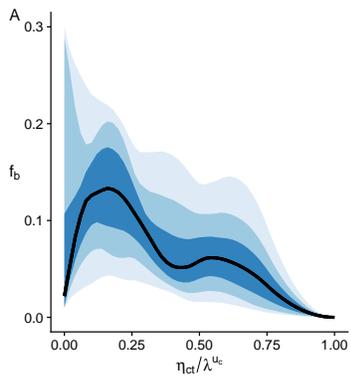 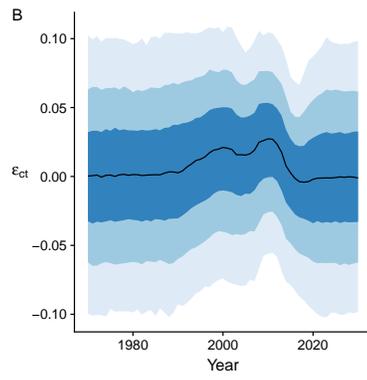 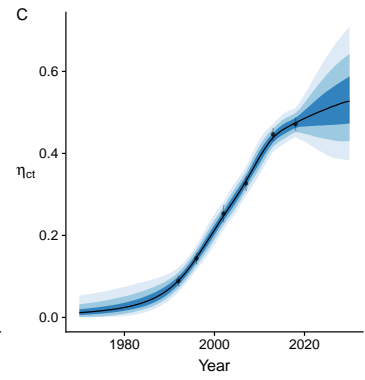

**Zimbabwe**

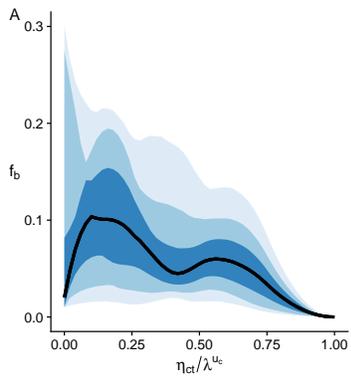 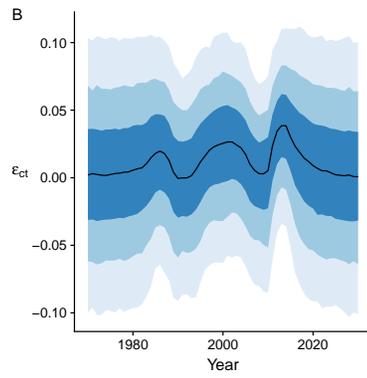 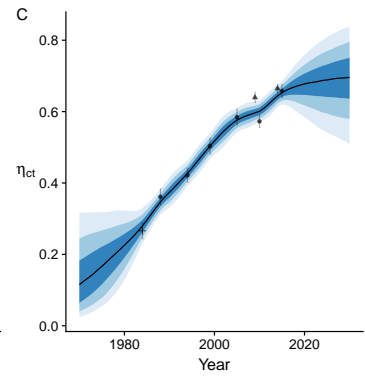



\end{document}